\documentclass[twocolumn]
{aastex631}
\let\splitbox\undefined
\usepackage{amsmath, amstext, natbib, verbatim, graphicx, xcolor, longtable, float
, hyperref, booktabs}
\usepackage[maxfloats=100]{morefloats}
\usepackage{adjustbox}

\newcommand{\OII}{\hbox{{\rm [O}\kern 0.1em{\sc ii}{\rm ]}}}
\newcommand{\NeIII}{\hbox{{\rm [Ne}\kern 0.1em{\sc iii}{\rm ]}}}
\newcommand{\OIII}{\hbox{{\rm [O}\kern 0.1em{\sc iii}{\rm ]}}}
\newcommand{\Hb}{\hbox{{\rm H}$\beta$}}
\newcommand{\Ha}{\hbox{{\rm H}$\alpha$}}
\newcommand{\NII}{\hbox{{\rm [N}\kern 0.1em{\sc ii}{\rm ]}}}
\newcommand{\SII}{\hbox{{\rm [S}\kern 0.1em{\sc ii}{\rm ]}}}

\newcommand{\HeI}{\hbox{{\rm He}\kern 0.1em{\sc i}}}
\newcommand{\HeII}{\hbox{{\rm He}\kern 0.1em{\sc ii}}}
\newcommand{\HII}{\hbox{{\rm H}\kern 0.1em{\sc ii}}}
\newcommand{\NeV}{\hbox{{\rm [Ne}\kern 0.1em{\sc v}{\rm ]}}}
\newcommand{\FeVII}{\hbox{{\rm [Fe}\kern 0.1em{\sc vii}{\rm ]}}}

\begin{document}

\title{\large \bf  
Extreme Emission Line Galaxies in CEERS Are Powered by Star Formation, not AGN}

\shorttitle{EELG AGN Contribution}
\shortauthors{Davis et al.}


\correspondingauthor{Kelcey Davis}
\email{kelcey.davis@uconn.edu}

\author[0000-0001-8047-8351]{Kelcey Davis}
\altaffiliation{NSF Graduate Research Fellow}
\affiliation{Department of Physics, 196A Auditorium Road, Unit 3046, University of Connecticut, Storrs, CT 06269, USA}
\affil{Los Alamos National Laboratory, Los Alamos, NM 87545, USA}

\author[0000-0001-5384-3616]{Madisyn Brooks}
\altaffiliation{NSF Graduate Research Fellow}
\affil{Department of Physics, 196A Auditorium Road, Unit 3046, University of Connecticut, Storrs, CT 06269, USA}

\author[0000-0002-1410-0470]{Jonathan R. Trump}
\affil{Department of Physics, 196A Auditorium Road, Unit 3046, University of Connecticut, Storrs, CT 06269, USA}

\author[0000-0003-0531-5450]{Vital Fern\'andez}
\affiliation{Michigan Institute for Data Science, University of Michigan, 500 Church Street, Ann Arbor, MI 48109, USA}

\author[0000-0001-6251-4988]{Taylor A. Hutchison}
\altaffiliation{NASA Postdoctoral Fellow}
\affiliation{Astrophysics Science Division, NASA Goddard Space Flight Center, 8800 Greenbelt Rd, Greenbelt, MD 20771, USA}

\author[0000-0003-2366-8858]{Rebecca L. Larson}
\altaffiliation{Giacconi Postdoctoral Fellow}
\affil{Space Telescope Science Institute, 3700 San Martin Drive, Baltimore, MD 21218, USA}

\author[0000-0003-1282-7454]{Anthony J. Taylor}
\affiliation{Department of Astronomy, The University of Texas at Austin, Austin, TX, USA}

\author[0000-0001-8688-2443]{Elizabeth J.\ McGrath}
\affiliation{Department of Physics and Astronomy, Colby College, Waterville, ME 04901, USA}

\author[0000-0001-6813-875X]{Guillermo Barro}
\affiliation{Department of Physics, University of the Pacific, Stockton, CA 90340 USA}

\author[0000-0002-6610-2048]{Anton M. Koekemoer}
\affiliation{Space Telescope Science Institute, 3700 San Martin Drive, Baltimore, MD 21218, USA}

\author[0000-0002-7959-8783]{Pablo Arrabal Haro}
\affiliation{Center for Space Sciences and Technology, UMBC, 5523 Research Park Dr, Baltimore, MD 21228 USA }
\affiliation{Astrophysics Science Division, NASA Goddard Space Flight Center, 8800 Greenbelt Rd, Greenbelt, MD 20771, USA}

\author[0000-0001-5414-5131]{Mark Dickinson}
\affiliation{NSF's National Optical-Infrared Astronomy Research Laboratory, 950 N. Cherry Ave., Tucson, AZ 85719, USA}

\author[0000-0001-8534-7502]{Bren E. Backhaus}
\affil{Department of Physics and Astronomy, University of Kansas, Lawrence, KS 66045, USA}

\author[0000-0001-7151-009X]{Nikko J. Cleri}
\affiliation{Department of Astronomy and Astrophysics, The Pennsylvania State University, University Park, PA 16802, USA}
\affiliation{Institute for Computational and Data Sciences, The Pennsylvania State University, University Park, PA 16802, USA}
\affiliation{Institute for Gravitation and the Cosmos, The Pennsylvania State University, University Park, PA 16802, USA}

\author[0000-0001-8519-1130]{Steven L. Finkelstein}
\affiliation{Department of Astronomy, The University of Texas at Austin, Austin, TX, USA}
\affiliation{Cosmic Frontier Center, The University of Texas at Austin, Austin, TX, USA}

\author[0009-0006-1252-206X]{Ananya Ganapathy}
\affiliation{Department of Physics \& Astronomy, Johns Hopkins University, Baltimore, MD 21218, USA}

\author[0000-0002-6386-7299]{Raymond C.\ Simons}
\affiliation{Department of Engineering and Physics, Providence College, 1 Cunningham Sq, Providence, RI 02918 USA}


\author[0000-0001-5758-1000]{Ricardo O. Amor\'{i}n} 
\affiliation{Instituto de Astrof\'{i}sica de Andaluc\'{i}a (CSIC), Apartado 3004, 18080 Granada, Spain}

\author[0000-0002-6219-5558]{Alexander de la Vega}
\affiliation{Department of Physics and Astronomy, University of California, 900 University Ave, Riverside, CA 92521, USA}

\author[0000-0001-9440-8872]{Norman A. Grogin}
\affiliation{Space Telescope Science Institute, 3700 San Martin Drive, Baltimore, MD 21218, USA}

\author[0000-0002-3301-3321]{Michaela Hirschmann}
\affiliation{Institute of Physics, Laboratory of Galaxy Evolution, Ecole Polytechnique Fédérale de Lausanne (EPFL), Observatoire de Sauverny, 1290 Versoix, Switzerland}

\author[0000-0003-3424-3230]{Weida Hu}
\affiliation{Department of Physics and Astronomy, Texas A\&M University, College Station, TX 77843-4242, USA}
\affiliation{George P. and Cynthia Woods Mitchell Institute for Fundamental Physics and Astronomy, Texas A\&M University, College Station, TX 77843-4242, USA}

\author[0000-0001-8002-2261]{Jarrett L. Johnson}
\affil{Los Alamos National Laboratory, Los Alamos, NM 87545, USA}

\author[0000-0001-9187-3605]{Jeyhan S. Kartaltepe}
\affiliation{Laboratory for Multiwavelength Astrophysics, School of Physics and Astronomy, Rochester Institute of Technology, 84 Lomb Memorial Drive, Rochester, NY 14623, USA}

\author[0000-0002-8360-3880]{Dale Kocevski}
\affiliation{Department of Physics and Astronomy, Colby College, Waterville, ME 04901, USA}

\author[0000-0003-1354-4296]{Mario Llerena}
\affiliation{INAF Osservatorio Astronomico di Roma, Via Frascati 33, 00078 Monte Porzio Catone, Rome, Italy}

\author[0000-0003-1581-7825]{Ray A. Lucas}

\affiliation{Space Telescope Science Institute, 3700 San Martin Drive, Baltimore, MD 21218, USA}

\author[0000-0001-6434-7845]{Madeline A. Marshall}
\affil{Los Alamos National Laboratory, Los Alamos, NM 87545, USA}

\author[0000-0001-9879-7780]{Fabio Pacucci}
\affiliation{Center for Astrophysics $\vert$ Harvard \& Smithsonian, 60 Garden St, Cambridge, MA 02138, USA}
\affiliation{Black Hole Initiative, Harvard University, 20 Garden St, Cambridge, MA 02138, USA}

\author[0000-0001-8940-6768]{Laura Pentericci}
\affiliation{INAF Osservatorio Astronomico di Roma, Via Frascati 33, 00078 Monte Porzio Catone, Rome, Italy}

\author[0000-0001-7968-3892]{Phoebe R. Upton Sanderbeck}
\affil{Los Alamos National Laboratory, Los Alamos, NM 87545, USA}





\begin{abstract}

    We present a spectroscopic study of photometrically identified extreme emission-line galaxies (EELGs) with observed-frame equivalent widths (EWs) $>5000$ \AA\ of either \Ha\ or \Hb\ + \OIII\ in the CEERS legacy deep field utilizing JWST NIRSpec spectroscopy from the CAPERS, RUBIES, THRILS and CEERS surveys. This master sample allows for performance tests of photometric selections and unveils what types of sources, either AGN or young star formation, were producing excessive ionizing radiation in the early Universe. We identify AGN through broad \Ha\ emission-lines and report 6 new broad-line AGN at $3.5<z<7$ identified by the deep ($\sim$8 hr) G395M THRILS survey. We investigate the photometrically selected EELGs in a color-color plot designed for ``Little Red Dot'' selection and demonstrate that it effectively removes AGN with non-extreme lines from the sample. EELGs with and without broad lines show similar optical line ratios. We compare emission-line morphology to EWs and continuum morphologies and find that \OIII\ morphology is more compact at higher EW. $\sim$10\% of photometrically selected EELGs have broad Balmer lines, jumping to 35\% in deep spectroscopy which indicates a significant fraction of photometrically selected EELGs may host AGN. However, many AGN selected as EELGs have incorrectly high photometric EWs. For sources with extreme emission-line EWs that pass our photometric criteria and host an AGN, we find that the narrow \Ha\ component dominates over the broad, especially in the highest-EW sources. This implies that even when an AGN is present, it does not dominate the extreme emission.

\end{abstract}

\section{Introduction}

The redshift ($z$) $\approx 4 \text{--} 9$ Universe is rife with sources exhibiting extremely strong rest-optical emission-line equivalent widths (EWs) \citep[e.g.,][]{Endsley2024, Mathee2023} of strengths rarer in nearby analogs \citep[e.g.,][]{Cardamone2009, Gimenez2025, vanderWel2011, Yang2017}. These lines are sufficiently strong in some cases that they can be identified from the JWST broadband photometry alone \citep{Davis2023, Endsley2024, Llerena2024, Mathee2023, Navarro-Carrera2024, Pirie2024,  Withers2023}.
Whether these early luminous sources, which occupied a Universe reionizing its intergalactic hydrogen, were driven by star formation (SF) in galaxies or active galactic nuclei (AGN) has been a subject of debate predating the launch of JWST \citep[e.g.,][]{Gardner2006, Gardner2023, Robertson2010, Madau2024, Atek2024}.

With JWST, we can begin to disentangle AGN at high-redshift from star-forming galaxies through direct observations of kinematic broadening in permitted optical emission lines \citep[e.g.,][]{Taylor2024, Hviding2025, Harikane2023, Kocevski2023}. 
AGN activity is clearest when the \Ha\ line is broadened with respect to forbidden lines, indicative of high orbital velocities in dense gas around AGN. We refer to systems with such broad \Ha\ or \Hb\ as broad-line AGN (BL AGN). \Ha\ emission from the broad line region of the AGN may drive the emission in early galaxies, or it may be coupled with narrow \Ha\ emission from the narrow-line region of an AGN and/or a young stellar population in a host galaxy. 

Extremely strong emission line strengths, when they are caused by SF, arise from short-lived O and B type stars that produce the necessary ionizing radiation to drive Balmer recombination emission lines \citep{Kennicutt2012}. Distinguishing between narrow \Ha\ emission lines arising from AGN or SF when a broad \Ha\ line is not detected is non-trivial at high-redshift \citep{ Mazzolari2024} and other methods must be employed to recover such narrow-line AGN (NLAGN).
However, commonly used optical strong line ratio diagnostics of ionizing sources (e.g., BPT \citep{Baldwin1981}, VO87 \citep{Veilleux1987}, and OHNO \citep{Backhaus2022}, etc.) are calibrated with low redshift sources ($z\lesssim1$), predating JWST data. Recent works \citep{Cleri2025, Scholtz2025} have shown that the observed line ratios of high-redshift AGN and star-forming galaxies are indistinguishable, due to lower metallicity, higher gas density, and higher ionization parameter in high-redshift star-forming galaxies.

The result is a potentially incomplete picture of AGN identification in emission-line galaxy populations. Deeper, higher resolution spectroscopy can recover additional galaxies with broad permitted emission-lines, but even a complete search for such lines misses AGN with faint or obscured broad-line regions. Morphology may be used to distinguish between extended emission associated with diffuse SF and compact emission associated with a central AGN, but this may be ambiguous with compact SF.



%



Here, we study sources with extremely strong \Ha\ and/or \Hb\ + \OIII\ emission, ``extreme emission line galaxies'' \citep[EELGs;][]{vanderWel2011}, focusing on galaxies whose emission lines are sufficiently strong to impact the broadband photometry, with observed-frame equivalent widths (EWs) $>5000$ \AA. Their strong lines can be explained either by galaxies undergoing recent SF or AGN activity.
We study the spectra of photometrically selected EELGs at $4<z<9$ from \citet{Davis2023} in the CEERS \citep[]{Finkelstein2025} field through JWST NIRSpec \citep[]{Jakobsen2022} surveys. We use this spectroscopy to probe whether AGN or SF dominate the extreme lines. 




In Section \ref{sec:Data}, we discuss the photometric and spectroscopic coverage of the EELG population derived from CEERS photometry in \citet{Davis2023} as well as our broad-line fitting and composite spectra methods. 
In Section \ref{sec:Selection}, we discuss the effectiveness of our photometric selection in terms of inferred redshift and emission-line EW as compared to the spectroscopic measurements.
In Section \ref{sec:Ident}, we discuss the BL AGN content of EELGs, present 6 new BL AGN, and confirm one previously identified. 
We demonstrate that broad \Ha\ lines are more effectively recovered from deeper, higher resolution NIRSpec spectroscopy.
In Section \ref{sec:Role}, we show that, when a BL AGN is present in an EELG, the narrow line dominates the emission pointing to a host SF galaxy driving the extreme emission. We investigate other methods of AGN identification including color-color selection, high-ionization line searches, and optical strong-line ratios to test the effectiveness of these selection methods. We also investigate morphology of the extreme emission lines through half-light radii ($R_e$) comparisons. 
We present the 7 BLAGN from the THRILS survey in the Appendix. 
We assume a flat $\Lambda$CDM cosmology consistent with measurements from Planck with $H_0 = 67.4$ $\mathrm{km} \mathrm{s}^{-1} \mathrm{Mpc}^{-1}$ and $\Omega_\mathrm{M}$ = 0.315 \citep{Planck2020}.
All logs are base ten and $z$ denotes object redshift.

\section{Data}
\label{sec:Data}

We present a spectroscopic study of photometrically identified EELGs in the CEERS legacy deep field. This study utilizes JWST NIRSpec spectroscopy from the CEERS, CAPERS, RUBIES, and THRILS programs. The surveys are described in Table \ref{tab:surv} in brief and in detail below. We additionally describe methods for broad-line fitting and stacking spectra.







\subsection{NIRSpec Spectroscopy}



We discuss data for the 181 EELGs with NIRSpec spectra. We required $>3 \sigma$ detections of \Ha\ and/or \Hb\ + \OIII\ as measured from \texttt{LiMe}, described in \citet{Ferandez2024}. 


\begin{table}[b]
\centering
\caption{EELGs with NIRSpec Coverage}
\label{tab:surv}
\begin{tabular}{lccc}
\hline
\hline
Survey & N (prism) & N (grating)  & $t_{exp}$ (hr)  \\
\hline
CEERS  & 9 & 5 & 0.86   \\
CAPERS  & 60 & -- &  1.58--4.74  \\
RUBIES  & 49 & 39 &   0.80  \\
THRILS  & -- & 19 &  8.36--8.85  \\
\hline
Total & 118 & 63 &  --  \\
\hline
\end{tabular}
\end{table}


Various NIRSpec surveys have successfully targeted 181 of the photometrically identified EELGs selected by \citet{Davis2023}.
Our final sample includes 60 prism targets from CAPERS \citep{Taylor2025, Kokorev2025}, 88 mixed prism and grating targets from RUBIES \citep{deGraaff2025}, 19 deep grating exposures from THRILS \citep{Hutchison2025}, and 14 spectra from the original CEERS survey \citep{ArrabalHaro23}. These spectra bring the spectroscopic coverage of the photometric sample up from 3\% when originally selected to now $16\%$ of EELGs. 

Surveys include data taken with both the prism low-resolution and medium-resolution grating spectral elements on NIRSpec. 
Prism spectra reach fainter continuum depths, but have lower spectral resolution of R$\sim 30-300$ that vary with observed-frame wavelength. Medium-resolution grating spectra have higher resolution $R \sim 1000$ and are more robust for emission line measurements, but are not as sensitive to continuum emission. To take advantage of all the available data, we utilize both grating and prism types in the combined stacks and data analysis. All surveys utilized 3-shutter slitlets and standard JWST reduction pipelines, described in detail in each survey's cited work.

For each EELG, we consider only one spectrum. We first prioritize data by deepest exposure time, regardless of instrument. 
We use grating data over prism when all three gratings are used to cover the full $0.97 - 5.27 \mu m$ wavelength range, and prism data when only the G395M grating is used.
This allows us to probe the full range of available wavelength coverage for stacking to be able to combine data from the different NIRSpec surveys. 
Each spectrum was visually inspected to ensure reliable identification of emission lines and remove spectra with artifacts (e.g., chip gaps or contamination from nearby sources). 
Spectroscopic redshift measurements were derived from peak locations of \Ha\ and/or \OIII\ $\lambda$5008. 
We give brief details for each survey below.

\noindent
\textbf{\underline{CEERS}} 
The complete description of the NIRSpec spectroscopic data reduction from the original Cosmic Evolution Early Release Science Survey (CEERS) [program ID 1345, PI: Steven L. Finkelstein]\citep{Finkelstein2024} coverage can be found in \citet{ArrabalHaro2023}. The original NIRSpec coverage of CEERS included prism spectroscopy and medium-resolution grating spectra with coverage from the G140M, G235M, and G395M filters which span observed-frame wavelengths of $0.97 - 5.27 \mu m$. CEERS included prism observations over 8 NIRSpec pointings with 0.86 hr exposure times \citep{Fujimoto2023} and 6 pointings of medium-resolution grating data, also with 0.86 hrs of exposure time. We recover 9 EELGs from the prism resolution and 5 EELGs from the grating resolution spectroscopy after removing duplicates in the spectroscopy sample.

\noindent
\textbf{\underline{CAPERS}} The CANDELS-Area Prism Epoch of Reionization Survey (CAPERS) [program ID 6368, PI: Mark Dickinson] target selection and reduction are described in \citet{Kokorev2025, Taylor2025}. CAPERS spectroscopy includes prism low-resolution data with 7 NIRSpec pointings targeting sources in the CEERS legacy deep field with 3 MSA configurations per pointing. CAPERS targets included coverage from 1-3 configurations and had exposures of 1.58--4.74 hours depending on the total number of configurations covering a given target. We recover 60 EELGs from the CAPERS spectroscopy after removing duplicates in the spectroscopy sample.

\noindent
\textbf{\underline{RUBIES}} We include NIRSpec data from the Red Unknowns: Bright Infrared Extragalactic Survey (RUBIES) [program ID 4233, PIs: A. de Graaff \& G. Brammer] \citep{deGraaff2025}, utilizing reductions from \citet{Taylor2024}. RUBIES included 6 NIRSpec pointings in the EGS field following the legacy CEERS footprint. The program observed with G395M, covering an observed frame wavelength range of $2.87 - 5.27 \mu m$ as well as prism-resolution spectroscopy in each pointing. Sources had exposure times of 0.8 hours. 
RUBIES targeted sources with both prism and G395M coverage. 
We identify 49 prism and 39 G395M EELGs after removing duplicates within the RUBIES survey and duplicates with other surveys.

\noindent
\textbf{\underline{THRILS}} The High-(Redshift+Ionization) Line Search (THRILS) [program ID 5507, PIs: T. Hutchison \& R. Larson] program \citep{Hutchison2025} utilized only G395M but with 8.36-8.85 hr exposures. THRILS EELG targets were selected from the EELG catalog in \citet{Davis2023}. 
The THRILS program produced 19 high-quality spectra for EELGs.
THRILS targeted three of the EELGs with presumed incorrect photometric redshifts of $z<1$, THRILS 49865, 105963, and 24975, and confirmed them as high-z EELGs as described in Section 3.2. We recover 19 EELGs from the THRILS spectroscopy after removing duplicates in the spectroscopy sample.






\subsection{Photometry}

 A complete description of the NIRCam photometric data reduction from CEERS \citep{Finkelstein2025} is described in \citet{Bagley2023, Cox2025}. The broad-band photometric filters utilized in the CEERS observing design as of this publication are described in Table \ref{tbl:filters}. Photometric data was utilized in the identification of EELGs, as described in Section 3.1. 
 Photometric redshifts are derived with \texttt{EAZY} \citep{Brammer2008} and presented in \citet{Finkelstein2024}. 

\begin{table}[h!]
   \begin{centering}
        
    \begin{tabular}{cccc}
          \hline
         Filter & Depth (mag) & Width ($\mu$m) & Band Center ($\mu$m) \\ [0.5ex] 
         \hline\hline
         F115W & 29.15 & 0.225 & 1.154 \\ 
         F150W & 28.90 & 0.318 & 1.501 \\
         F200W & 28.97 & 0.461 & 1.990 \\
         F277W & 29.15 & 0.672 & 2.786 \\
         F356W & 28.95 & 0.787 & 3.553 \\
         F410M & 28.40 & 0.436 & 4.092 \\
         F444W & 28.60 & 1.024 & 4.421 \\ [1ex] 
         \hline
    \end{tabular}

    \end{centering}
    \caption{Photometric data in CEERS utilizing NIRCam photometric filters. Depths in AB Mag defined in \citet[]{Oke1983} are for a 5$\sigma$ point source.}
    \label{tbl:filters}
\end{table}

\subsection{Broad Line Fitting}

For a positive identification of a broad-line, we require that the broad component of the line have a $>3 \sigma$ detection with a full-width half-max (FWHM) of $>1000$ km/s. We follow the \Ha\ broad line fitting procedure described in \cite{Brooks2025}. In brief, we model the \Ha\ emission line profile with two Gaussian functions: one broad and one narrow. We also model the \NII$\lambda\lambda$6550, 6585 doublet with a fixed amplitude ratio of 1:2.94, and fix the \NII\ line width and redshift to the narrow \Ha\ component. We implement the following priors in our fitting routine: positive line fluxes for both the narrow and broad line component, $\rm{FWHM_{narrow}} < 700~km~s^{-1}$ and $\rm{FWHM_{broad}} > 700~km~s^{-1}$. We additionally let the \Ha\ line center vary within $\rm{250~km~s^{-1}}$ of the previously-found line center described in Section 2.1.1.

To calculate BH masses, we use the following prescription from \cite{ReinesVolonteri2015}: \begin{multline}
    M_{\rm{BH}} = 10^{6.57} \times \left(\frac{L_{\mathrm{H} \alpha, \rm{broad}}}{10^{42} \mathrm{~ergs} \mathrm{~s}^{-1}}\right)^{0.47} \\ 
    \times \left(\frac{\mathrm{FWHM}_{\mathrm{H} \alpha, \rm{broad}}}{10^{3} \mathrm{~km} \mathrm{~s}^{-1}}\right)^{2.06} M_{\odot}
\end{multline}
where $L_{\rm{H\alpha,\rm{broad}}}$ is the luminosity of the broad \Ha\ component and $\rm{FWHM}_{\rm{H\alpha, broad}}$ is the FWHM of the \Ha\ broad component. We use the full flux and FWHM posteriors from the best-fit broad \Ha\ Gaussian component to compute our BH masses, which allows us to propagate the uncertainties from the \Ha\ emission-line fit and results in a full BH mass posterior distribution. We note that the BH mass errors reported in the Appendix are the observational uncertainties and do not include the $\sim$0.5~dex uncertainty that accounts for the intrinsic scatter in single-epoch BH mass recipes \citep{Shen2010}. The masses reported here are in agreement with masses reported in Ganapathy et al (in prep).

\subsection{Stacked Spectra}
We stack the spectra from our EELG sample after duplicates have been eliminated. We stack the spectra of the EELGs after duplicates have been removed for a deeper study of average population properties and recover weak features undetected in individual spectra. Stacked spectra include both medium-resolution grating spectra and low-resolution prism spectra from NIRSpec. Prism resolution varies as a function of wavelength, reaching maximum at the reddest end of the spectrum at a resolution of $ \lambda / \Delta \lambda  =300$ and a minimum resolution near the middle of the detecting range of $ \lambda / \Delta \lambda = 31$. 
The medium-resolution grating spectra have a resolution of $ \lambda / \Delta \lambda = 600$--$1000$, and we degrade these grating spectra to a lower resolution of $ \lambda / \Delta \lambda = 300$ when stacked with the prism spectra to ensure we are stacking comparable data.

We combine the degraded medium-resolution grating spectra and prism spectra by interpolating them onto a common wavelength grid and calculating a median pixel flux. 
The interpolated spectra were normalized to their \Ha\ integrated flux. Sources without \Ha\ are more difficult to search for emission line broadening associated with AGN activity and were not included in stacked spectra. These high-z stacks will be the subject of future work.
Errors on the composite spectra are reported as statistical uncertainties through standard error as $\sigma / \sqrt{N}$. 

\begin{figure*}[t]
    \centering
    \includegraphics[width = 1\linewidth]{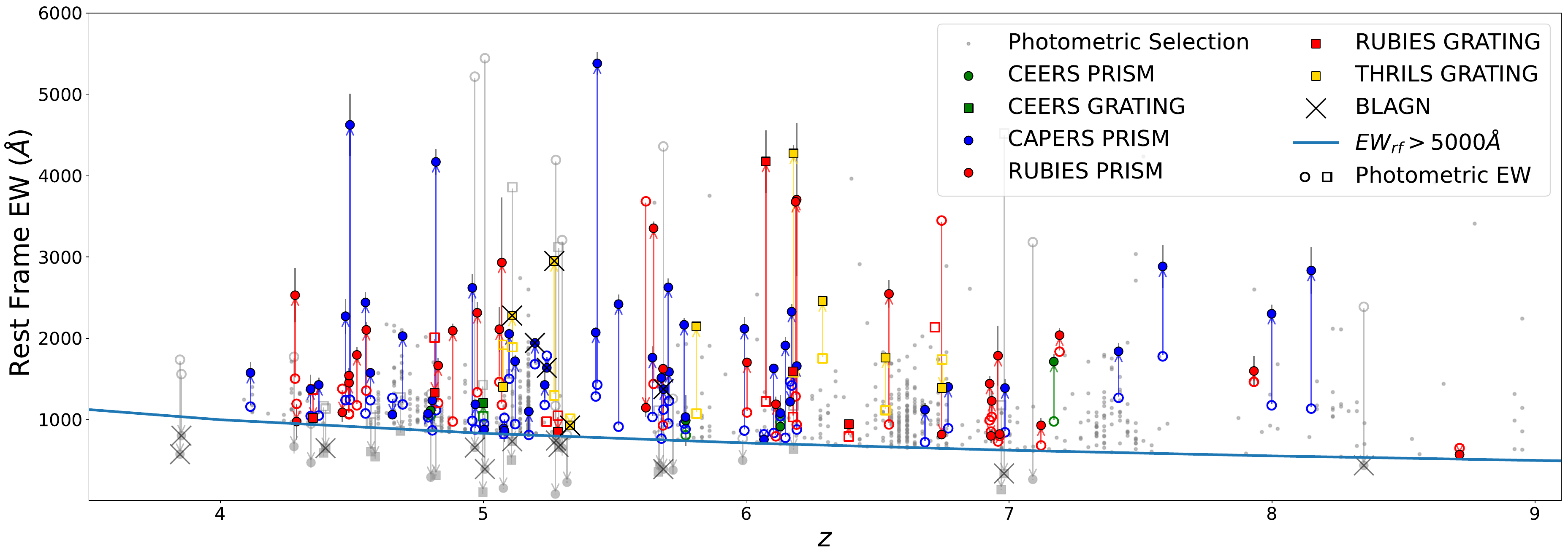}
    \caption{
    Selection criteria for the photometric sample. The original photometric selection required an observed-frame EW of 5000\AA\ for either \Ha\ or \Hb\ +\OIII, indicated here in the rest-frame by a solid blue line. Spectroscopically targeted sources are plotted with spectroscopic redshifts while sources with only photometric coverage are plotted with photometric redshifts. The photometric selection is indicated in gray. Spectroscopic EWs are indicated by solid markers while photometric EWs are indicated with open markers. We note that some spectroscopic measurements remove some photometrically selected EELGs out of the detection threshold, especially prevalent in EELGs that host AGN. The objects with inconsistent spectroscopic and photometric EWs are discussed in Section 3.3.  }
    \label{fig:samplefig}
\end{figure*}

Emission lines are fit from the composite spectra by fitting single Gaussian profiles with fixed  line centers. For \Hb\ + \OIII\, we fit three Gaussians and keep the \OIII$\lambda5008$/\OIII$\lambda4960$ line ratio set at 2.98 while requiring a shared line width. We fit the \SII\ doublet as two Gaussians with a shared line width and the \NeIII\ triplet and blended H$\epsilon$ as four Gaussians with shared line width, removing contribution from potentially blended H$\epsilon$ to the \NeIII lines. We fit the \OII\ doublet as a single Gaussian because it is completely blended at the NIRSpec spectral resolution. 
We integrate these Gaussians to report line fluxes. 
Signal to noise ratios (SNRs) are reported as the ratio of integrated flux to integrated error.

\section{Effectiveness of Photometric Extreme Emission Line Galaxy Identification}
\label{sec:Selection}

We interpret the 1165 photometrically selected EELGs from \citet{Davis2023} through the 181 EELGs with spectroscopic coverage to test the effectiveness of our EW and photometric redshift calculations. 

\subsection{Photometric EELG Sample Selection}

We investigate the spectroscopic coverage of a photometrically selected sample of EELGs in CEERS from \citet{Davis2023} to test selection performance. 
\citet{Davis2023} selected EELGs by searching for flux excess corresponding to \Ha\ or blended \Hb\ + \OIII\ in one to two of the four reddest NIRCam filters covering CEERS: F277W, F356W, F410M, and F444W. The \Ha\ and \Hb\ + \OIII\ emission lines fall in these filters within an approximate redshift range of $4<z<9$. The sample was selected in the observed frame to minimize reliance on SED template modeling introduced by redshift fitting. The final sample was required to have an inferred EW$_{\rm phot}$ $>5000$ \AA\ in the observed frame, with EW defined as:

\begin{equation}
    EW = \left( \frac{F_\nu - C_\nu}{C_\nu} \right) \Delta \lambda 
    \label{eqn:ew}
\end{equation}

\noindent where $F_{\nu}$ is the flux density in a filter that includes a putative emission line, $\Delta \lambda$ is the width of the photometric filter capturing the emission line, and $C_\nu$ is the continuum flux density. 

The continuum was defined as the flux density equal to the outlier-rejected mean flux ($>1\sigma$ from the mean total flux) from the NIRCam filters described in Table \ref{tbl:filters}. We also exclude the NIRCam filters capturing \Ha\ and/or \Hb\ + \OIII\ from the mean flux calculation. This simple definition assumes that the continuum is flat in $F_\nu$. We require that a line fitting the continuum filters from an SED with flux in $F_{\nu}$ (nJy) and wavelength in $\mu$m have slope $>0$ to avoid sources with reddened continua. The selection and sample properties are described in detail in \citet{Davis2023} and the full photometric and spectroscopic sample is shown in Figure \ref{fig:samplefig}.

\subsection{Redshift Recovery}

\begin{figure}[h!]
    \centering
    \includegraphics[width=1\linewidth]{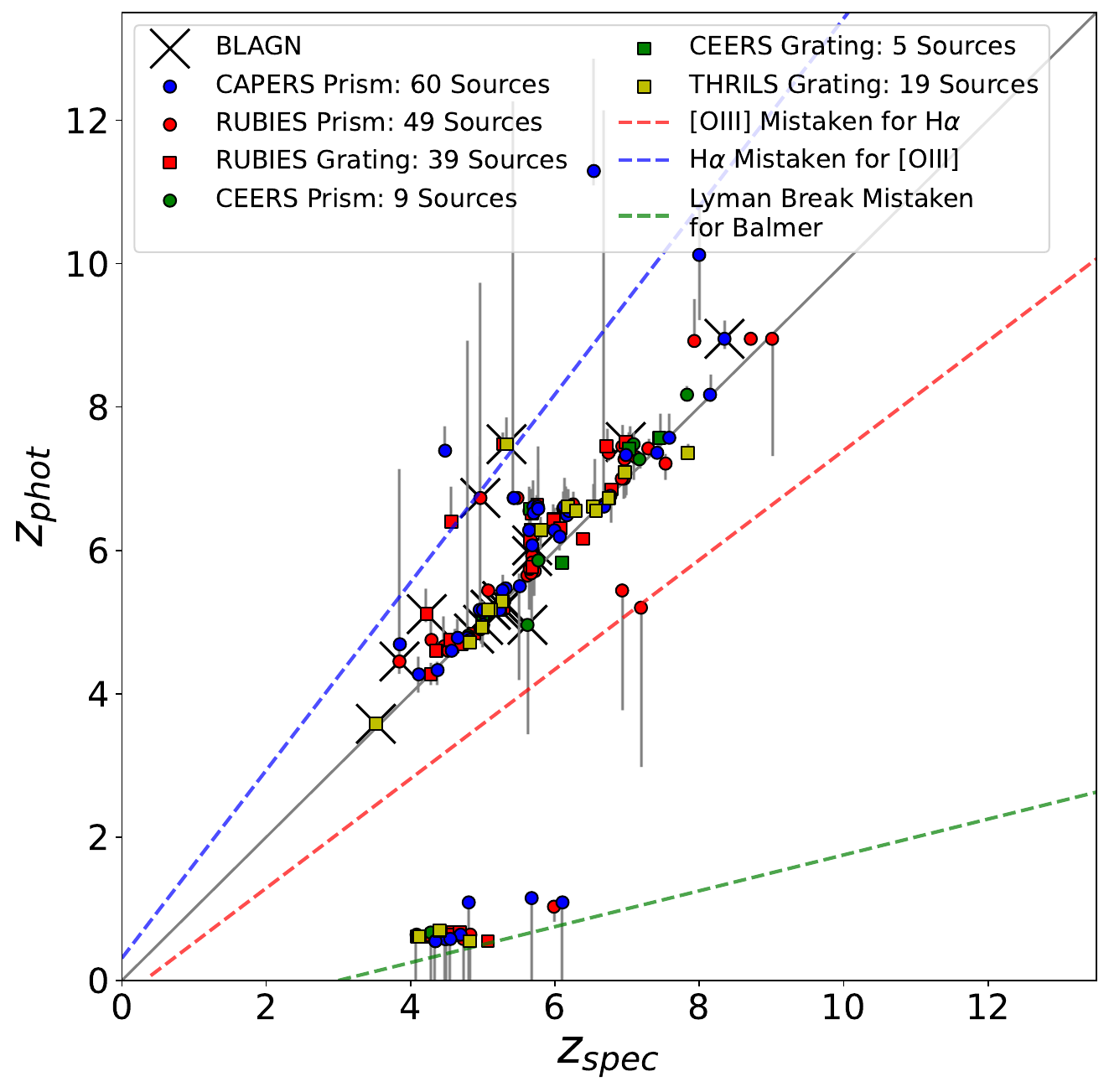}
    \includegraphics[width=1\linewidth]{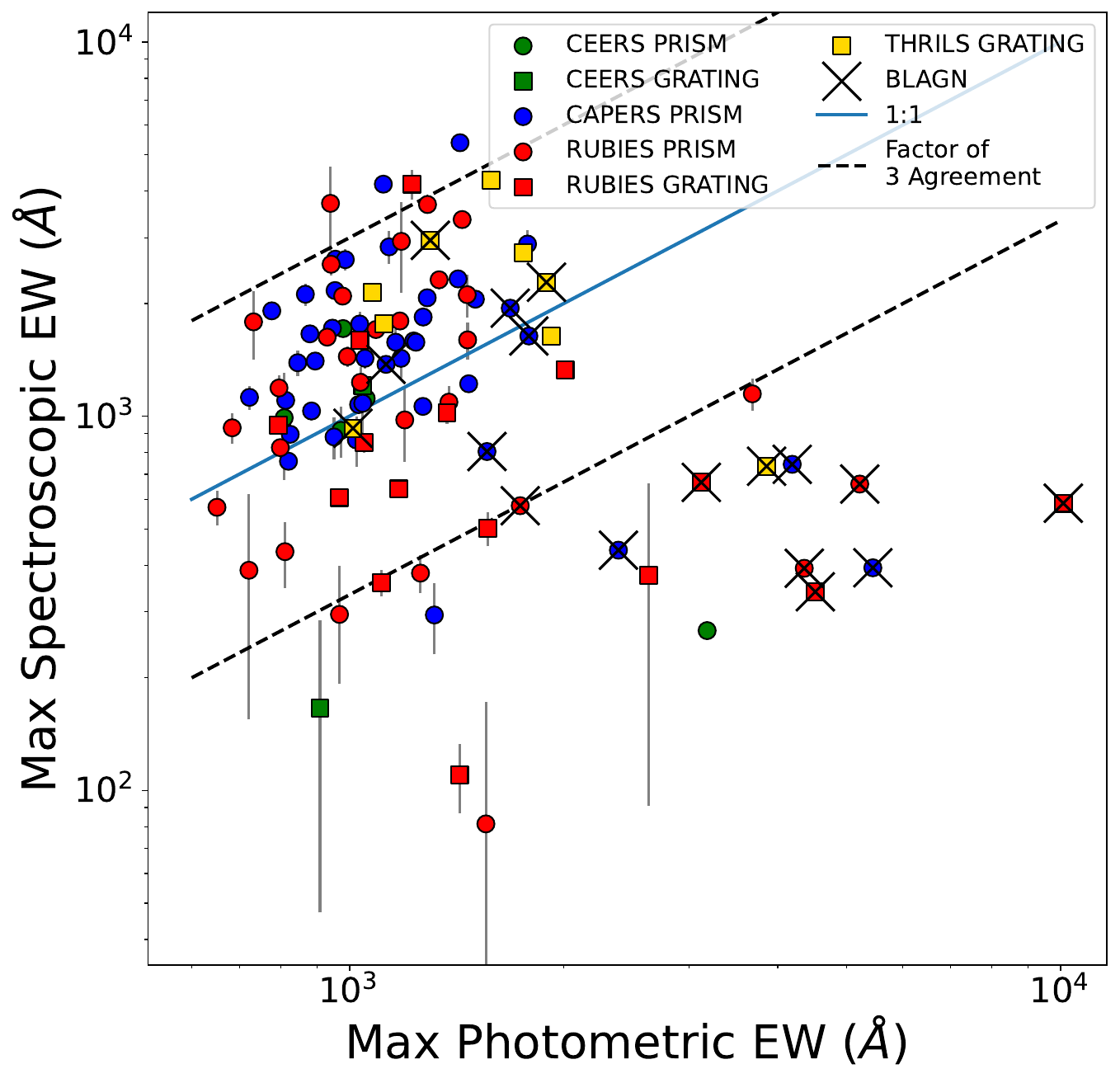}
    \caption{ \textbf{Top:}
    Comparison between spectroscopic and photometric redshifts for the EELG sample with confident \Ha\ and/or \Hb\ + \OIII\ fits. Error bars represent 68 \% confidence ranges from \texttt{EAZY}. The solid gray line indicates 1:1 agreement. The dashed blue line traces sources with redshifts consistent with confusion of \Ha\ for \OIII. The dashed red line traces sources with \OIII\ mistaken for \Ha. The dashed green line traces sources with Lyman breaks mistaken for Balmer breaks. \textbf{Bottom:} Comparison between photometric and spectroscopic EWs. We note general agreement within a factor of 3, consistent with agreement noted in the original selection presented in \citet{Davis2023} with the exception of BL AGN. Many of the AGN have overestimated EWs due to confusion from the blue UV and red optical continua, as discussed in Section 3.3.  }
    \label{fig:phot_ew}
\end{figure}

The photometric EELG sample comprised $1165$ sources. Of these, $\sim 30\%$ had photometric redshifts that disagreed with the redshifts implied by the flux excess selection. These represent sources that may be lost in traditional SED fitting selections of EELGs. Most had suspected \Ha\ emission in F277W and \Hb\ + \OIII\ in F356W, associated with a redshift range of $\sim 4<z<5$. However, these EELGs had photometric redshifts of $z<1$, representing a catastrophic difference between emission-line implied and SED-fitted redshifts. 

\citet{Davis2023} predicted that the lack of coverage from medium-band photometric filters in CEERS contributed to this discrepancy and that the galaxies were true EELGs. In the original CEERS NIRSpec coverage, two such sources were targeted and confirmed to be at a redshift consistent with our photometric selection. The expanded coverage from other NIRSpec surveys adds spectroscopy for 26 more EELGs with incorrect SED-fit redshifts and recovers them as true, high-z EELGs. This implies pre-JWST low-$z$ derived SED templates are insufficient to describe all $4<z<9$ galaxies and gives added confidence to the simple flux excess selection method.

Figure \ref{fig:phot_ew} shows the spectroscopic and photometric redshift agreement. We find that most EELGs hug the line of agreement between spectroscopic and photometric redshifts. 
We include 3 dashed lines in the top panel of Figure \ref{fig:phot_ew} to represent possible sources of confusion in template fitting. The dashed red line indicates redshift solutions associated with mistaking an extreme \OIII\ line for an \Ha\ line. Only 3 EELGs fall on this line within their photometric redshift error, indicating that this emission line confusion is rarer but possible. The dashed blue line indicates redshift solutions associated with confusing an extreme \Ha\ line for \OIII. Eight of our EELGs fall on this line within their error, notably including a few BL AGN (see Section 4, Appendix). 
The lack of medium-band coverage that contributes to these outliers may also contribute to redshift solution degeneracy for sources with extreme \Ha\ and \OIII\ that is either sub-extreme or not recovered in the photometry. 

We also include a green dashed line in Figure \ref{fig:phot_ew} that traces the redshift solutions for galaxies with Lyman breaks mistaken for Balmer breaks. In the case of EELGs in CEERS, when no medium-band photometric filter probes the continuum between \Ha\ and \Hb\ + \OIII, the extreme emission lines are blended in such a way that mimics a 2 $\mu m$ stellar bump in a low-z galaxy. Inconveniently, the Balmer break in a low-z galaxy with a stellar bump lies where a high-z EELG has a Lyman break in an SED. This introduces a chronic problem for SED fitting through degeneracy between a low-z galaxy with a stellar bump and a high-z EELG. In \citet{Davis2023}, we noted that the flux associated with these 2$\mu m$ stellar bumps was too strong to represent a true stellar bump and we confirm here that we correctly identified the galaxies as high-z EELGs. This simple flux-excess EELG selection correctly identifies many EELGs, as confirmed by spectroscopy, that are mischaracterized by SED fitting. 

\subsection{Equivalent Width Recovery}

Our simple flux-excess selection assumed that galaxies at $4<z<9$ could be represented by continua flat in $F_\nu$. 
The simple flat-continuum prior is effective for correctly identifying emission-line redshifts (indeed, it is more effective than SED fitting, as discussed in Section 3.2), but can incorrectly estimate EWs for non-flat continua.
The spectroscopic and photometric EWs for the maximum of either \Ha\ or \Hb\ + \OIII\  are compared in Figure \ref{fig:phot_ew}. 

    

The most dramatically overestimated photometric EWs are for BL AGN.
We identify AGN in the sample through searches for broad \Ha\ lines described in Sections 5.1 and 5.2. In the case that a BL AGN is present, the source typically also had a red optical continuum. Our method in \citet{Davis2023} assumed that AGN with red optical continuum would also have red UV continua and successfully removed these sources. However, this assumption is challenged by the red optical and blue UV continua in ``Little Red Dots'' (LRDs; \citet{Mathee2024, Kocevski2023}) at $z>4$. 
These LRDs had extreme-emission photometric filters falsely boosted by the red optical continuum and so their photometric EWs were over-estimated in contrast to their lower spectroscopic EWs. However, many still satisfy our spectroscopic EW threshold and can be classified as EELGs.

This assumption of a flat $F_\nu$ continuum also underestimates emission-line EWs in galaxies with extremely blue UV continua. In these cases, the extreme emission-line EWs were underestimated and the sources have more extreme emission-lines than the strengths implied photometrically.

We investigate these EELGs under the lens of our photometric selection criteria. Figure \ref{fig:samplefig} highlights the difference between photometric and spectroscopic EWs. We find that BL AGN are the most impacted by comparison to spectroscopic EWs with just 35\% of EELGs with broad lines still passing the selection criteria compared to 82\% of EELGs without evidence for broad Balmer lines. Our original sample selection identified galaxies from their maximum implied photometric EW of either \Ha\ or \Hb\ + \OIII. In the case that a galaxy has spectroscopic coverage from just G395M, as is the case for many of our sources, we can only compare to \Ha\ spectroscopic EWs. This further reduces our sample of high-confidence, high-EW spectroscopic sources and can introduce under-estimations of spectroscopic EWs. While most EELGs generally have photometric and spectroscopic EWs that agree, the photometric estimates tend to slightly underestimate the spectroscopic EWs in EELGs without AGN and catastrophically overestimate the spectroscopic EWs in the case that a BL AGN is present. This creates a high-EW tail in the photometry that is not present in the distribution of spectroscopic EWs in Figure \ref{fig:samplefig} driven by the BL AGN.




\section{BL AGN Identification }
\label{sec:Ident}

We identify AGN contribution to the sample first by searching for broad Balmer emission lines which imply high-velocity orbits of ionized gas near a supermassive black hole. These sources are identified through published searches when available and \Ha\ broad-line fitting for new data. We discuss the completeness of these selections in the context of their observations. Our final sample identifies 23 EELGs as BL AGN, or about 10\% of EELGs with spectroscopic coverage.
We note, as detailed in Section 3.3, that only 12 
of these EELGs hosting BL AGN have a spectroscopic EW that passes our $>5000$ \AA\ observed-frame cut which we refer to hereafter as EELG-AGN. 





\subsection{Previous Identification of BL AGN in the EELG Sample}

To identify broad-line AGN in our sample, we first turn to existing searches for broad-lines in the CEERS \citep{Harikane2023, Kocevski2023} and RUBIES \citep{Taylor2024, Kocevski2024} NIRSpec data. 
We find 9 BL AGN from \citet{Taylor2024} met our photometric criteria for EELGs: RUBIES 42232, 37124, and 37032 with prism coverage and RUBIES 55604, 50052, 42046, 60935, 28812, and 6411 with G395M coverage identified in \citet{Taylor2024}, 6 of which are also identified in \citet{Hviding2025}. We also find overlap with the sample of LRDs in \citet{Kocevski2024} that identifies broad \Ha\ emission in RUBIES 927271 and 926125. 

Both sources discussed in \citet{Kocevski2023}, CEERS 746 and 2782, which were selected for their extreme photometric EWs, met our photometric EELG criteria and were included in our sample. 
We note that CEERS 2782 was also observed as RUBIES 50052, with better centered coverage of the same target as discussed in \citet{Taylor2024}. We include only the CEERS 2782 spectra in our data to utilize the larger wavelength coverage in the prism-resolution spectroscopy. 
\citet{Harikane2023} spectroscopically selected BL AGN in the original CEERS NIRSpec coverage, three of which we identified as EELGs, CEERS 746 and 2782, discussed in \citet{Kocevski2023}, and CEERS 397. 
There are an additional 7 BLAGN identified from the CAPERS prism spectroscopy (Taylor et al. in prep.).



\subsection{Newly Identified BL AGN in This Work}

The THRILS survey \citep{Hutchison2025} targeted 19 EELGs identified in our parent sample. Of these, 7 satisfy our criteria for broad \Ha\ emission associated with AGN activity described in Section 2.2.1. RUBIES 50812/THRILS 46155 was previously identified as a BL AGN in \citet{Taylor2024} but the 6 other BL AGN are presented here for the first time. Their properties and emission-line fits are described in Table \ref{table:thrils} in the Appendix. This brings our total BL AGN count in the photometric EELG sample to 23 unique BL AGN. 

When revisiting the photometrically derived EWs, we find that 2 of these 7 THRILS BL AGN have spectroscopic EWs that satisfy our photometrically defined criteria of EW $>$ 5000 \AA.  THRILS 44774, 46155, and 24975  did not satisfy these criteria. Of the 23 BL AGN across the full spectroscopy sample, 12 total EELGs with broad Balmer emission lines pass our spectroscopic EW threshold.

\subsection{Spectral Resolution Selection Effects}

This lower resolution prism spectroscopy is ideal for probing UV continuum emission in high-z galaxies \citep{ArrabalHaro2023} and fills in the wavelength coverage missed by G395M-only surveys.
However, it is difficult to recover broad \Ha\ components at the resolutions possible with prism data. 

Prism has a variable spectral resolution that reaches a minimum of R$\sim31$ at 1.2 $\mu m$ and R$\sim300$ near the red-most detector edge at $5.3\mu m$. We take the instrumental FWHM to represent the minimum velocity of a hypothetical detectable broad line. Taking $R = \Delta \lambda / \lambda$ and assuming the broadening is purely due to Doppler shifted gas kinematics, we can express the instrumental FWHM as $\rm{FWHM}_{inst} = \rm{c}/R$ where c is the speed of light in a vacuum. This implies a minimum detectable velocity at the lowest prism resolution of $\sim10,000 \rm{km}/\rm{s}$ and  $\sim700 \rm{km}/\rm{s}$ at the highest resolution red end. 
At our sample's lower range, $z = 4$, prism resolution requires a 2500 km/s broad component for a detection. However, at $z = 7$, prism spectroscopy can detect broad components $>950 $ km/s. For G395M, we see a minimum instrumental FWHM range between 360 and 225 km/s. 

We raise caution for BL AGN identification in EELGs with spectroscopic coverage from prism-resolution spectroscopy only, representing 30\% of our sample. There may be EELGs with intrinsic but undetected broad \Ha\ emission and should be interpreted in the context of the instrumental resolution.



\subsection{Observational Depth Selection Effects}

\begin{figure}
\begin{centering}

    \includegraphics[width=1\linewidth]{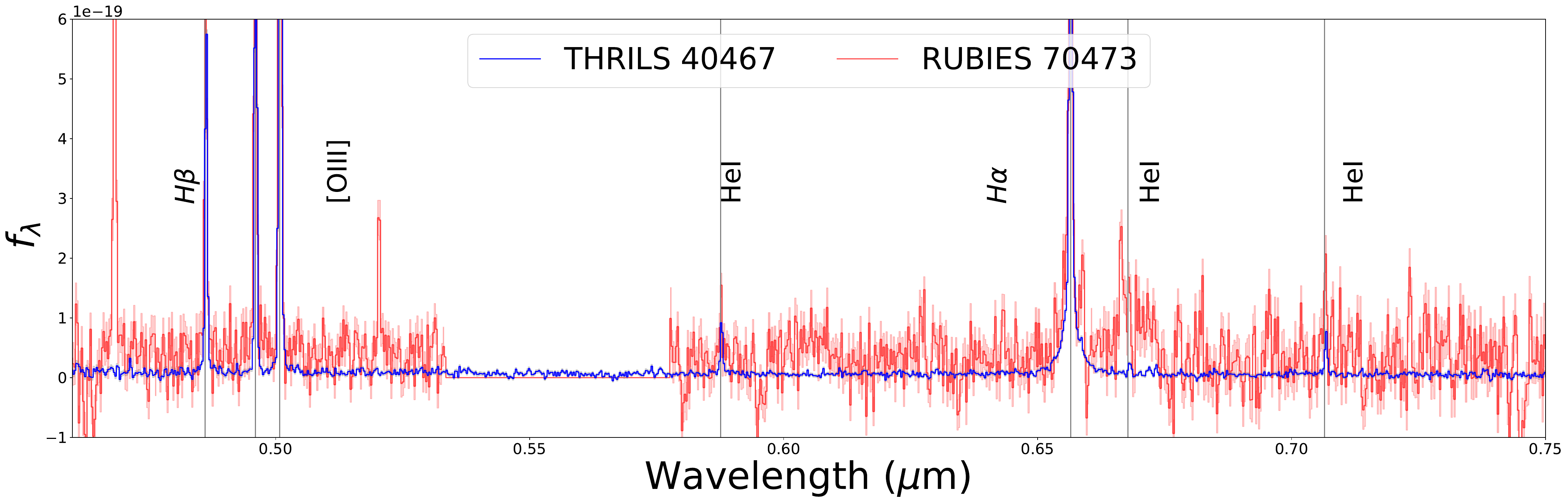}
    \includegraphics[width=.49\linewidth]{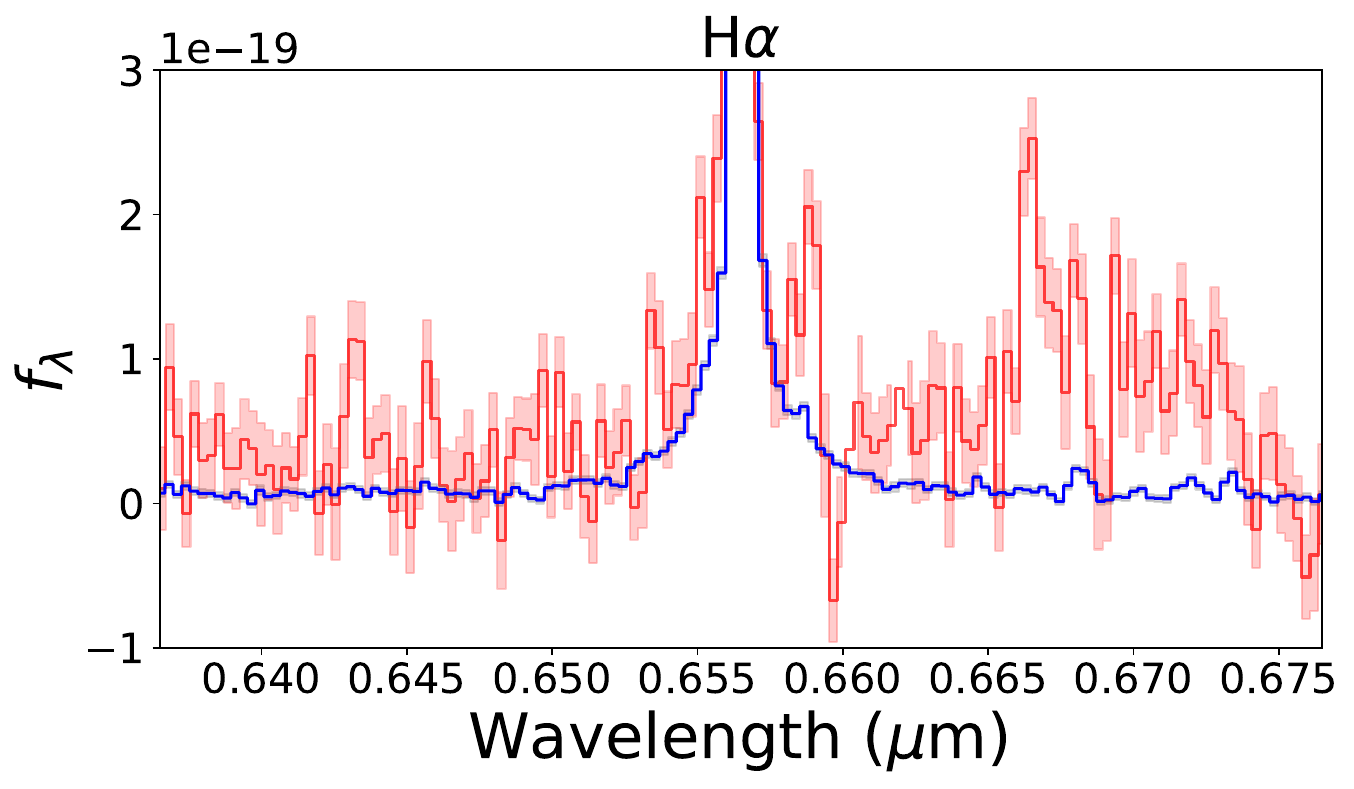} 
    \includegraphics[width=.49\linewidth]{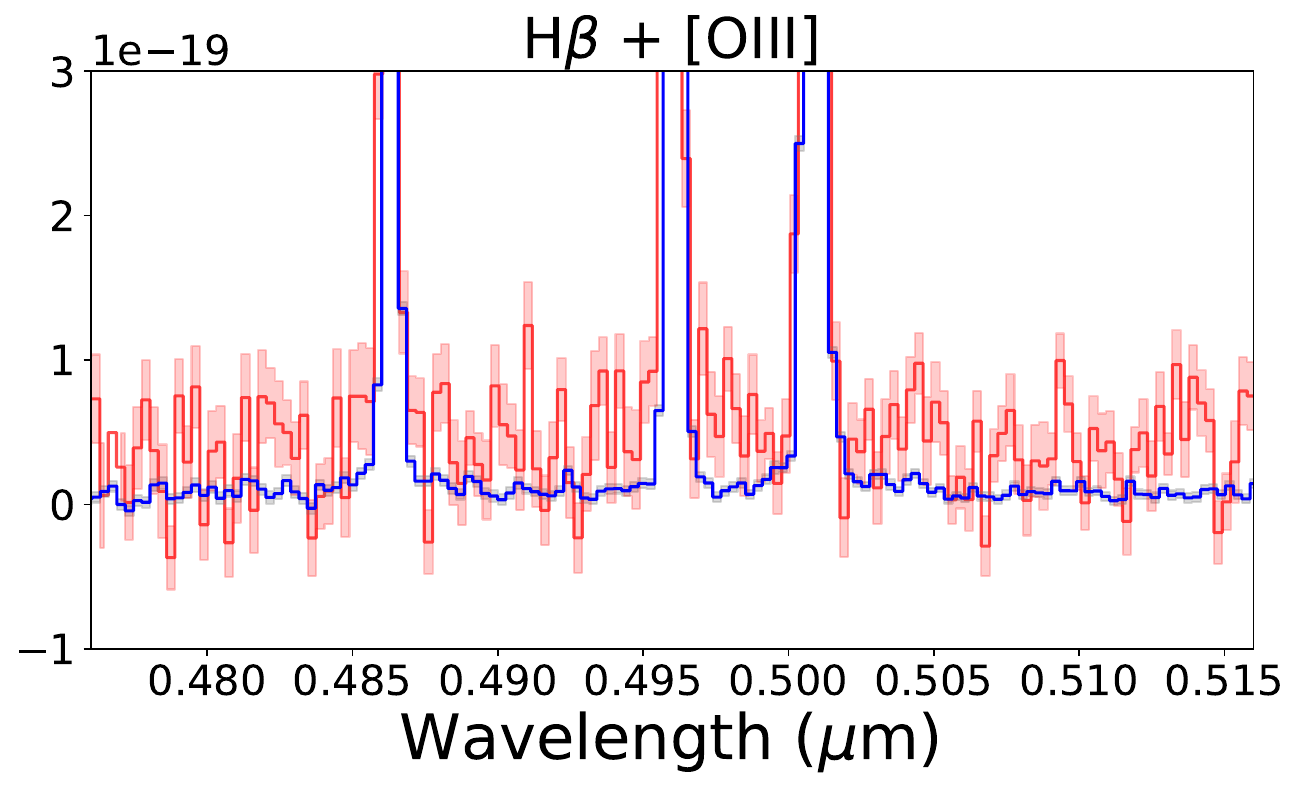} 
    \includegraphics[width=.49\linewidth]{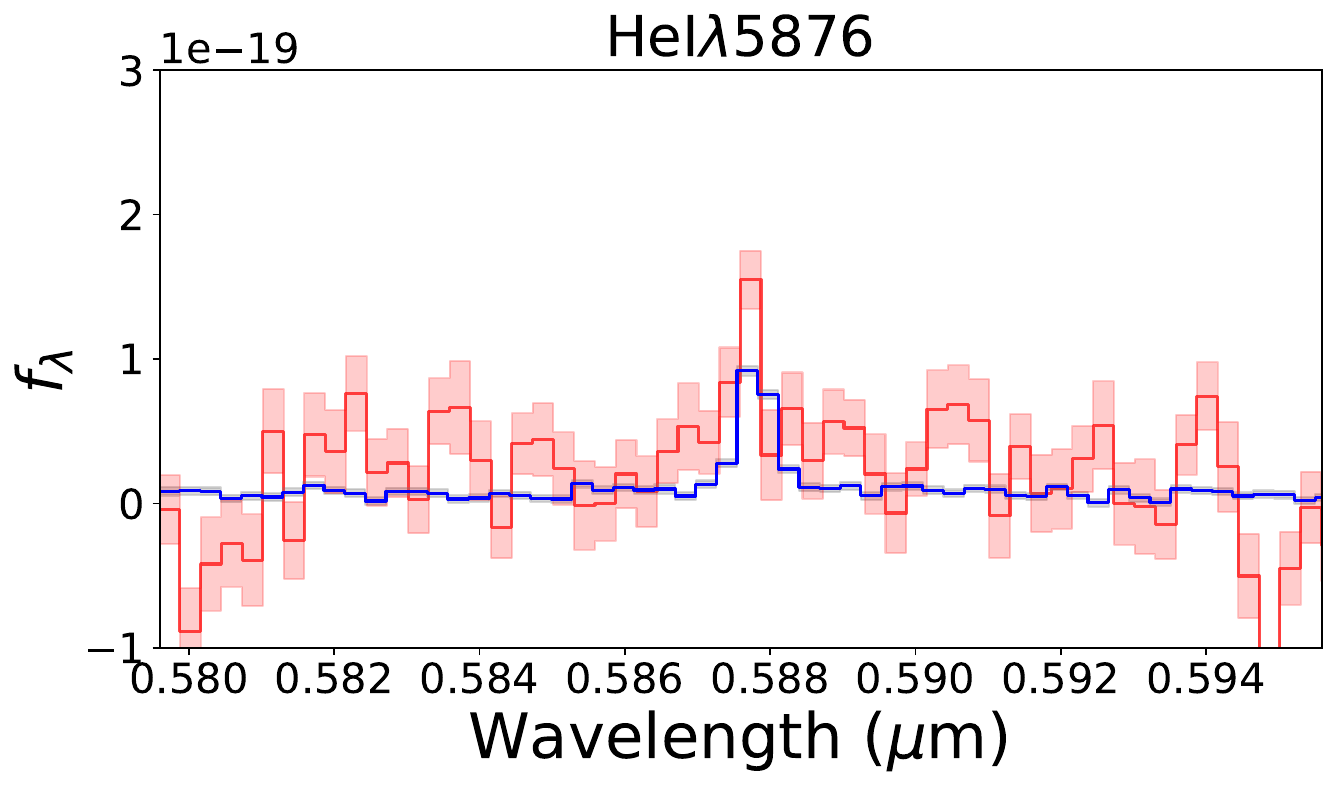} 
    \includegraphics[width=.49\linewidth]{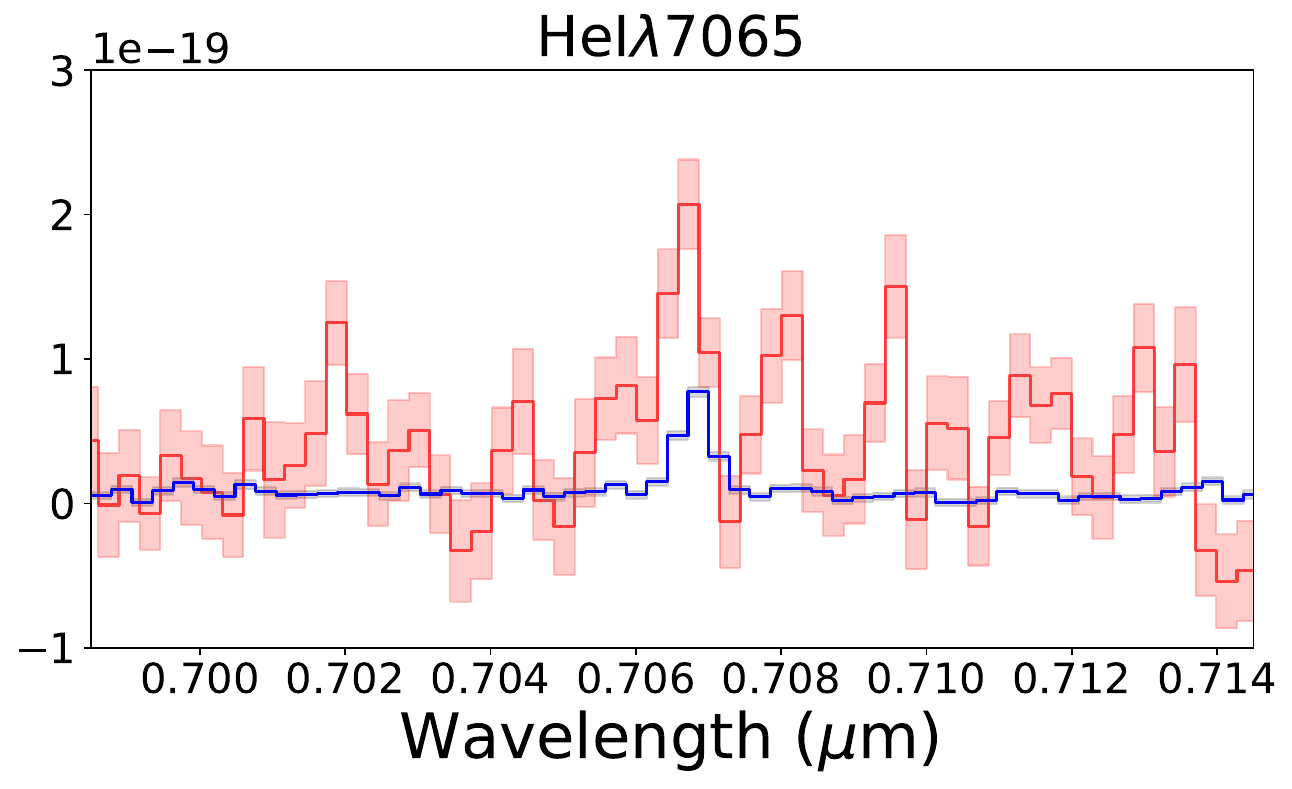} 
    \caption{An example of one of the THRILS AGN reported for the first time here but not detected in a previous RUBIES observation. We plot the RUBIES observation, using the same instrument but with a much shorter exposure time, in red and the THRILS observation in blue. We note the recovery of \HeI$\lambda$5876 and \HeI$\lambda$7065 in the THRILS observation. 
    }
\end{centering}
    \label{fig:overlap}
\end{figure}

The THRILS survey \citep{Hutchison2025} targeted a total of 19 EELGs identified by \citet{Davis2023} with deep ($\sim 8$ hour exposure) G395M coverage. Seven of the EELGs observed in THRILS were observed previously by the RUBIES survey with both low-resolution prism and medium-resolution G395M spectral elements over 0.8 hour exposures. Two were targeted by the original CEERS configuration, also with $\sim0.8$ hour exposures. Of these 9, only one target from the RUBIES survey, THRILS 46155/RUBIES 50812, was previously identified as a BL AGN \citep{Taylor2024}. We confirm this BL AGN and report 6 additional BL AGN that satisfy our criteria for a 3$\sigma$ detection of a broad component and an implied FWHM velocity $>$1000~km/s. We present the fits for these BL AGN alongside their line measurements and masses in the Appendix.

We plot both the THRILS and RUBIES observations for one of these missed BL AGN in Figure \ref{fig:overlap}, where the RUBIES G395M spectrum is plotted in red and the THRILS observation of the same source in blue. 
We measure a broad-component FWHM of $1696\pm51$ km/s from the THRILS 40467 spectra implying that signal to noise, rather than detector resolution, is a dominating factor for broad-line recovery. 
However, the broad \Ha\ component was not recovered in the RUBIES observation. The lower signal to noise ratio (SNR) in the RUBIES spectra is insufficient to detect the broad \Ha\ component. The longer exposure also recovers multiple emission lines lost to noise in the shallower observation. Notably, we recover multiple \HeI\ emission lines that are undetected in the RUBIES observation.  

THRILS 40467 has \Ha\ emission dominated by a narrow \Ha\ line with a ratio of $H\alpha_{narrow}/H\alpha_{broad} = 0.46$, see Section 5.1. As discussed there, this is consistent with a host star-forming galaxy, rather than the AGN, dominating the \Ha\ emission. We recover 7 BLAGN from the THRILS survey ($36\%$) which is our highest BL AGN identification rate among the NIRSpec surveys compared to 21\% of CEERS spectra, 12\% of CAPERS spectra, and 10\% of RUBIES spectra. 
This may imply that more AGN with broad components could be revealed in previously targeted sources through deep spectroscopy, especially in the case that the broad component does not dominate the emission line. 

\section{The Role of AGN in EELGs}
\label{sec:Role}

We next investigate the contribution of BL AGN in the EELGs where they are present.
We demonstrate that when an AGN is identified, it does not dominate the extreme emission by looking at relative emission-line contribution from narrow and broad \Ha\ components. 

\subsection{BL AGN Contribution to Extreme Lines}





To build a complete understanding of the AGN contribution to EELGs, we consider both the EELGs with extreme spectroscopic EWs and contaminants in the photometric selection with sub-extreme EWs that were incorrectly selected as EELGs. We refer to these sources as ``non-EELG BL AGN''. 

\begin{figure}
    \centering
    \includegraphics[width=1\linewidth]{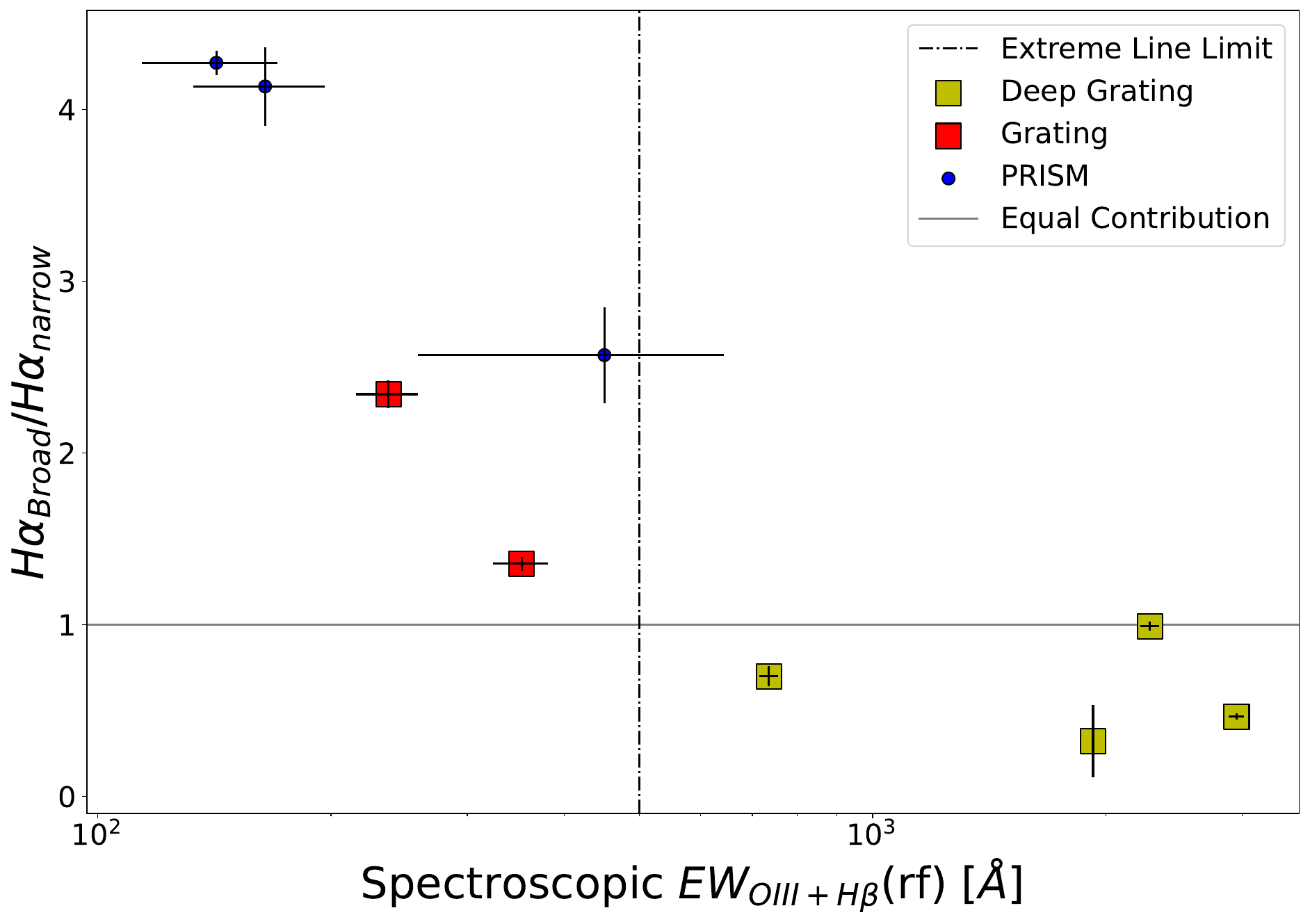}
    \includegraphics[width=1\linewidth]{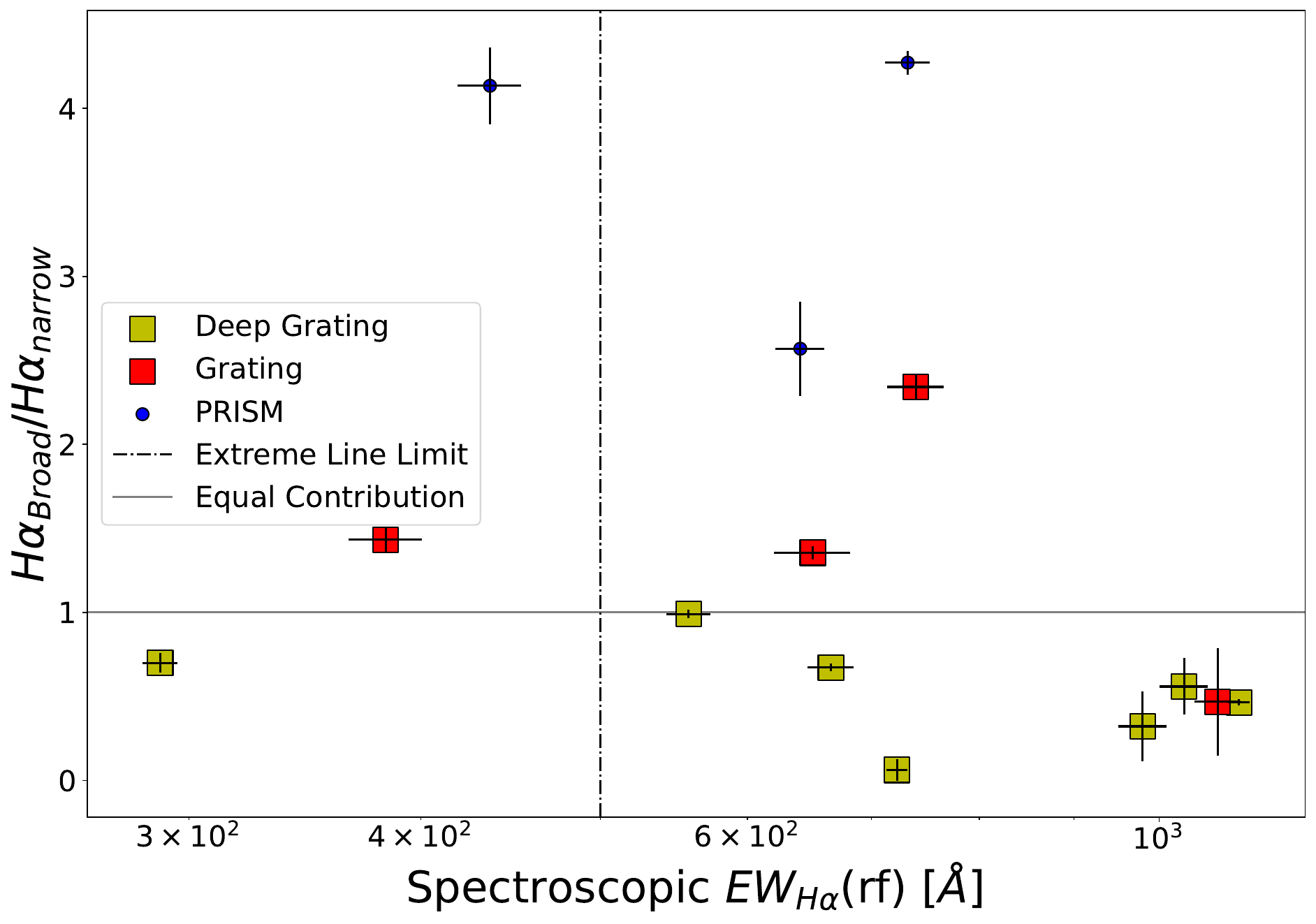}
    \caption{ Emission line contribution from the broad component for BL AGN in the EELG sample plotted against \Hb\ + \OIII\ EW (\textbf{top}) or \Ha\ EW (\textbf{bottom}). We find that most EELGs with AGN that pass our spectroscopic EW threshold have \Ha\ emission dominated by a narrow, rather than broad, line component. }
    \label{fig:linecont}
\end{figure}

We investigate the AGN contribution to the extreme emission lines by taking the ratio of broad to narrow \Ha\ line flux. We plot these ratios against the rest-frame spectroscopic EW of either \Ha\ or \Hb\ + \OIII\ in Figure \ref{fig:linecont}. We find that a few EELGs with broad line contribution are dominated by broad emission, mostly the non-EELG BL AGN which fall below the vertical detection threshold in Figure \ref{fig:linecont}. Others, most prevalent in deep spectroscopy, exhibit \Ha\ emission dominated by the narrow component, illustrated in Figure \ref{fig:linecont}. We note that higher \OIII\ EW sources are strongly associated with EELGs whose narrow \Ha\ dominates the emission, indicating that extreme \OIII\ emitters are dominated by non-AGN processes, such as SF, even in the case that an AGN is present. 
The most extreme \Ha\ emitters are similarly dominated by their narrow components.

These narrow-line-dominated EELGs with broad lines may represent a population of SF galaxies that host BL AGN while the broad-component-dominated EELGs are more likely to be AGN-dominated sources. These broad \Ha\ dominated sources, the non-EELG BL AGN, are falsely selected as EELGs and do not have truly extreme emission-lines. The other EELGs, with truly extreme emission-lines and sub-dominant BL AGN imply an emission source external to the AGN is required to generate the extreme narrow lines. Massive O and B type stars are the only stars with sufficient ionizing radiation to drive hydrogen recombination. Such stars are short lived, on the order of $\sim10s - 100s$ of Myrs \citep{Kennicutt2012}. The over-abundance of such stars implied by excessive narrow \Ha\ emission indicates the galaxies in narrow-line-dominated EELGs are undergoing a recent burst of SF. AGN may also contribute to the narrow-line emission, but we assume the emission is dominated by this recent SF.


\subsection{LRD Color-Color Selection}


AGN without detected broad \Ha\ regions, or NLAGN, are likely present in this and other EELG samples in CEERS \citep{Mazzolari2025} and require follow up interpretation to disentangle NLAGN from strong SF. We explore simple color-color selection as a method to disentangle AGN from SF populations. We include the AGN selected as photometric EELGs that do not pass our spectroscopic EW threshold in this section to explore whether their continuum-dominated emission can be recovered from the photometry. We refer to these incorrectly selected EELGS as ``non-EELG BL AGN'' but remove them from subsequent sections of this paper.

\begin{figure}[h!]
    \centering
    \includegraphics[width=1\linewidth]{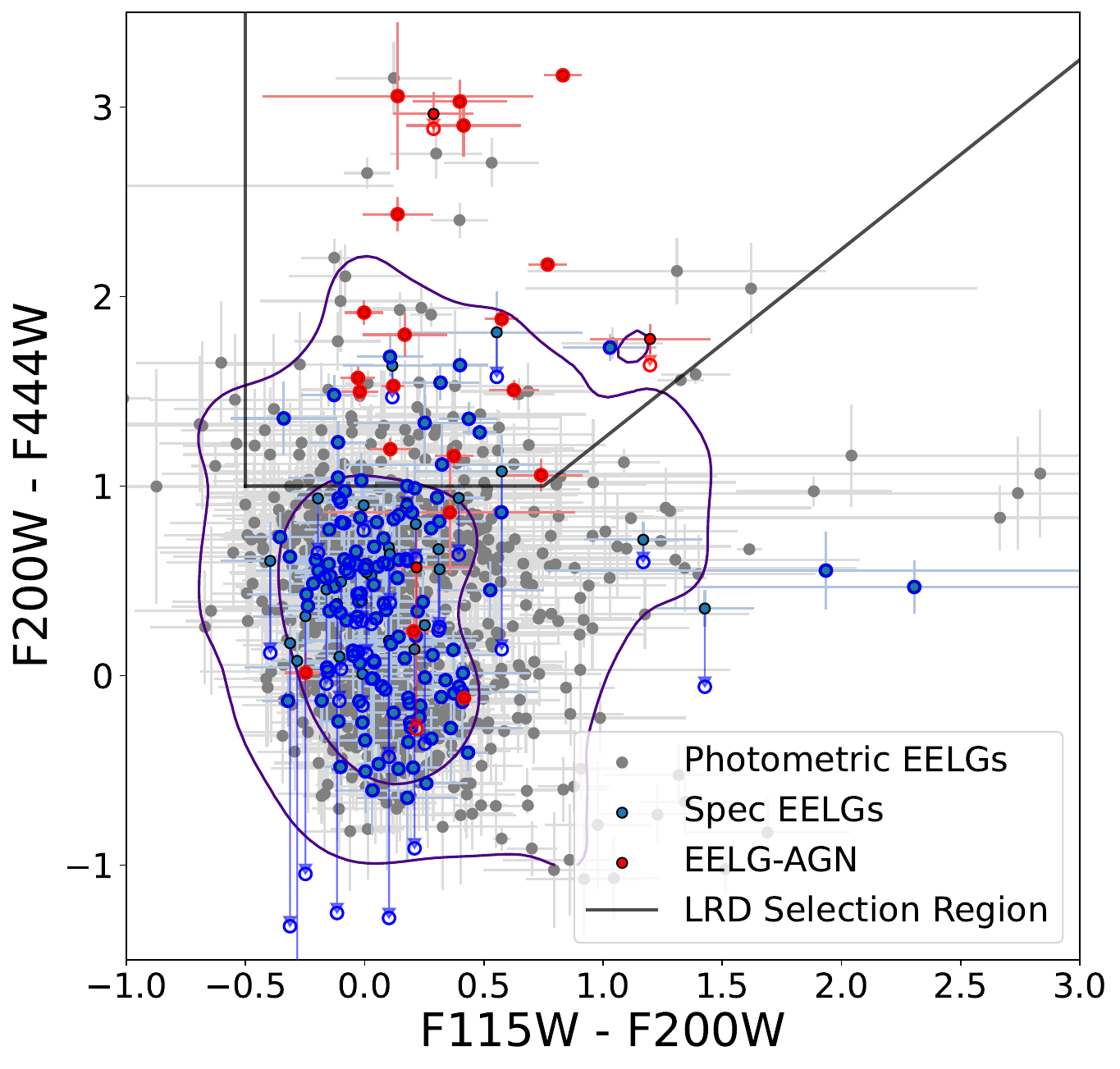}
    \caption{Color-color selection for LRDs from \citet{Barro2024}, solid black line, applied to our EELG population. Grey points trace our full photometric EELG sample, blue points indicate EELGs with spectroscopic coverage but no broad Balmer lines, and red points indicate the EELG-BL AGN. Purple contours represent 2 $\sigma$ and 3 $\sigma$ population contours for all CEERS galaxies at $4<z<9$. Open circles are EW corrected photometry and represent the source's colors from continuum emission only. Arrows point from true photometric colors to these EW corrected colors of the same source. We find that EELG-BL AGN are more common within the LRD selection threshold, that these BL AGN also have LRD colors. }
    \label{fig:color}
\end{figure}

We investigate how our EELG sample presents in color-color space to test AGN recovery from photometry. Most photometrically selected LRDs have broad lines \citep{Kocevski2024} and so a color-color selection targeting LRDs may be an effective way to recover AGN with broad lines that are too faint to be detected at prism resolution or in shallow G395M spectra.
We explore the EELGs in color-color space used to select LRDs in Figure \ref{fig:color}. 
We follow the methodology in \citet{Barro2024} which required $\rm{F200W} - \rm{F444W > 1}$, $(\rm{F200W} - \rm{F444W}) > (\rm{F115W} - \rm{F200W})+0.25$, $\rm{F115W} - \rm{F200W > -0.5}$, and  $\rm{F444W <27 }$ mag. This set of LRD color-color criterion avoids contamination from \Hb + \OIII\ and selects for the bluest and reddest objects at $z>4$, allowing for direct comparison with our EELG sample. We plot spectroscopic EELGs in blue, photometric EELGs in gray, and EELGs with BL AGN in red.

We test this color-color selection against the available spectroscopy. In the case that an EELG has a line falling in a color selection filter, we correct the photometry to subtract the line EW, indicated by an open circle. These sources are traced by arrows on the plot, beginning on the true photometric colors and terminating on a point representing the emission-line subtracted photometry. 
EELGs with colors in the LRD space have only small emission-line corrections, indicating that these choices of colors are effective at selecting LRD continua and are not strongly contaminated by emission lines, even for EELGs. Some EELGs have larger emission-line corrections but these tend to lie in the non-LRD color space (lower left of Figure \ref{fig:color}).


We find that 108 of our photometrically selected EELGs (9 \%) meet the LRD selection criteria. Of the EELGs with broad \Ha\ components, which we refer to as EELG-BL AGN, with spectroscopic coverage, 18 of 23 (78\%) fall in this region while 17 of the remaining 158 EELGs without broad Balmer lines (10\%) meet the criteria. All EELGs with spectroscopic coverage and  $\rm{F200W} - \rm{F444W}>2$ are BL AGN. 
The LRD color-color selection of \citet{Barro2024} is highly effective at selecting EELGs with BL AGN, and the filters used in the color selection are not strongly contaminated by emission lines (even for these extreme, high-EW galaxies). The 17 EELGs that meet the LRD color selection but are not BL AGN are likely AGN with broad lines too faint for detection in the prism and shallow G395M data.
We remove EELGs and EELGs with AGN that do not meet our spectroscopic EW criteria from subsequent discussion.



\begin{figure*}
    \centering
    \includegraphics[width = 1 \linewidth]{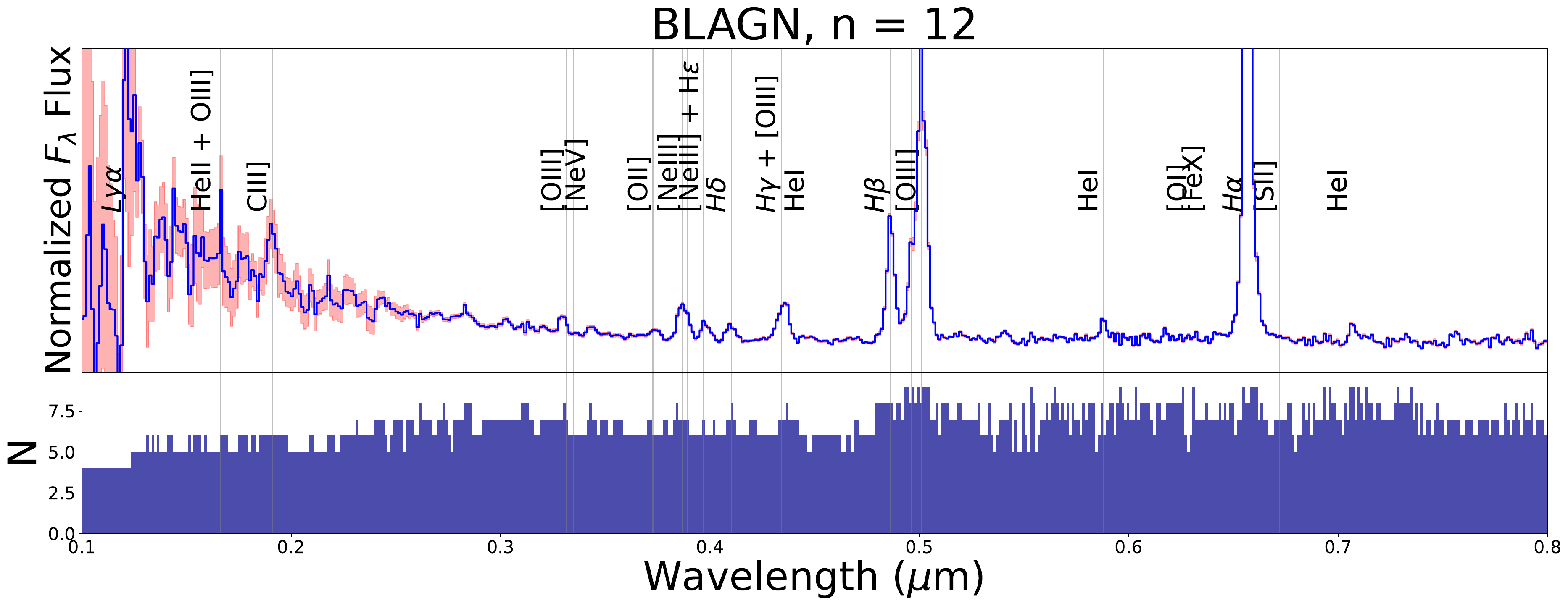}
    \caption{
    Stack of all 12 known BL AGN in the EELG spectroscopic sample with EWs that pass both our photometric and spectroscopic definitions, normalized in $F_\lambda$ to \Ha\ before stacking. 
    We note a detection of [NeV]$\lambda3427$, (S/N = 4.05) promising evidence that broad-line selections identify AGN. However, this detection lies in a region where just 9 of 23.
    We fit the stacked BL AGN and recover a $>3\sigma$ detection of a broad \Ha\ emission line component. The histogram in the lower panel counts source contribution to each pixel. Red shaded regions denote error as described in Section 2.4.}
    \label{fig:allstack}
\end{figure*}


\subsection{AGN Indicators in Composite Spectra}

We evaluate the effectiveness of broad Balmer line detections as AGN indicators by looking for AGN signatures in composite spectra. We find high-ionization emission associated with AGN in the AGN composite but find no such signatures in EW binned composite spectra. In this section, we exclude EELGs that did not have sufficient spectroscopic EWs to pass our original photometric selection criteria, the ``non-EELG BLAGN'' discussed in the previous subsections.

We stack the 12 EELG-BL AGN with extreme EWs that pass both our spectroscopic and photometric EELG criteria in Figure \ref{fig:allstack}. 
These sources are primarily stacked to search for faint high-ionization lines that signpost AGN activity, such as \NeV.  We measure SNRs for the available emission lines and report them in Table \ref{tab:emission_linesBL AGN}. We successfully fit a broad \Ha\ component to the AGN stack, verifying our ability to recover broad components in composite spectra.

We report significant \NeV$\lambda3427$ emission with a SNR of 4.05. The high ionization \NeV\ line is strongly indicative of ionization harder than almost all stellar population models \citep{Berg2021, Chisholm2024, Cleri2023, Cleri2023b, Cleri2025, Izotov2012, Izotov2021, Mingozzi2025, Negus2023,Olivier2022} with an ionization energy of 97.19 eV. Being a forbidden line, \NeV\ emission probes the area outside the broad-line region of the AGN, the ``coronal line'' region and so is independent of broad-line selections \citep{McKaig2024, Smith2025}.
%

The \NeV\ emission is a doublet, with a bluer line at $\lambda=$3347 \AA. We expect this bluer line to be weaker and so a non-detection of the bluer line is observationally consistent. We note that \NeV$\lambda3427$ lies in a redshift range not probed by the G395M filter spectra of these sources.  Only prism spectroscopy contributes to the \NeV$\lambda3427$ stack bins.

Although this detection of \NeV$\lambda3427$ is a promising indicator of AGN activity, we note no higher ionization coronal lines in the composite. This may imply that the AGN are present but not dominating the ionizing radiation field. This result may be biased by the wavelength coverage selection effects discussed in Section 5.1. We note that \NeV$\lambda3427$ lies in a redshift range not probed by the G395M filter spectra of these sources. Only prism-resolution spectroscopy, from  9 of the BL AGN, contributes to the \NeV$\lambda3427$ stack bins. However, it may also reflect the resolution of the composite spectra. We note a few marginal detections of [FeX] in the Appendix for the THRILS BL AGN. However, the emission is weak and blended with the [OI] doublet and so requires follow up observations for a full interpretation. Both the [FeX] and [OI] lines are not detected in the BL AGN composite.  


\begin{table}[h!]
\centering

\begin{tabular}{lcc}
\hline
\hline
Line & SNR & Creation Energy \\
\hline
He\textsc{ii} $\lambda$1640 + OIII]$\lambda1663$ &\textit{1.82} & 54.42 eV, 35.12 eV \\
{C\textsc{iii}]} $\lambda$1909 & \textit{2.38} & 24.38 eV \\
{[O\textsc{iii}]} $\lambda$3313 & 8.49 & 35.12 eV \\
{[Ne\textsc{v}]} $\lambda$3427 & 4.05 & \textbf{97.19 eV} \\
{[O\textsc{ii}]} $\lambda$3726 & 4.11 & 13.62 eV \\
{[Ne\textsc{iii}]} $\lambda$3869 & 17.24  & 40.96 eV\\
H$\delta$  & 10.45 & 13.6 eV \\
H$\gamma$ + [O\textsc{iii}] $\lambda$4363 & 24.81 & 13.6 eV , 35.12 eV\\
He\textsc{i} $\lambda$4473 & \textit{2.24} & 24.59 eV, \\
H$\beta$  & 50.89& 13.6 eV \\
{[O\textsc{iii}]} $\lambda$4960 & 15.45 & 35.12 eV \\
{[O\textsc{iii}]}$\lambda$5008 & 41.97 & 35.12 eV \\
He\textsc{i} $\lambda$5877 & 15.03 & 24.59 eV, \\
H$\alpha$  & 131.99 & 13.6 eV \\
{[O\textsc{i}]} $\lambda$6302 & \textit{0.66 } & 13.13 eV\\
{[S\textsc{ii}]} $\lambda$6718 & $--$ & 10.36 eV \\
He\textsc{i} $\lambda$7067 & 12.00& 24.59 eV, \\
\hline
\end{tabular}
\caption{Emission line SNRs and ionization energies for BL AGN stack. We italicize lines that fall below our S/N = 3 detection threshold and bold our highest ionization detection. Creation energies are from \texttt{pyneb}.}
\label{tab:emission_linesBL AGN}
\end{table}

\begin{figure*}
    \centering
    \includegraphics[width = 1\linewidth]{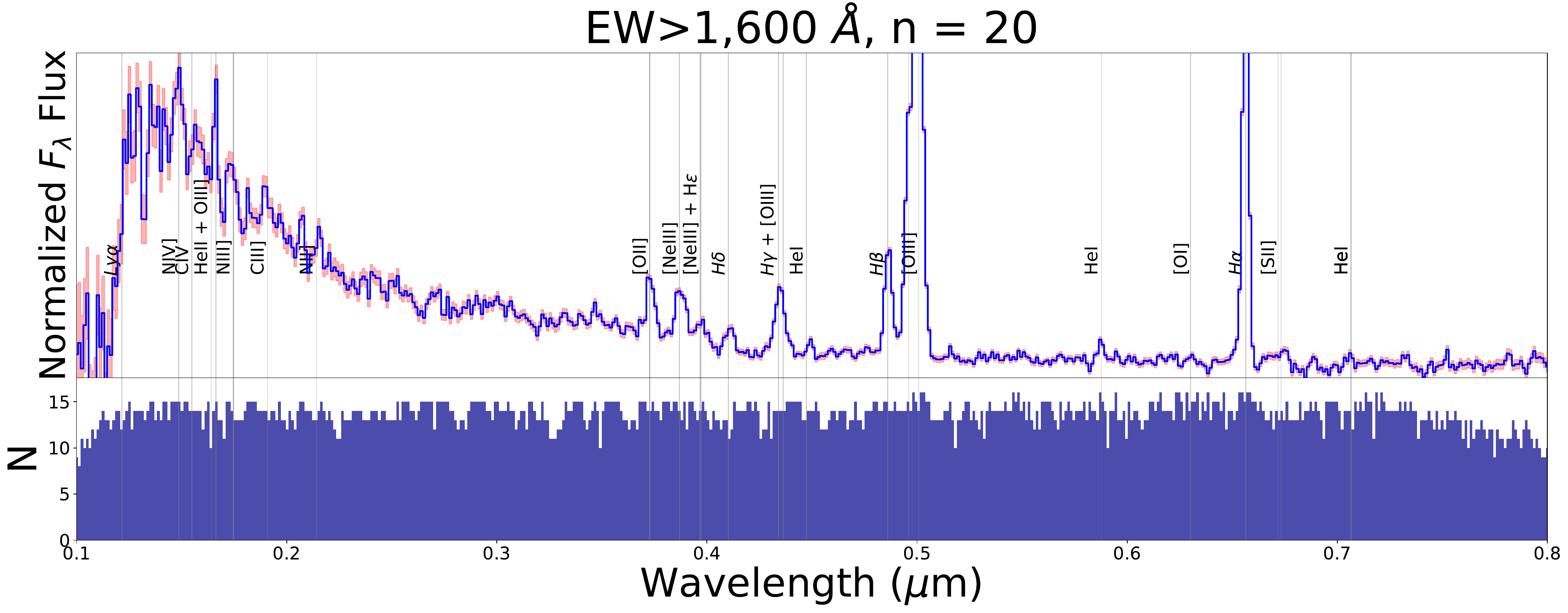}
    \includegraphics[width = 1\linewidth]{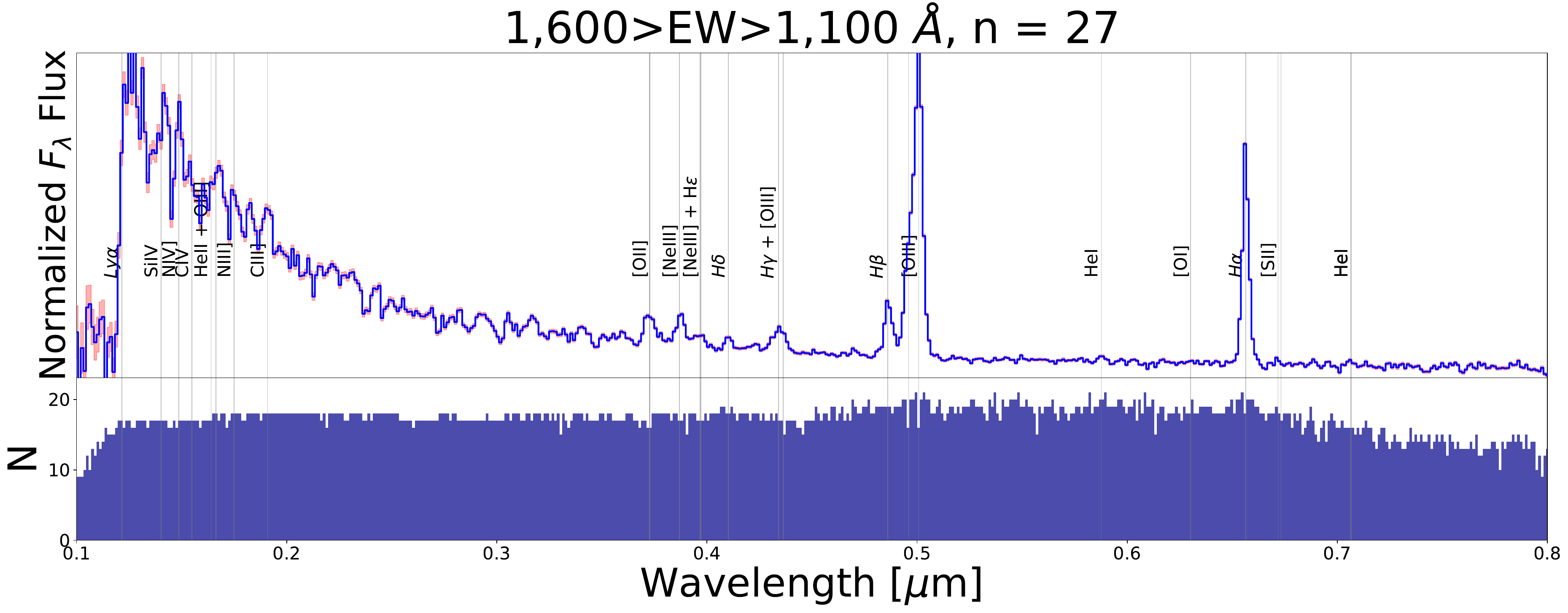}
        \includegraphics[width = 1\linewidth]{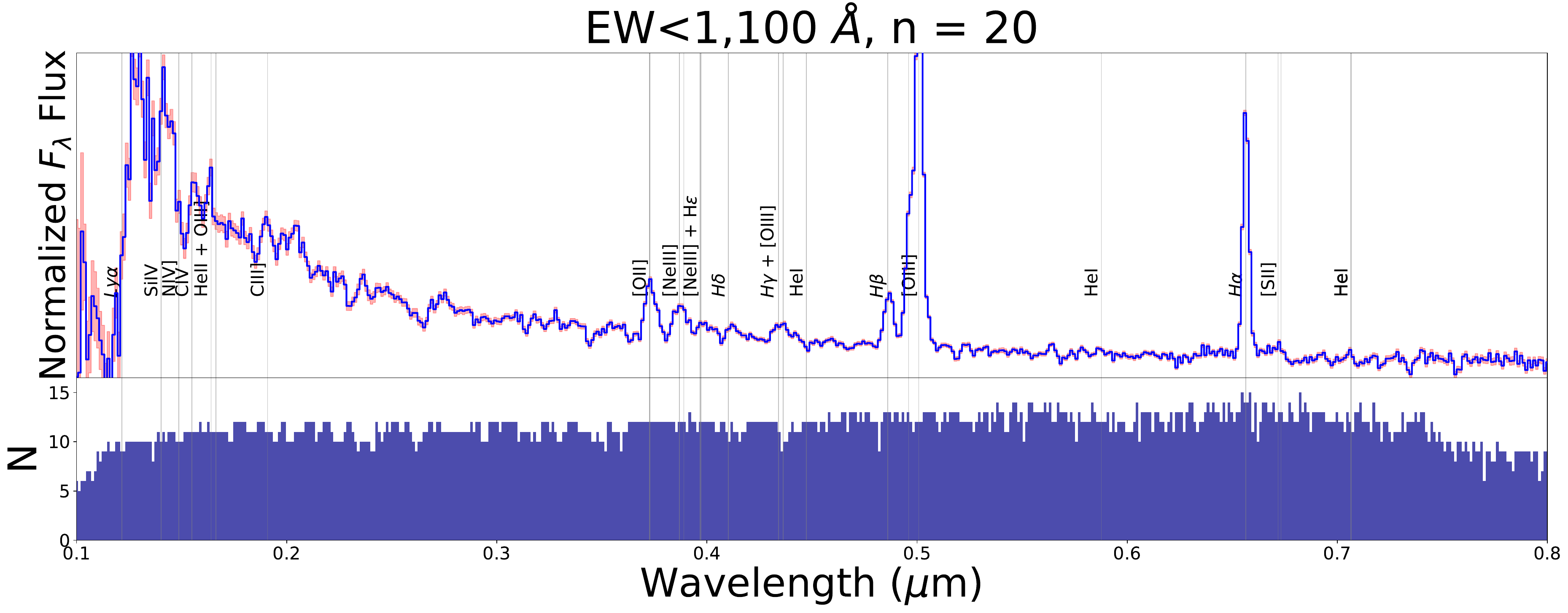}
    \caption{EELG stacked spectra binned by rest-frame EW normalized in $F_\lambda$ to \Ha\ peak emission. \textbf{Top:} Our high-EW tail of EELGS at EW$>$1600 \AA. 
    We note several high-ionization lines, but none that signpost AGN activity. 
    \textbf{Middle:} Our stack of sources below the high-EW tail but above the sample's mean EW (1600 \AA\ $>EW>1100$ \AA\ ). 
    \textbf{Bottom:} Stack of EELGs with EWs below the population mean (EW$<$1100 \AA\ ). 
    }
    \label{fig:ewstacks}
\end{figure*}

We begin our search for AGN signatures in the remaining EELGs by stacking the sources without broad \Ha\ emission to search for high-ionization lines.
We stack our full sample over three EW bins to further investigate the presence of AGN in the high EW sample, shown in Figure \ref{fig:ewstacks}. 

We report a mean spectroscopic sample rest-frame EW of 1100\AA\ and use this as a reference point to investigate EW-related properties. Our composite spectra have EW ranges of EW$>$1600 \AA, 1600 \AA\ $>$EW$>$1100\AA\, and EW$<$1100 \AA\ which yield similar EELG counts in each bin and therefore similar SNRs. The composites are shown in Figure \ref{fig:ewstacks}. 

We search the composites for AGN signatures. We do not detect a significant broad-line component in any of the EW-binned stacked spectra and so we turn to other signatures. We detect \ion{N}{4}], with an ionization energy of 47.45 eV, in all EW bins and  \ion{C}{4}, with an ionization energy of 47.89 eV, strongest in the low EW stack. These are our highest energy line detections. While these imply a highly ionized environment, they have been observed in lensed high EW systems to occur in clumps of bursting star formation \citep{Topping2024} and are not sufficiently high ionization to rule out SF as a source of ionization. Detections of emission lines such as \NeV, with a creation energy of 97 eV or [FeX] with a creation energy of 234 eV are more traditionally used as signposts for intense radiation from AGN \citep{Doan2025, Gelbord2009, Molina2021} and we report no such detections here. 

\begin{table}[h!]
\begin{centering}
\setlength{\tabcolsep}{2pt} 
\begin{tabular}{lcccc}
\hline
\hline
Line & High EW & Mid EW  & Low EW  & Creation \\ 
     &SNR            &SNR            &SNR            & Energy \\
\hline
Si\textsc{iv}$\lambda$1403 & -- & 6.39 & 6.24 & 33.49 eV \\
N\textsc{iv}]$\lambda$1486 & 4.40 & 10.14 & 6.04  & \textbf{47.45 eV}\\
C\textsc{iv}$\lambda$1549 & 3.14 & \textit{2.15}& 7.29 & \textbf{47.89 eV} \\

He\textsc{ii} $\lambda$1640 + & 33.43 & 5.24 & 6.52  & 54.42 eV, 35.12\\
\quad OIII]$\lambda1663$ & & & & \\
N\textsc{iii}]$\lambda$1746 &\textit{ 2.27} & 20.69 & -- & 29.60 eV \\
C\textsc{iii}]$\lambda$1909 & 3.89 & 8.54 & 3.34 & 24.38 eV \\
N\textsc{ii}]$\lambda$2142 & 3.72 & -- & -- & 14.53 eV\\
{[O\textsc{ii}]}$\lambda$3726 & 15.27 & 13.96 & 21.03 & 13.62 eV\\
{[Ne\textsc{iii}]}$\lambda$3869 & 12.61 & 14.72 & 9.21 & 40.96 eV\\
H$\delta$  & 6.04 & 6.47 & \textit{2.95} & 13.6 eV\\
H$\gamma$ + [O\textsc{iii}]$\lambda$4363 & 20.22 & 13.11 & 6.76 & 13.6 eV \\
He\textsc{i}$\lambda$4473 & 3.63 & -- & 50.96 &  24.59  \\
H$\beta$  & 35.83 & 37.10 & 25.63 & 13.6 eV\\
{[O\textsc{iii}]} $\lambda$4960 & 65.50 & 63.32 & 47.82 & 35.12 eV \\
{[O\textsc{iii}]}$\lambda$5008 & 186.60 & 148.56 & 129.30 & 35.12 eV\\
He\textsc{i}$\lambda$5877 & 6.91 & 3.76 & 3.43 &  24.59  \\
{[O\textsc{i}]}$\lambda$6302 & \textit{2.19 }& 6.29 & 0.55 & 13.13 eV \\
H$\alpha$  & 116.11 & 132.88 & 88.90 & 13.6 eV\\
{[S\textsc{ii}]}$\lambda$6718 & 3.93 & 27.47 & 4.04 & 10.39 eV \\
He\textsc{i}$\lambda$7067 & \textit{2.90} & 3.64 & 3.71 &  24.59 \\
\hline
\end{tabular}
\caption{Emission line SNRs for the EW binned composites. We bold our highest ionization detections and italicize detections that fall bellow out SNR = 3 detection threshold. Creation energies are from \texttt{pyneb}.}
    
\end{centering}
\label{tab:ew_stack_lines}
\end{table}




\subsection{Optical Emission-Line Diagnostics }

We evaluate the composites with optical line ratio diagnostic plots. We show that EELGs have line ratios that are ambiguous between high-redshift AGN and SF galaxies and demonstrate through MAPPINGS V models that the EELGs trace areas on the line diagnostic plots consistent with high-ionization, low-metallicity conditions. 


\begin{figure}
    \centering
    \includegraphics[width=1\linewidth]{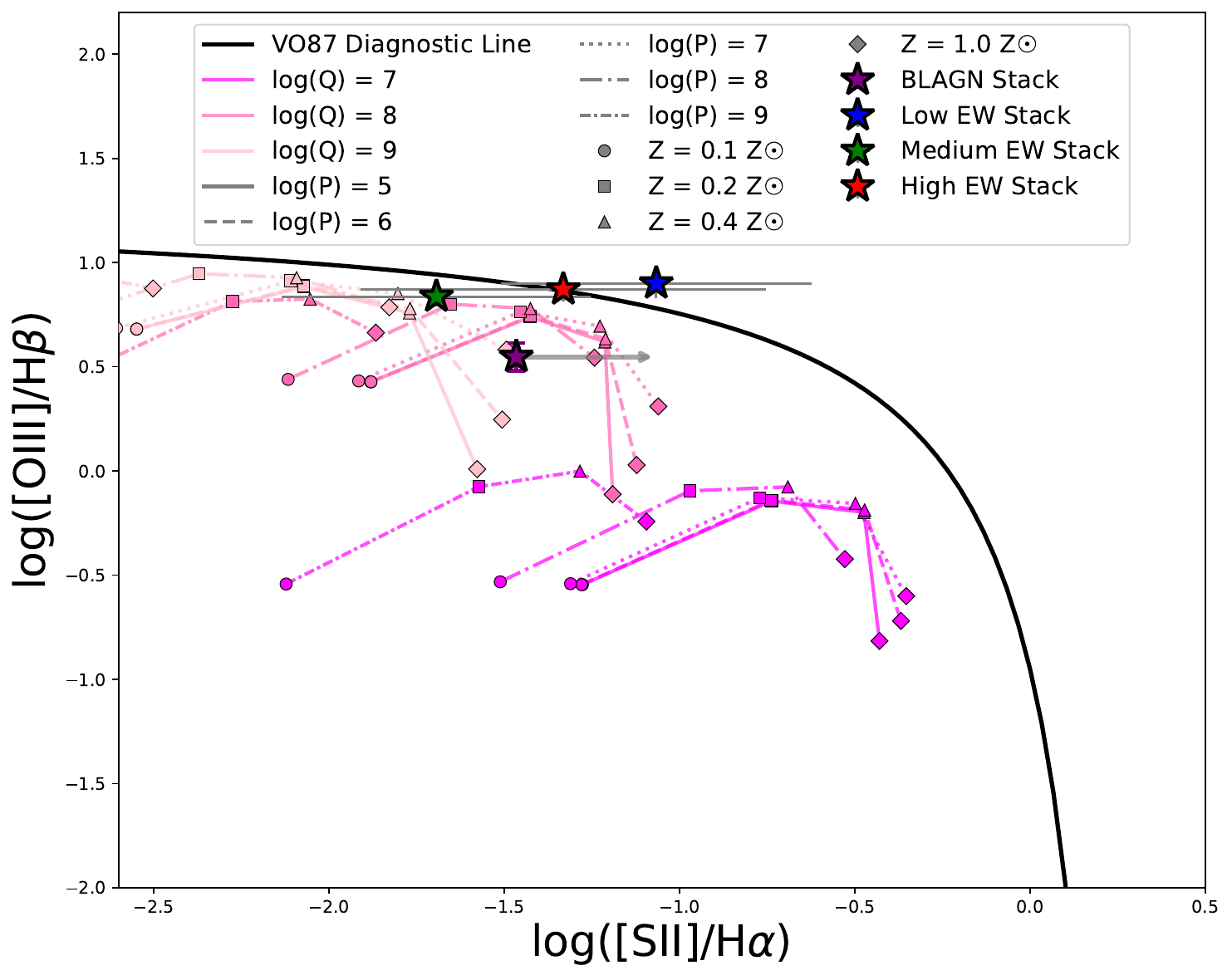}
    \caption{VO87 diagram and our EELG stacks with the diagnostic line from \citet{Kewley2006} and photoionization models from MAPPINGS V \citep{Sutherland2018}. We plot only the narrow \Ha\ emission for the BL AGN stack. All stacks fall in a region that is ambiguous within the diagnostic, especially due to the large error on the \SII\ detection. \SII\ is undetected in the BL AGN stack, shown in Figures \ref{fig:allstack}, so we report this \SII\ /\Ha\ ratio as an upper limit.}
    \label{fig:vo87}
\end{figure}


AGN and SF galaxies at $z<3$ were disentangled via emission line ratio plots \citep{Kewley2006}. For example, the ``BPT'' \citep{Baldwin1981, Kewley2013}, ``VO87'' \citep{Veilleux1987}, and ``OHNO'' \citep{Backhaus2022} diagrams exhibit a convenient bimodality for SF and AGN populations at $z<3$.
However, results from JWST spectroscopy have suggested that rest-frame optical strong line ratios are insufficient separators of SF and black hole accretion at high redshifts \citep{Ubler2024, Scholtz2025, Backhaus2025, Larson2023, Cleri2025}. EELGs in CEERS, even with more conservative selection criteria, are non-trivial to interpret through these optical emission-line plots \citep{Llerena2025}. This is due in part to the low metallicity, high-ionization parameter gas conditions at high redshift \citep{Cleri2025}. 

Early galaxies have lower metallicity \citep{Curti2024, Sanders2024}, especially galaxies with extreme lines \citep{Davis2023}. This implies the presence of hotter stars with higher energy ionizing spectra that power the higher ionization lines whose absence distinguishes them from AGN at lower redshift. It may also point to different initial mass functions (IMFs) for early galaxies, especially an increasingly top-heavy, as opposed to constant Salpeter \citep{Kroupa}, IMF which would increase abundance of young, massive stars \citep{Cueto2024}.

\begin{figure}
    \centering
    \includegraphics[width=1\linewidth]{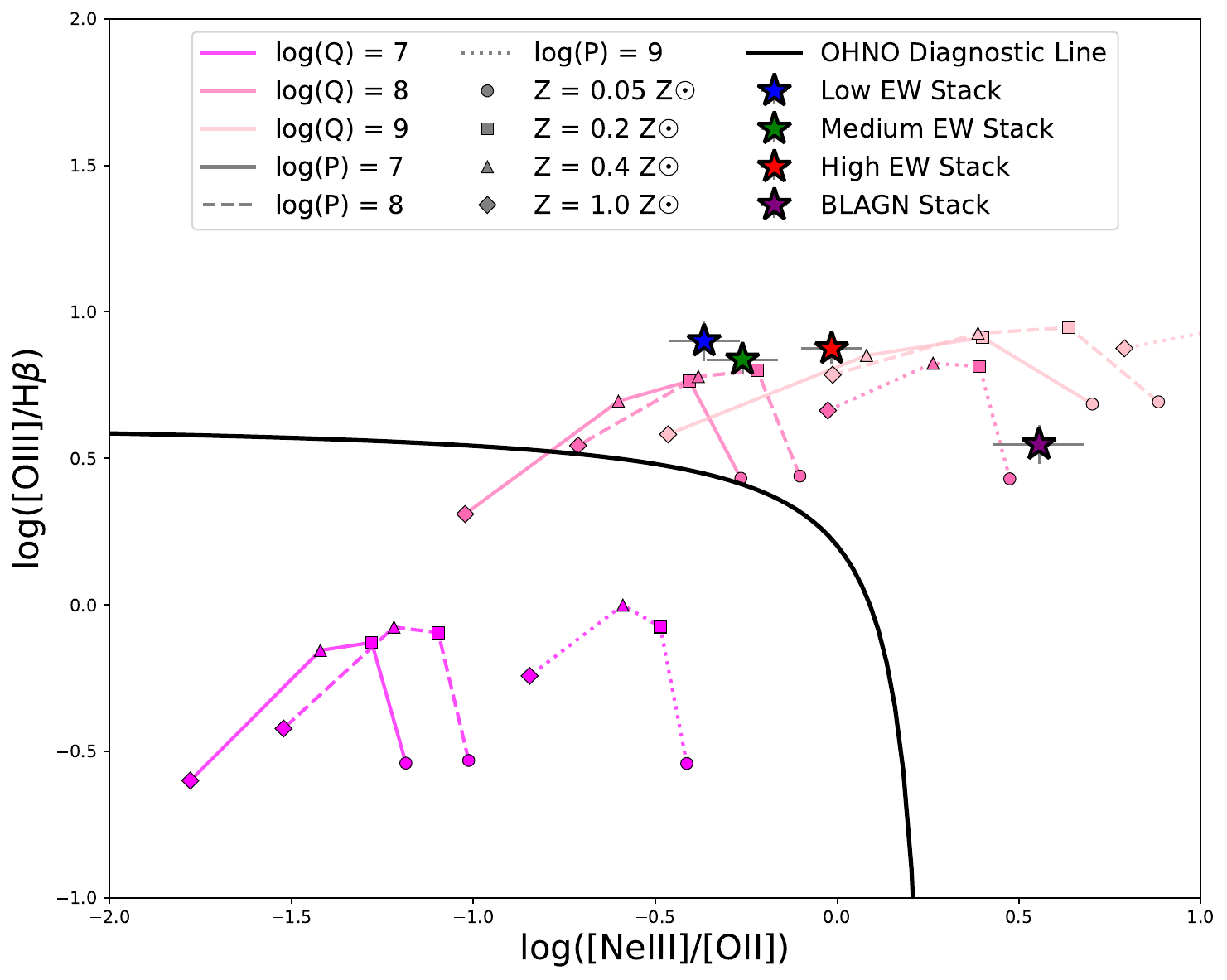}
    \caption{OHNO diagram with diagnostic line in black from \citet{Backhaus2022} and MAPPINGS models from \citet{Sutherland2018, Kewley2019a}. MAPPINGS lines have log($\rm{q}$) indicated by color, log($\rm{P}$) indicated by line style, and $\rm{Z}$ indicated by marker shape. All our stacks fall in the AGN section of the diagnostic plot.}
    \label{fig:ohno}
\end{figure}
\begin{figure}
    \centering
    \includegraphics[width=1\linewidth]{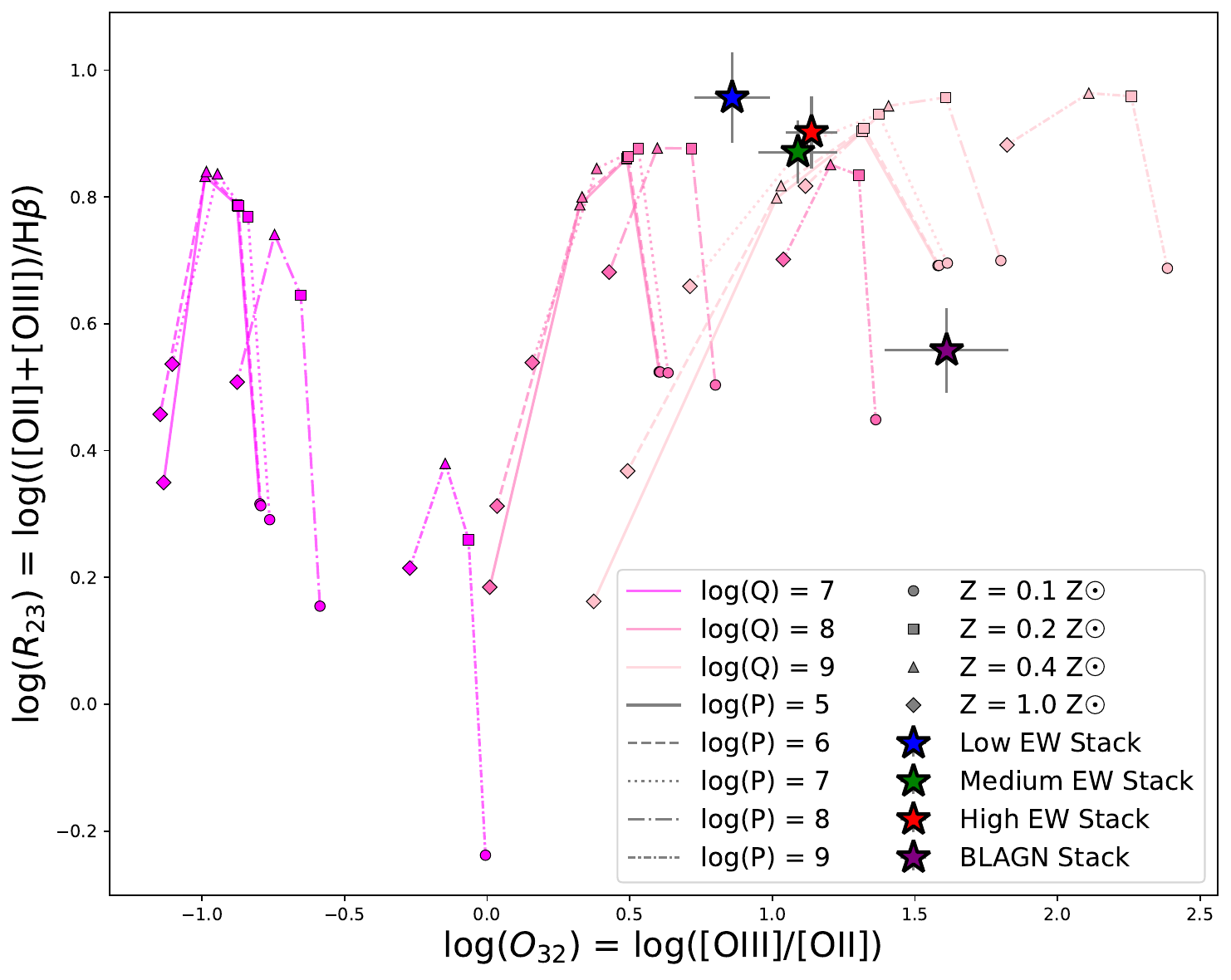}
    \caption{$O_{32}$ vs $R_{23}$ diagram from  \cite{Sanders2016} and MAPPINGS models from \citep{Sutherland2018}. The $O_{32}$ axis is highly ionization dependent while the $R_{23}$ axis is highly metallicity dependent, confirmed here by the MAPPINGS models. We find that all our stacked spectra are consistent with high-ionization and low metallicity models.    }
    \label{fig:r23o32}
\end{figure}

We place the EW binned and BL AGN composite spectra in AGN diagnostic plots and discuss them in the context of their implied ionization conditions. 
Emission lines in each ratio are close in wavelength, except in the final diagnostic, to mitigate the impact of dust attenuation. 
They also often probe permitted and forbidden emission line regions which are sensitive to different temperatures and densities.
Photoionization modeling is from MAPPINGS V models described in \citet{Sutherland2018, Kewley2019}. We consider pressure, $\rm{P}$, in units of $k \rm{cm}^{-3}$ where $k$ is the Boltzman constant, metallicity, $\rm{Z}$, and ionization parameter $\rm{q}$ where $\rm{q} = \rm{Uc}$, $\rm{c}$ is the speed of light, $\rm{U}$ is the dimensionless ionization parameter, and $\rm{q}$ has units of velocity. For all line diagnostic plots, Figures \ref{fig:vo87}, \ref{fig:ohno}, and \ref{fig:r23o32}, we plot the MAPPINGS models as lines of constant pressure. Line styles indicate log($\rm{P}$), colors indicate log($\rm{q}$), and line marker shapes indicate $\rm{Z}$. In all plots, $\rm{q}$ and $\rm{Z}$ differences are responsible for the most dramatic changes while $\rm{P}$ has little impact.

We first plot the stack line ratios in the ``VO87'' \citep{Veilleux1987} diagram in Figure \ref{fig:vo87}. We note that the BL AGN composite spectrum in Figure \ref{fig:allstack} has a non-detection of \SII\ and we report the \SII/\Ha\ ratio as an upper limit. We remove the broad flux from the composite spectra in Figure \ref{fig:allstack} and consider only narrow \Ha\ flux. The location of all our stacks on the VO87 diagram hugs the diagnostic line from \citep{Kewley2006}. Their errors overlap each side making them statistically ambiguous within the AGN/SF diagnostic. We argue, in this and the following diagnostics, that the position of these sources on the diagnostic plot at high-$z$ is reflective of gas conditions rather than AGN or SF contribution.
The VO87 diagnostic plot is most sensitive to density changes along the \SII/\Ha\ axis and temperature/ionization parameter sensitive along the \OIII/\Hb\ axis. We find that all stacks are consistent with the higher ionization (log($\rm{q}$) = 8,9) MAPPINGS models.

We also place the stacks on the ``OHNO'' diagram \citep{Backhaus2022}. 
The line ratios of our sources are associated with high-ionization parameter (log($\rm{q}$)= 8,9) models, consistent with \citet{Kocevski2023} and \citet{Trump2023} (see also \citealt{Pacucci_2026}). This implies that the line ratios are of high-ionization sources rather than definitive signposts of AGN activity. Rather than drawing conclusions about AGN content, we instead can draw that the stacks are best modeled by MAPPINGS models with higher ionization and lower metallicity as in \citet{Cleri2025}. This is consistent with the photometrically implied $\rm{Z}$ $\sim$ 0.2 Z$\odot$ from MAPPINGS V model comparisons in \citet{Davis2023}.

Composite spectra are placed in an $R_{23} O_{32}$ diagram \citep{Sanders2016} in Figure \ref{fig:r23o32}. We begin our analysis of the diagram by studying the $R_{23}$ = (\OIII $\lambda$4960 + \OIII $\lambda$5008 + \OII$\lambda3726$ )/\Hb\ ratio. This ratio is highly dependent on temperature (and consequently, on metallicity). We find our EW binned stacks are consistent with metallicities of $0.2-0.4 \rm{Z}\odot$, again consistent with photometric comparisons in \citet{Davis2023}. Our BL AGN composite ratios, however, are best fit by the lowest metallicity $0.1\rm{Z}\odot$ models. This diagnostic plot is more sensitive to dust attenuation due to the \OIII\ and \OII\ separation of 1234 \AA. We neglect dust corrections here because our stacked sources have been vetted to have extreme rest-optical lines and blue optical continua and so are unlikely to be impacted by significant attenuation.   

The $O_{32}$ axis of the $R_{23} O_{32}$ plot measures oxygen abundances and is highly dependent on ionization parameters. Our EW-binned stacks have oxygen ratios of $0.9-1.4$, implying that almost all the oxygen is in the doubly-ionized state. The BL AGN composite has a $O_{32}$ line ratio of 1.6, representing the extreme of our composite ratios. This requires either the highest pressure in the MAPPINGS models (log($\rm{P}$) = 9) or the highest ionization parameter (log($\rm{q}$) = 9) to reproduce. This is consistent with the upper right corner of the  $R_{23} O_{32}$ being associated with higher ionization parameters and lower metallicity \citep{Sanders2016}. 

High $O_{32}$ ratios in SF galaxies are associated with high specific SF rates \citep{Llerena2024}, 
which is consistent with SF, rather than AGN activity, being the primary driver of the emission. We note, as in the original BPT classification, \citep{Baldwin1981, Veilleux1987}, that high $O_{32}$ ratios are also associated with AGN activity. The lack of other evidence for AGN activity in the composite spectra implies these high line ratios instead imply SF drivers. In all diagnostic plots, we find little differentiation between the locations of EELGs with and without broad \Ha\ emission. This similarity arises from similar gas conditions and may imply that the extreme emission arises from the same place in EELGs with and without an AGN. While AGN are present in these systems, they do not appear to be the drivers of the extreme emission-line EWs, especially extreme \OIII\ emission, and likely arise from the same physical location as star forming galaxies with no AGN present.

\subsection{Morphology Selection}

We turn to morphology as an additional method for identifying AGN that may be missed in spectroscopy. We compare the morphological sizes of EELGs in filters with strong emission-line contribution and in filters dominated by rest-UV continuum emission and show they are increasingly compact at higher \Hb\ + \OIII\ EWs.

We utilize \texttt{GALFIT} measurements from \citet{McGrath2026} to investigate emission line morphology as potential AGN/SF diagnostic. We take the broad-band image in which \Ha\ or \Hb\ + \OIII\ falls as the emission-line image and select the broad-band image tracing the $2800\lambda$ \AA\ continuum as our reference continuum filter. This represents the reddest available broad-band filter that is not emission-line contaminated for the EELGs in CEERS.

The \Ha\ and/or \Hb\ + \OIII\ luminosity in these systems dominate the photometry in the broad-band filters that capture them because their EWs are much greater than the photometric filter widths \citep{Davis2023}. We assume that the broad-band image capturing the emission line is dominated by the emission line. We also assume that AGN are more likely to cause centrally-concentrated emission-line images, while emission lines from SF are more likely to be extended in a similar fashion in the continuum image. Thus, \Ha\ emission-line images that are more concentrated than the continuum images may suggest AGN as the dominant extreme-emission driver.


\begin{figure}[h!]
    \centering
    \includegraphics[width=1\linewidth]{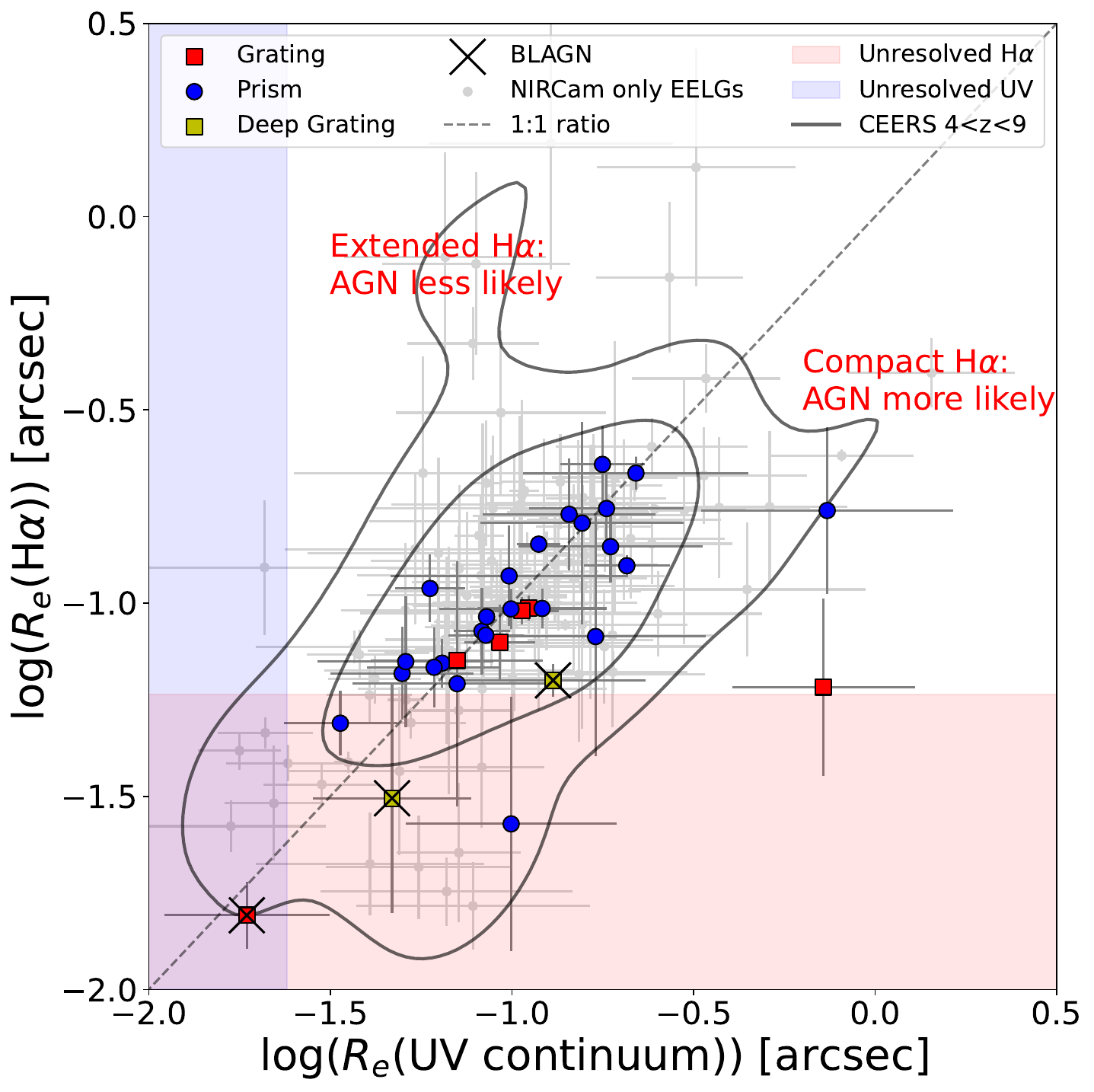}
    \caption{\texttt{GALFIT} effective radius measurements ($R_e$) \citep{McGrath2026} for the image containing \Ha\ and the image probing the $2800\lambda$ \AA\ continuum. EELGs with only NIRCam coverage and high confidence photometric redshifts are plotted in gray. Contours represent 2$\sigma$ and 3$\sigma$ population curves for CEERS background galaxies. BL AGN are marked with black ``X'' markers.}
    \label{fig:compactha}
\end{figure}


This is supported by simulated sizes of high-EW star-forming galaxies \citep{McClymont2025} which show that a galaxy's \Ha\ size relative to its UV size increases more rapidly as a galaxy leaves the SF main sequence and increases in SF rate. HST observations also found star-forming galaxies exhibit larger half-light radii for \Ha\ \citep{Nelson2012} with larger studies showing consistently extended \Ha\ with respect to the stellar continuum \citep{Nelson2016}.
\citet{Morishita2024} found that galaxies in CEERS with high star-formation surface density may also appear centrally compact in \Ha\ images, which introduces some degeneracy for \Ha-compact EELGs. However, \citet{Morishita2024} used direct observational signatures to identify AGN contaminants and may be impacted by AGN detection thresholds. Similarly, \citet{Matharu2024} noted a trend of increasing \Ha\ compactness with \Ha\ EW but concluded this was due to central SF. While this may introduce some ambiguity to the interpretation of \Ha-compact sources, \Ha-diffuse EELGs are unlikely to originate from AGN. We utilize our EELG sample to investigate any apparent emission-line compactness.



The \Ha\ and UV continuum size comparison, Figure \ref{fig:compactha}, shows that all known BL AGN lie in the compact-\Ha\ region of the plot. \Ha\ compactness in EELGs is a potential signpost for AGN activity \citep{Guo2025} and may point to AGN as drivers of the extreme line. The EELGs are otherwise evenly distributed along the 1:1 ratio line, with the exception of a few outliers that are relatively compact in \Ha, implying that the EELG sample is otherwise evenly distributed. However, we note that all AGN are consistent within error with being unresolved in \Ha\ images. All but one are resolved in their UV image and this indicates a tendency towards \Ha\ compactness for EELGs that host AGN. This compact emission points to either AGN activity or central SF driving the extreme lines.

\begin{figure}
    \centering
    \includegraphics[width=1\linewidth]{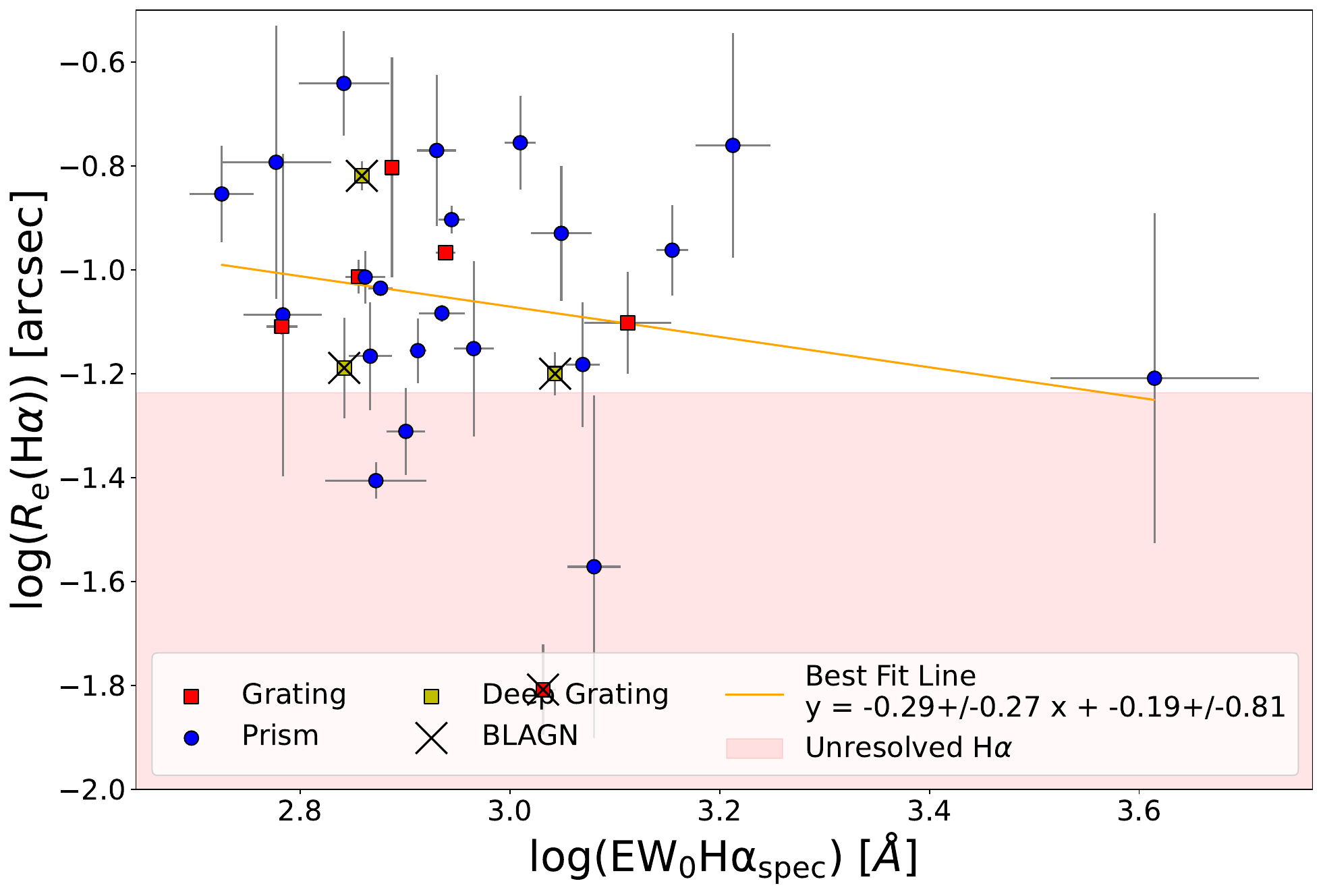}
    \includegraphics[width=1\linewidth]{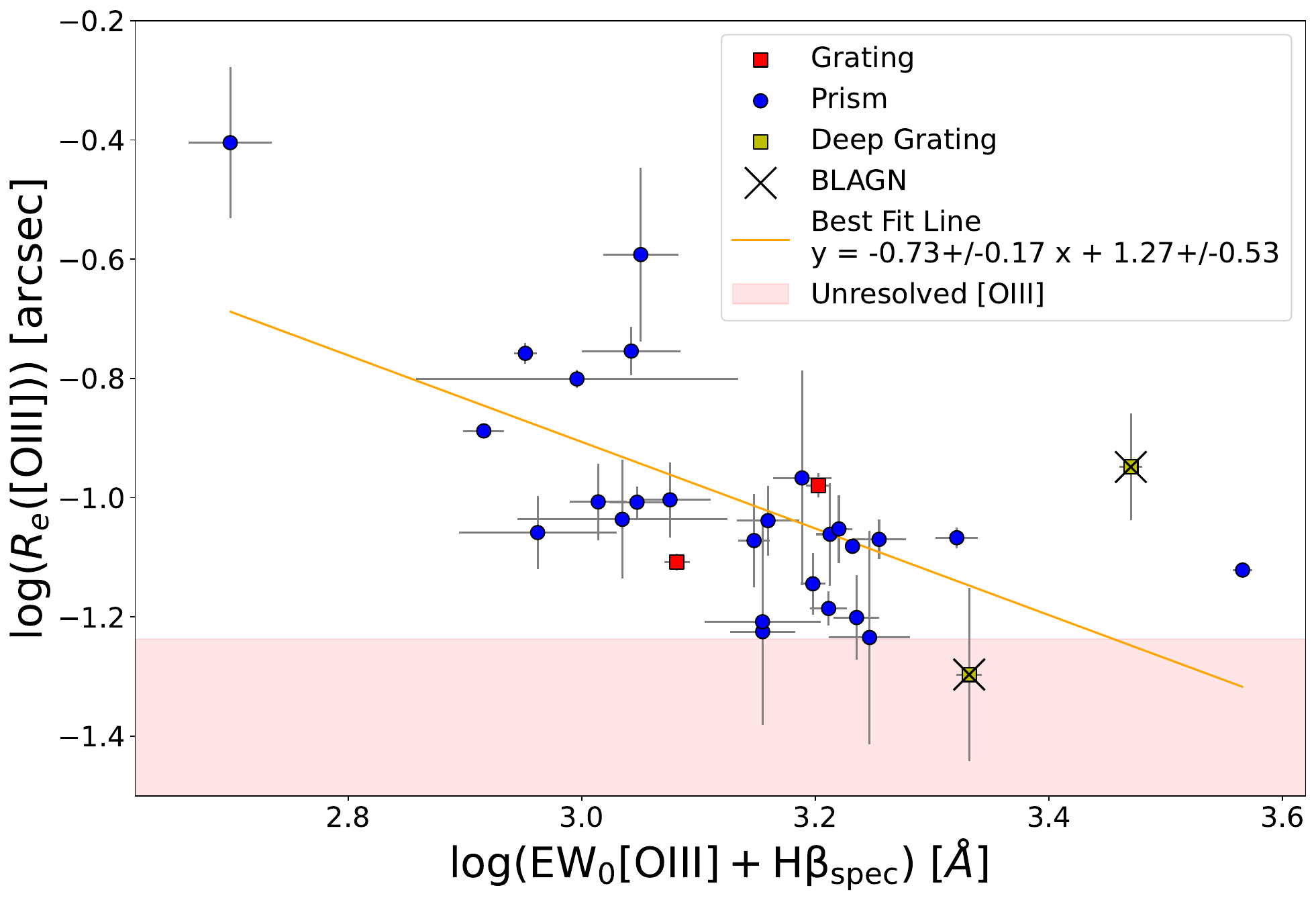}
    \caption{\texttt{GALFIT} effective radius measurements ($R_e$) \citep{McGrath2026} for the image containing \Ha\ (\textbf{top}) or \Hb\ + \OIII\ (\textbf{bottom}) and rest-frame EW. AGN are marked with black ``X'' markers. Red shaded areas indicate unresolved $R_e$.}
    \label{fig:reew}
\end{figure}

The top panel of Figure \ref{fig:reew} investigates a potential relationship between \Ha\ compactness and EW. We find no statistically significant relationship among the sources. This is likely due to scatter among a small sample of spectroscopically observed EELGs since \Ha\ compactness has been noted to correlate with EW by previous studies \citep{Guo2025, Matharu2024, Morishita2024}. 
However, this ambiguity supports our hypothesis that the \Ha\ emission arises from the star-forming regions of the galaxy regardless of emission source.

The bottom panel of Figure \ref{fig:reew} investigates a relationship with \Hb\ + \OIII\ EW and compactness and does find a statistically significant ($>$3$\sigma$) negative corelation. This \OIII\ compactness trend was originally noted in \citet{Davis2023} and is now verified with the spectroscopic EWs. We note the AGN in this plot all occupy the lower right corner of the plot, where we originally predicted them in \citet{Davis2023}. This trend implies that EELGs that host AGN may have more compact \Hb\ + \OIII\ images with respect to their UV continuum, indicating that the \OIII\ emission is highly compact in EELG-AGN.

\section{Summary and Future Work}
\label{sec:sum}

We explore the spectroscopic properties of a photometrically derived EELG sample presented in \citet{Davis2023} from the CEERS legacy field. We verify the presence of EELGs at $4<z<9$, especially when photometric redshift fitting fails to produce the correct redshift solution. 
Optical continuum slopes are especially difficult to extract photometrically in the presence of extreme lines and poor continuum estimations can lead to incorrect photometric EWs, especially in the presence of BL AGN. Photometric searches for extreme line emitters independent of SED template fitting are, however, highly effective at recovering correct object redshifts.

We identify a total of 23 BL AGN in our original EELG sample, 12 of which have spectroscopic EWs that passes our photometric EW criteria. Of these true EELGS that host BL AGN, we find that the narrow component of the \Ha\ line dominates over the broad, especially at higher EW, implying that even when an AGN is present it does not dominate the extreme emission. We find better recovery of BL AGN in deep G395M grating observations and recover several BL AGN missed by previous surveys.
We report coverage of 19 EELGs with deep G395M spectroscopy from the THRILS \citep{Hutchison2025} survey and recover 7 EELGs as BL AGN. One of these has been previously identified as a BL AGN and we report 6 new BL AGN in this work for the first time. We report line fits for these AGN in the appendix.

Remaining spectra are searched for AGN presence through stacks. We identify no high-ionization lines associated with AGN activity in our EW-binned stacks and only recover a \NeV\ emission line in our BL AGN composite spectra. 
The EELGS are not cleanly classified by the ``OHNO'' \citep{Backhaus2022} and VO87 \citep{Veilleux1987} diagnostics. Instead, we demonstrate that the location of EELGs in these diagrams are consistent with MAPPINGS models of high ionization, low metallicity galaxies. We also place the EELGs in an $R_{23} O_{32}$ diagram which points to low-metallicity, high-ionization conditions in all EELGs.

We also place our EELG sample in a color-color selection diagram used to identify LRDs. We correct the colors to remove the emission line in the cases that we resolve an EW and show EELGs selected by this diagram typically have colors dominated by continuum emission rather than emission lines (which dominate in EELGs not selected by these colors). Further, we contextualize the EELGs by investigating the morphology in the broadband filter capturing the emission lines and the UV continuum. We find that AGN tend to be more compact in their \Ha\ filter than their UV filter and find evidence for a significant negative correlation between \Hb\ + \OIII\ equivalent width and \Hb\ + \OIII $R_e$. We note that the AGN in this plot lie at the higher EW, compact \Hb\ + \OIII\ end of the plot as predicted in \citet{Davis2023}.

Upcoming coverage of CEERS in JWST Cycle 4, namely the SPAM (proposal ID 8559, PIs Kelcey Davis \& Rebecca Larson) and MINERVA (proposal ID 7814, PI Adam Muzzin) \citep{Muzzin2025} programs, will augment CEERS with coverage from all medium-band and wide-band photometric filters on NIRCam. This unprecedented coverage will allow for better measurement of EWs, continua, and galaxy properties at $4<z<9$. We will leverage this coverage alongside the lessons learned from this study to further our understanding of EELG properties in the early universe.



\facility{JWST (NIRCam, NIRSpec)}

\software{\texttt{Astropy} \citep{astropy:2013, astropy:2018, astropy:2022} , \texttt{LIME} \citep{Ferandez2024}, \texttt{Matplotlib} \citep{Hunter2007}, \texttt{NumPy} \citep{numpy},
 \texttt{Pandas} \citep{pandas1, pandas2}, \texttt{Scipy} \citep{scipy}} 

\begin{acknowledgments}

 We acknowledge the work of our colleagues in the CEERS, CAPERS, and THRILS collaborations and everyone involved in the JWST mission. KD acknowledges support from a NSF Graduate Research Fellowship award number 2040433. MB acknowledges support from a NSF Graduate Research Fellowship.  JRT, KD, and MB acknowledge support from NSF CAREER-1945546, NASA JWST-GO-06368 and JWST-GO-05718. RA acknowledges financial support from grants PID2023-147386NB-I00 ``XTREM'' funded by MICIU AEI 10.13039 501100011033 and ERDF EU, and the Severo Ochoa grant CEX2021-001131-S funded by MCIN AEI 10.13039 501100011033. KD acknowledges support from Los Alamos National Laboratory. This work was released under LA-UR-25-30978. 

\end{acknowledgments}


\bibliographystyle{aasjournal}
\bibliography{main}{}

\begin{thebibliography}{}
\expandafter\ifx\csname natexlab\endcsname\relax\def\natexlab#1{#1}\fi
\providecommand{\url}[1]{\href{#1}{#1}}
\providecommand{\dodoi}[1]{doi:~\href{http://doi.org/#1}{\nolinkurl{#1}}}
\providecommand{\doeprint}[1]{\href{http://ascl.net/#1}{\nolinkurl{http://ascl.net/#1}}}
\providecommand{\doarXiv}[1]{\href{https://arxiv.org/abs/#1}{\nolinkurl{https://arxiv.org/abs/#1}}}

\bibitem[{{Arrabal Haro} {et~al.}(2023{\natexlab{a}}){Arrabal Haro}, {Dickinson}, {Finkelstein}, {Fujimoto}, {Fern{\'a}ndez}, {Kartaltepe}, {Jung}, {Cole}, {Burgarella}, {Chworowsky}, {Hutchison}, {Morales}, {Papovich}, {Simons}, {Amor{\'\i}n}, {Backhaus}, {Bagley}, {Bisigello}, {Calabr{\`o}}, {Castellano}, {Cleri}, {Dav{\'e}}, {Dekel}, {Ferguson}, {Fontana}, {Gawiser}, {Giavalisco}, {Harish}, {Hathi}, {Hirschmann}, {Holwerda}, {Huertas-Company}, {Koekemoer}, {Larson}, {Lucas}, {Mobasher}, {P{\'e}rez-Gonz{\'a}lez}, {Pirzkal}, {Rose}, {Santini}, {Trump}, {de la Vega}, {Wang}, {Weiner}, {Wilkins}, {Yang}, {Yung}, \& {Zavala}}]{ArrabalHaro23}
{Arrabal Haro}, P., {Dickinson}, M., {Finkelstein}, S.~L., {et~al.} 2023{\natexlab{a}}, \apjl, 951, L22, \dodoi{10.3847/2041-8213/acdd54}

\bibitem[{{Arrabal Haro} {et~al.}(2023{\natexlab{b}}){Arrabal Haro}, {Dickinson}, {Finkelstein}, {Kartaltepe}, {Donnan}, {Burgarella}, {Carnall}, {Cullen}, {Dunlop}, {Fern{\'a}ndez}, {Fujimoto}, {Jung}, {Krips}, {Larson}, {Papovich}, {P{\'e}rez-Gonz{\'a}lez}, {Amor{\'\i}n}, {Bagley}, {Buat}, {Casey}, {Chworowsky}, {Cohen}, {Ferguson}, {Giavalisco}, {Huertas-Company}, {Hutchison}, {Kocevski}, {Koekemoer}, {Lucas}, {McLeod}, {McLure}, {Pirzkal}, {Seill{\'e}}, {Trump}, {Weiner}, {Wilkins}, \& {Zavala}}]{ArrabalHaro2023}
---. 2023{\natexlab{b}}, \nat, 622, 707, \dodoi{10.1038/s41586-023-06521-7}

\bibitem[{{Astropy Collaboration} {et~al.}(2013){Astropy Collaboration}, {Robitaille}, {Tollerud}, {Greenfield}, {Droettboom}, {Bray}, {Aldcroft}, {Davis}, {Ginsburg}, {Price-Whelan}, {Kerzendorf}, {Conley}, {Crighton}, {Barbary}, {Muna}, {Ferguson}, {Grollier}, {Parikh}, {Nair}, {Unther}, {Deil}, {Woillez}, {Conseil}, {Kramer}, {Turner}, {Singer}, {Fox}, {Weaver}, {Zabalza}, {Edwards}, {Azalee Bostroem}, {Burke}, {Casey}, {Crawford}, {Dencheva}, {Ely}, {Jenness}, {Labrie}, {Lim}, {Pierfederici}, {Pontzen}, {Ptak}, {Refsdal}, {Servillat}, \& {Streicher}}]{astropy:2013}
{Astropy Collaboration}, {Robitaille}, T.~P., {Tollerud}, E.~J., {et~al.} 2013, \aap, 558, A33, \dodoi{10.1051/0004-6361/201322068}

\bibitem[{{Astropy Collaboration} {et~al.}(2018){Astropy Collaboration}, {Price-Whelan}, {Sip{\H{o}}cz}, {G{\"u}nther}, {Lim}, {Crawford}, {Conseil}, {Shupe}, {Craig}, {Dencheva}, {Ginsburg}, {VanderPlas}, {Bradley}, {P{\'e}rez-Su{\'a}rez}, {de Val-Borro}, {Aldcroft}, {Cruz}, {Robitaille}, {Tollerud}, {Ardelean}, {Babej}, {Bach}, {Bachetti}, {Bakanov}, {Bamford}, {Barentsen}, {Barmby}, {Baumbach}, {Berry}, {Biscani}, {Boquien}, {Bostroem}, {Bouma}, {Brammer}, {Bray}, {Breytenbach}, {Buddelmeijer}, {Burke}, {Calderone}, {Cano Rodr{\'\i}guez}, {Cara}, {Cardoso}, {Cheedella}, {Copin}, {Corrales}, {Crichton}, {D'Avella}, {Deil}, {Depagne}, {Dietrich}, {Donath}, {Droettboom}, {Earl}, {Erben}, {Fabbro}, {Ferreira}, {Finethy}, {Fox}, {Garrison}, {Gibbons}, {Goldstein}, {Gommers}, {Greco}, {Greenfield}, {Groener}, {Grollier}, {Hagen}, {Hirst}, {Homeier}, {Horton}, {Hosseinzadeh}, {Hu}, {Hunkeler}, {Ivezi{\'c}}, {Jain}, {Jenness}, {Kanarek}, {Kendrew}, {Kern}, {Kerzendorf}, {Khvalko}, {King}, {Kirkby}, {Kulkarni},
  {Kumar}, {Lee}, {Lenz}, {Littlefair}, {Ma}, {Macleod}, {Mastropietro}, {McCully}, {Montagnac}, {Morris}, {Mueller}, {Mumford}, {Muna}, {Murphy}, {Nelson}, {Nguyen}, {Ninan}, {N{\"o}the}, {Ogaz}, {Oh}, {Parejko}, {Parley}, {Pascual}, {Patil}, {Patil}, {Plunkett}, {Prochaska}, {Rastogi}, {Reddy Janga}, {Sabater}, {Sakurikar}, {Seifert}, {Sherbert}, {Sherwood-Taylor}, {Shih}, {Sick}, {Silbiger}, {Singanamalla}, {Singer}, {Sladen}, {Sooley}, {Sornarajah}, {Streicher}, {Teuben}, {Thomas}, {Tremblay}, {Turner}, {Terr{\'o}n}, {van Kerkwijk}, {de la Vega}, {Watkins}, {Weaver}, {Whitmore}, {Woillez}, {Zabalza}, \& {Astropy Contributors}}]{astropy:2018}
{Astropy Collaboration}, {Price-Whelan}, A.~M., {Sip{\H{o}}cz}, B.~M., {et~al.} 2018, \aj, 156, 123, \dodoi{10.3847/1538-3881/aabc4f}

\bibitem[{{Astropy Collaboration} {et~al.}(2022){Astropy Collaboration}, {Price-Whelan}, {Lim}, {Earl}, {Starkman}, {Bradley}, {Shupe}, {Patil}, {Corrales}, {Brasseur}, {N{\"o}the}, {Donath}, {Tollerud}, {Morris}, {Ginsburg}, {Vaher}, {Weaver}, {Tocknell}, {Jamieson}, {van Kerkwijk}, {Robitaille}, {Merry}, {Bachetti}, {G{\"u}nther}, {Aldcroft}, {Alvarado-Montes}, {Archibald}, {B{\'o}di}, {Bapat}, {Barentsen}, {Baz{\'a}n}, {Biswas}, {Boquien}, {Burke}, {Cara}, {Cara}, {Conroy}, {Conseil}, {Craig}, {Cross}, {Cruz}, {D'Eugenio}, {Dencheva}, {Devillepoix}, {Dietrich}, {Eigenbrot}, {Erben}, {Ferreira}, {Foreman-Mackey}, {Fox}, {Freij}, {Garg}, {Geda}, {Glattly}, {Gondhalekar}, {Gordon}, {Grant}, {Greenfield}, {Groener}, {Guest}, {Gurovich}, {Handberg}, {Hart}, {Hatfield-Dodds}, {Homeier}, {Hosseinzadeh}, {Jenness}, {Jones}, {Joseph}, {Kalmbach}, {Karamehmetoglu}, {Ka{\l}uszy{\'n}ski}, {Kelley}, {Kern}, {Kerzendorf}, {Koch}, {Kulumani}, {Lee}, {Ly}, {Ma}, {MacBride}, {Maljaars}, {Muna}, {Murphy}, {Norman},
  {O'Steen}, {Oman}, {Pacifici}, {Pascual}, {Pascual-Granado}, {Patil}, {Perren}, {Pickering}, {Rastogi}, {Roulston}, {Ryan}, {Rykoff}, {Sabater}, {Sakurikar}, {Salgado}, {Sanghi}, {Saunders}, {Savchenko}, {Schwardt}, {Seifert-Eckert}, {Shih}, {Jain}, {Shukla}, {Sick}, {Simpson}, {Singanamalla}, {Singer}, {Singhal}, {Sinha}, {Sip{\H{o}}cz}, {Spitler}, {Stansby}, {Streicher}, {{\v{S}}umak}, {Swinbank}, {Taranu}, {Tewary}, {Tremblay}, {de Val-Borro}, {Van Kooten}, {Vasovi{\'c}}, {Verma}, {de Miranda Cardoso}, {Williams}, {Wilson}, {Winkel}, {Wood-Vasey}, {Xue}, {Yoachim}, {Zhang}, {Zonca}, \& {Astropy Project Contributors}}]{astropy:2022}
{Astropy Collaboration}, {Price-Whelan}, A.~M., {Lim}, P.~L., {et~al.} 2022, \apj, 935, 167, \dodoi{10.3847/1538-4357/ac7c74}

\bibitem[{{Atek} {et~al.}(2024){Atek}, {Labb{\'e}}, {Furtak}, {Chemerynska}, {Fujimoto}, {Setton}, {Miller}, {Oesch}, {Bezanson}, {Price}, {Dayal}, {Zitrin}, {Kokorev}, {Weaver}, {Brammer}, {Dokkum}, {Williams}, {Cutler}, {Feldmann}, {Fudamoto}, {Greene}, {Leja}, {Maseda}, {Muzzin}, {Pan}, {Papovich}, {Nelson}, {Nanayakkara}, {Stark}, {Stefanon}, {Suess}, {Wang}, \& {Whitaker}}]{Atek2024}
{Atek}, H., {Labb{\'e}}, I., {Furtak}, L.~J., {et~al.} 2024, \nat, 626, 975, \dodoi{10.1038/s41586-024-07043-6}

\bibitem[{{Backhaus} {et~al.}(2022){Backhaus}, {Trump}, {Cleri}, {Simons}, {Momcheva}, {Papovich}, {Estrada-Carpenter}, {Finkelstein}, {Matharu}, {Ji}, {Weiner}, {Giavalisco}, \& {Jung}}]{Backhaus2022}
{Backhaus}, B.~E., {Trump}, J.~R., {Cleri}, N.~J., {et~al.} 2022, \apj, 926, 161, \dodoi{10.3847/1538-4357/ac3919}

\bibitem[{{Backhaus} {et~al.}(2025){Backhaus}, {Cleri}, {Trump}, {Kirkpatrick}, {Simons}, {Arrabal Haro}, {Bagley}, {Brooks}, {Calabr{\`o}}, {Davis}, {Dickinson}, {Finkelstein}, {Hirschmann}, {Kartaltepe}, {Koekemoer}, {Llerena}, {Pacucci}, {Pirzkal}, {Papovich}, \& {Wilkins}}]{Backhaus2025}
{Backhaus}, B.~E., {Cleri}, N.~J., {Trump}, J.~R., {et~al.} 2025, arXiv e-prints, arXiv:2502.03519, \dodoi{10.48550/arXiv.2502.03519}

\bibitem[{{Bagley} {et~al.}(2023){Bagley}, {Finkelstein}, {Koekemoer}, {Ferguson}, {Arrabal Haro}, {Dickinson}, {Kartaltepe}, {Papovich}, {P{\'e}rez-Gonz{\'a}lez}, {Pirzkal}, {Somerville}, {Willmer}, {Yang}, {Yung}, {Fontana}, {Grazian}, {Grogin}, {Hirschmann}, {Kewley}, {Kirkpatrick}, {Kocevski}, {Lotz}, {Medrano}, {Morales}, {Pentericci}, {Ravindranath}, {Trump}, {Wilkins}, {Calabr{\`o}}, {Cooper}, {Costantin}, {de la Vega}, {Hilbert}, {Hutchison}, {Larson}, {Lucas}, {McGrath}, {Ryan}, {Wang}, \& {Wuyts}}]{Bagley2023}
{Bagley}, M.~B., {Finkelstein}, S.~L., {Koekemoer}, A.~M., {et~al.} 2023, \apjl, 946, L12, \dodoi{10.3847/2041-8213/acbb08}

\bibitem[{{Baldwin} {et~al.}(1981){Baldwin}, {Phillips}, \& {Terlevich}}]{Baldwin1981}
{Baldwin}, J.~A., {Phillips}, M.~M., \& {Terlevich}, R. 1981, \pasp, 93, 5, \dodoi{10.1086/130766}

\bibitem[{{Barro} {et~al.}(2024){Barro}, {P{\'e}rez-Gonz{\'a}lez}, {Kocevski}, {McGrath}, {Trump}, {Simons}, {Somerville}, {Yung}, {Arrabal Haro}, {Akins}, {Bagley}, {Cleri}, {Costantin}, {Davis}, {Dickinson}, {Finkelstein}, {Giavalisco}, {G{\'o}mez-Guijarro}, {Hathi}, {Hirschmann}, {Holwerda}, {Huertas-Company}, {Kartaltepe}, {Koekemoer}, {Lucas}, {Papovich}, {Pirzkal}, {Seill{\'e}}, {Tacchella}, {Wuyts}, {Wilkins}, {de la Vega}, {Yang}, \& {Zavala}}]{Barro2024}
{Barro}, G., {P{\'e}rez-Gonz{\'a}lez}, P.~G., {Kocevski}, D.~D., {et~al.} 2024, \apj, 963, 128, \dodoi{10.3847/1538-4357/ad167e}

\bibitem[{{Berg} {et~al.}(2021){Berg}, {Chisholm}, {Erb}, {Skillman}, {Pogge}, \& {Olivier}}]{Berg2021}
{Berg}, D.~A., {Chisholm}, J., {Erb}, D.~K., {et~al.} 2021, \apj, 922, 170, \dodoi{10.3847/1538-4357/ac141b}

\bibitem[{{Brammer} {et~al.}(2008){Brammer}, {van Dokkum}, \& {Coppi}}]{Brammer2008}
{Brammer}, G.~B., {van Dokkum}, P.~G., \& {Coppi}, P. 2008, \apj, 686, 1503, \dodoi{10.1086/591786}

\bibitem[{{Brooks} {et~al.}(2025){Brooks}, {Simons}, {Trump}, {Taylor}, {Bagley}, {Backhaus}, {Davis}, {Buat}, {Cleri}, {de la Vega}, {Finkelstein}, {Hirschmann}, {Holwerda}, {Kocevski}, {Koekemoer}, {Lucas}, {Pacucci}, \& {Seill{\'e}}}]{Brooks2025}
{Brooks}, M., {Simons}, R.~C., {Trump}, J.~R., {et~al.} 2025, \apj, 986, 177, \dodoi{10.3847/1538-4357/addac4}

\bibitem[{{Cardamone} {et~al.}(2009){Cardamone}, {Schawinski}, {Sarzi}, {Bamford}, {Bennert}, {Urry}, {Lintott}, {Keel}, {Parejko}, {Nichol}, {Thomas}, {Andreescu}, {Murray}, {Raddick}, {Slosar}, {Szalay}, \& {Vandenberg}}]{Cardamone2009}
{Cardamone}, C., {Schawinski}, K., {Sarzi}, M., {et~al.} 2009, \mnras, 399, 1191, \dodoi{10.1111/j.1365-2966.2009.15383.x}

\bibitem[{{Chisholm} {et~al.}(2024){Chisholm}, {Berg}, {Endsley}, {Gazagnes}, {Richardson}, {Lambrides}, {Greene}, {Finkelstein}, {Flury}, {Guseva}, \& et~al.}]{Chisholm2024}
{Chisholm}, J., {Berg}, D.~A., {Endsley}, R., {et~al.} 2024, \mnras, 534, 2633, \dodoi{10.1093/mnras/stae2199}

\bibitem[{{Cleri} {et~al.}(2023{\natexlab{a}}){Cleri}, {Olivier}, {Hutchison}, {Papovich}, {Trump}, {Amor{\'\i}n}, {Backhaus}, {Berg}, {Fern{\'a}ndez}, {Finkelstein}, \& et~al.}]{Cleri2023}
{Cleri}, N.~J., {Olivier}, G.~M., {Hutchison}, T.~A., {et~al.} 2023{\natexlab{a}}, \apj, 953, 10, \dodoi{10.3847/1538-4357/acde55}

\bibitem[{{Cleri} {et~al.}(2023{\natexlab{b}}){Cleri}, {Yang}, {Papovich}, {Trump}, {Backhaus}, {Estrada-Carpenter}, {Finkelstein}, {Giavalisco}, {Hutchison}, {Ji}, \& et~al.}]{Cleri2023b}
{Cleri}, N.~J., {Yang}, G., {Papovich}, C., {et~al.} 2023{\natexlab{b}}, \apj, 948, 112, \dodoi{10.3847/1538-4357/acc1e6}

\bibitem[{{Cleri} {et~al.}(2025){Cleri}, {Olivier}, {Backhaus}, {Leja}, {Papovich}, {Trump}, {Arrabal Haro}, {Buat}, {Burgarella}, {Burnham}, {Calabro}, {Cohn}, {Cole}, {Davis}, {Dickinson}, {Finkelstein}, {Garner}, {Hirschmann}, {Hu}, {Hutchison}, {Kocevski}, {Koekemoer}, {Larson}, {Lewis}, {Maseda}, {Seille}, \& {Simons}}]{Cleri2025}
{Cleri}, N.~J., {Olivier}, G.~M., {Backhaus}, B.~E., {et~al.} 2025, arXiv e-prints, arXiv:2506.21660, \dodoi{10.48550/arXiv.2506.21660}

\bibitem[{{Cox} {et~al.}(2025){Cox}, {Kartaltepe}, {Bagley}, {Finkelstein}, {Rose}, {Khostovan}, {Chworowsky}, {Ilbert}, {Koekemoer}, {Ferguson}, {Arrabal Haro}, {Backhaus}, {Dickinson}, {Fontana}, {Guo}, {Grazian}, {Grogin}, {Harish}, {Hathi}, {Holwerda}, {Iyer}, {Kewley}, {Kirkpatrick}, {Kocevski}, {Larson}, {Lotz}, {Lucas}, {Pacucci}, {Papovich}, {Pentericci}, {P{\'e}rez-Gonz{\'a}lez}, {Pirzkal}, {Ravindranath}, {Somerville}, {Trump}, {Wilkins}, {Yang}, \& {Yung}}]{Cox2025}
{Cox}, I.~G., {Kartaltepe}, J.~S., {Bagley}, M.~B., {et~al.} 2025, arXiv e-prints, arXiv:2510.08743, \dodoi{10.48550/arXiv.2510.08743}

\bibitem[{{Cueto} {et~al.}(2024){Cueto}, {Hutter}, {Dayal}, {Gottl{\"o}ber}, {Heintz}, {Mason}, {Trebitsch}, \& {Yepes}}]{Cueto2024}
{Cueto}, E.~R., {Hutter}, A., {Dayal}, P., {et~al.} 2024, \aap, 686, A138, \dodoi{10.1051/0004-6361/202349017}

\bibitem[{{Curti} {et~al.}(2024){Curti}, {Maiolino}, {Curtis-Lake}, {Chevallard}, {Carniani}, {D'Eugenio}, {Looser}, {Scholtz}, {Charlot}, {Cameron}, {{\"U}bler}, {Witstok}, {Boyett}, {Laseter}, {Sandles}, {Arribas}, {Bunker}, {Giardino}, {Maseda}, {Rawle}, {Rodr{\'\i}guez Del Pino}, {Smit}, {Willott}, {Eisenstein}, {Hausen}, {Johnson}, {Rieke}, {Robertson}, {Tacchella}, {Williams}, {Willmer}, {Baker}, {Bhatawdekar}, {Egami}, {Helton}, {Ji}, {Kumari}, {Perna}, {Shivaei}, \& {Sun}}]{Curti2024}
{Curti}, M., {Maiolino}, R., {Curtis-Lake}, E., {et~al.} 2024, \aap, 684, A75, \dodoi{10.1051/0004-6361/202346698}

\bibitem[{{Davis} {et~al.}(2024){Davis}, {Trump}, {Simons}, {McGrath}, {Wilkins}, {Arrabal Haro}, {Bagley}, {Dickinson}, {Fern{\'a}ndez}, {Amor{\'\i}n}, {Backhaus}, {Cleri}, {Llerena}, {Brunker}, {Barro}, {Bisigello}, {Brooks}, {Costantin}, {de La Vega}, {Dekel}, {Finkelstein}, {Hathi}, {Hirschmann}, {Kartaltepe}, {Koekemoer}, {Lucas}, {Papovich}, {P{\'e}rez-Gonz{\'a}lez}, {Pirzkal}, {Rodighiero}, {Rose}, {Yung}, \& {Ceers Collaborators}}]{Davis2023}
{Davis}, K., {Trump}, J.~R., {Simons}, R.~C., {et~al.} 2024, \apj, 974, 42, \dodoi{10.3847/1538-4357/ad6865}

\bibitem[{{de Graaff} {et~al.}(2025){de Graaff}, {Brammer}, {Weibel}, {Lewis}, {Maseda}, {Oesch}, {Bezanson}, {Boogaard}, {Cleri}, {Cooper}, {Gottumukkala}, {Greene}, {Hirschmann}, {Hviding}, {Katz}, {Labb{\'e}}, {Leja}, {Matthee}, {McConachie}, {Miller}, {Naidu}, {Price}, {Rix}, {Setton}, {Suess}, {Wang}, {Whitaker}, \& {Williams}}]{deGraaff2025}
{de Graaff}, A., {Brammer}, G., {Weibel}, A., {et~al.} 2025, \aap, 697, A189, \dodoi{10.1051/0004-6361/202452186}

\bibitem[{{Doan} {et~al.}(2025){Doan}, {Satyapal}, {Reefe}, {Sexton}, {Matzko}, {McKaig}, {Secrest}, {Cann}, {Laor}, \& {Canalizo}}]{Doan2025}
{Doan}, S., {Satyapal}, S., {Reefe}, M., {et~al.} 2025, \apjs, 280, 57, \dodoi{10.3847/1538-4365/ade304}

\bibitem[{{Endsley} {et~al.}(2024){Endsley}, {Stark}, {Whitler}, {Topping}, {Johnson}, {Robertson}, {Tacchella}, {Alberts}, {Baker}, {Bhatawdekar}, {Boyett}, {Bunker}, {Cameron}, {Carniani}, {Charlot}, {Chen}, {Chevallard}, {Curtis-Lake}, {Danhaive}, {Egami}, {Eisenstein}, {Hainline}, {Helton}, {Ji}, {Looser}, {Maiolino}, {Nelson}, {Pusk{\'a}s}, {Rieke}, {Rieke}, {Rix}, {Sandles}, {Saxena}, {Simmonds}, {Smit}, {Sun}, {Williams}, {Willmer}, {Willott}, \& {Witstok}}]{Endsley2024}
{Endsley}, R., {Stark}, D.~P., {Whitler}, L., {et~al.} 2024, \mnras, 533, 1111, \dodoi{10.1093/mnras/stae1857}

\bibitem[{{Fern{\'a}ndez} {et~al.}(2024){Fern{\'a}ndez}, {Amor{\'\i}n}, {Firpo}, \& {Morisset}}]{Ferandez2024}
{Fern{\'a}ndez}, V., {Amor{\'\i}n}, R., {Firpo}, V., \& {Morisset}, C. 2024, arXiv e-prints, arXiv:2405.15072, \dodoi{10.48550/arXiv.2405.15072}

\bibitem[{{Finkelstein} {et~al.}(2024){Finkelstein}, {Leung}, {Bagley}, {Dickinson}, {Ferguson}, {Papovich}, {Akins}, {Arrabal Haro}, {Dav{\'e}}, {Dekel}, {Kartaltepe}, {Kocevski}, {Koekemoer}, {Pirzkal}, {Somerville}, {Yung}, {Amor{\'\i}n}, {Backhaus}, {Behroozi}, {Bisigello}, {Bromm}, {Casey}, {Ch{\'a}vez Ortiz}, {Cheng}, {Chworowsky}, {Cleri}, {Cooper}, {Davis}, {de la Vega}, {Elbaz}, {Franco}, {Fontana}, {Fujimoto}, {Giavalisco}, {Grogin}, {Holwerda}, {Huertas-Company}, {Hirschmann}, {Iyer}, {Jogee}, {Jung}, {Larson}, {Lucas}, {Mobasher}, {Morales}, {Morley}, {Mukherjee}, {P{\'e}rez-Gonz{\'a}lez}, {Ravindranath}, {Rodighiero}, {Rowland}, {Tacchella}, {Taylor}, {Trump}, \& {Wilkins}}]{Finkelstein2024}
{Finkelstein}, S.~L., {Leung}, G. C.~K., {Bagley}, M.~B., {et~al.} 2024, \apjl, 969, L2, \dodoi{10.3847/2041-8213/ad4495}

\bibitem[{{Finkelstein} {et~al.}(2025){Finkelstein}, {Bagley}, {Arrabal Haro}, {Dickinson}, {Ferguson}, {Kartaltepe}, {Kocevski}, {Koekemoer}, {Lotz}, {Papovich}, {P{\'e}rez-Gonz{\'a}lez}, {Pirzkal}, {Somerville}, {Trump}, {Yang}, {Yung}, {Fontana}, {Grazian}, {Grogin}, {Kewley}, {Kirkpatrick}, {Larson}, {Pentericci}, {Ravindranath}, {Wilkins}, {Almaini}, {Amor{\'\i}n}, {Barro}, {Bhatawdekar}, {Bisigello}, {Brooks}, {Buat}, {Buitrago}, {Burgarella}, {Calabr{\`o}}, {Castellano}, {Cheng}, {Cleri}, {Cole}, {Cooper}, {Cooper}, {Costantin}, {Cox}, {Croton}, {Daddi}, {Davis}, {Dekel}, {Elbaz}, {Fern{\'a}ndez}, {Fujimoto}, {Gandolfi}, {Gardner}, {Gawiser}, {Giavalisco}, {G{\'o}mez-Guijarro}, {Guo}, {Gupta}, {Hathi}, {Harish}, {Henry}, {Hirschmann}, {Hu}, {Hutchison}, {Iyer}, {Jaskot}, {Jha}, {Jung}, {Kassin}, {Kokorev}, {Kurczynski}, {Leung}, {Llerena}, {Long}, {Lucas}, {Lu}, {McGrath}, {McIntosh}, {Merlin}, {Mobasher}, {Morales}, {Napolitano}, {Pacucci}, {Pandya}, {Rafelski}, {Rodighiero}, {Rose}, {Santini},
  {Seill{\'e}}, {Simons}, {Shen}, {Straughn}, {Tacchella}, {Taylor}, {Vanderhoof}, {Vega-Ferrero}, {Weiner}, {Willmer}, {Zhu}, {Bell}, {Wuyts}, {Holwerda}, {Wang}, {Wang}, {Zavala}, \& {CEERS Collaboration}}]{Finkelstein2025}
{Finkelstein}, S.~L., {Bagley}, M.~B., {Arrabal Haro}, P., {et~al.} 2025, \apjl, 983, L4, \dodoi{10.3847/2041-8213/adbbd3}

\bibitem[{{Fujimoto} {et~al.}(2023){Fujimoto}, {Arrabal Haro}, {Dickinson}, {Finkelstein}, {Kartaltepe}, {Larson}, {Burgarella}, {Bagley}, {Behroozi}, {Chworowsky}, {Hirschmann}, {Trump}, {Wilkins}, {Yung}, {Koekemoer}, {Papovich}, {Pirzkal}, {Ferguson}, {Fontana}, {Grogin}, {Grazian}, {Kewley}, {Kocevski}, {Lotz}, {Pentericci}, {Ravindranath}, {Somerville}, {Wilkins}, {Amor{\'\i}n}, {Backhaus}, {Calabr{\`o}}, {Casey}, {Cooper}, {Fern{\'a}ndez}, {Franco}, {Giavalisco}, {Hathi}, {Harish}, {Hutchison}, {Iyer}, {Jung}, {Lucas}, \& {Zavala}}]{Fujimoto2023}
{Fujimoto}, S., {Arrabal Haro}, P., {Dickinson}, M., {et~al.} 2023, \apjl, 949, L25, \dodoi{10.3847/2041-8213/acd2d9}

\bibitem[{{Gardner} {et~al.}(2006){Gardner}, {Mather}, {Clampin}, {Doyon}, {Greenhouse}, {Hammel}, {Hutchings}, {Jakobsen}, {Lilly}, {Long}, {Lunine}, {McCaughrean}, {Mountain}, {Nella}, {Rieke}, {Rieke}, {Rix}, {Smith}, {Sonneborn}, {Stiavelli}, {Stockman}, {Windhorst}, \& {Wright}}]{Gardner2006}
{Gardner}, J.~P., {Mather}, J.~C., {Clampin}, M., {et~al.} 2006, \ssr, 123, 485, \dodoi{10.1007/s11214-006-8315-7}

\bibitem[{{Gardner} {et~al.}(2023){Gardner}, {Mather}, {Abbott}, {Abell}, {Abernathy}, {Abney}, {Abraham}, {Abraham}, {Abul-Huda}, {Acton}, \& et~al.}]{Gardner2023}
{Gardner}, J.~P., {Mather}, J.~C., {Abbott}, R., {et~al.} 2023, \pasp, 135, 068001, \dodoi{10.1088/1538-3873/acd1b5}

\bibitem[{Gelbord {et~al.}(2009)Gelbord, Mullaney, \& Ward}]{Gelbord2009}
Gelbord, J.~M., Mullaney, J.~R., \& Ward, M.~J. 2009, Monthly Notices of the Royal Astronomical Society, 397, 172, \dodoi{10.1111/j.1365-2966.2009.14961.x}

\bibitem[{{Gim{\'e}nez-Alc{\'a}zar} {et~al.}(2025){Gim{\'e}nez-Alc{\'a}zar}, {Amor{\'\i}n}, {V{\'\i}lchez}, {Hern{\'a}n-Caballero}, {Gonz{\'a}lez-Otero}, {Arroyo-Polonio}, {Iglesias-P{\'a}ramo}, {Lumbreras-Calle}, {Fern{\'a}ndez-Ontiveros}, {Bonatto}, {Gonz{\'a}lez Delgado}, {Kehrig}, {Torralba}, {Rahna}, {Jim{\'e}nez-Teja}, {M{\'a}rquez}, {Breda}, {{\'A}lvarez-Candal}, {Abramo}, {Alcaniz}, {Benitez}, {Bonoli}, {Carneiro}, {Cenarro}, {Crist{\'o}bal-Hornillos}, {Dupke}, {Ederoclite}, {Hern{\'a}ndez-Monteagudo}, {Mar{\'\i}n-Franch}, {Mendes de Oliveira}, {Moles}, {Sodr{\'e}}, {Taylor}, {Varela}, \& {V{\'a}zquez Rami{\'o}}}]{Gimenez2025}
{Gim{\'e}nez-Alc{\'a}zar}, A., {Amor{\'\i}n}, R., {V{\'\i}lchez}, J.~M., {et~al.} 2025, arXiv e-prints, arXiv:2512.08484, \dodoi{10.48550/arXiv.2512.08484}

\bibitem[{{Guo} {et~al.}(2025){Guo}, {Onoue}, {Inayoshi}, {Kocevski}, {Finkelstein}, {Bagley}, \& {McGrath}}]{Guo2025}
{Guo}, J., {Onoue}, M., {Inayoshi}, K., {et~al.} 2025, \apj, 991, 74, \dodoi{10.3847/1538-4357/adee0f}

\bibitem[{{Harikane} {et~al.}(2023){Harikane}, {Zhang}, {Nakajima}, {Ouchi}, {Isobe}, {Ono}, {Hatano}, {Xu}, \& {Umeda}}]{Harikane2023}
{Harikane}, Y., {Zhang}, Y., {Nakajima}, K., {et~al.} 2023, \apj, 959, 39, \dodoi{10.3847/1538-4357/ad029e}

\bibitem[{Harris {et~al.}(2020)Harris, Millman, van~der Walt, Gommers, Virtanen, Cournapeau, Wieser, Taylor, Berg, Smith, Kern, Picus, Hoyer, van Kerkwijk, Brett, Haldane, del R{\'{i}}o, Wiebe, Peterson, G{\'{e}}rard-Marchant, Sheppard, Reddy, Weckesser, Abbasi, Gohlke, \& Oliphant}]{numpy}
Harris, C.~R., Millman, K.~J., van~der Walt, S.~J., {et~al.} 2020, Nature, 585, 357, \dodoi{10.1038/s41586-020-2649-2}

\bibitem[{Hunter(2007)}]{Hunter2007}
Hunter, J.~D. 2007, Computing in Science \& Engineering, 9, 90, \dodoi{10.1109/MCSE.2007.55}

\bibitem[{{Hutchison} {et~al.}(2025){Hutchison}, {Larson}, {Arrabal Haro}, {Lambrides}, {Chworowsky}, {Khullar}, {Davis}, {Finkelstein}, {Rigby}, {Barro}, {Cleri}, {Kocevski}, {Antwi-Danso}, {Bagley}, {Berg}, {Bromm}, {Chavez Ortiz}, {Chisholm}, {Coffin}, {Cooper}, {Cooper}, {Cox}, {Dickinson}, {Ferguson}, {Franco}, {Gardner}, {Ganapathy}, {Grogin}, {Hirschmann}, {Huertas-Company}, {Jung}, {Kartaltepe}, {Koekemoer}, {Lucas}, {McGrath}, {Morales}, {Olivier}, {Papovich}, {Perez-Gonzalez}, {Pirzkal}, {Somerville}, {Taylor}, {Trump}, {Vanderhoof}, {Weiner}, {Welch}, {Yung}, {Zavala}, \& {the THRILS collaboration}}]{Hutchison2025}
{Hutchison}, T.~A., {Larson}, R.~L., {Arrabal Haro}, P., {et~al.} 2025, arXiv e-prints, arXiv:2512.12509, \dodoi{10.48550/arXiv.2512.12509}

\bibitem[{{Hviding} {et~al.}(2025){Hviding}, {de Graaff}, {Miller}, {Setton}, {Greene}, {Labb{\'e}}, {Brammer}, {Bezanson}, {Boogaard}, {Cleri}, {Leja}, {Maseda}, {McConachie}, {Matthee}, {Naidu}, {Oesch}, {Wang}, {Whitaker}, \& {Williams}}]{Hviding2025}
{Hviding}, R.~E., {de Graaff}, A., {Miller}, T.~B., {et~al.} 2025, arXiv e-prints, arXiv:2506.05459, \dodoi{10.48550/arXiv.2506.05459}

\bibitem[{{Izotov} {et~al.}(2012){Izotov}, {Thuan}, \& {Guseva}}]{Izotov2012}
{Izotov}, Y.~I., {Thuan}, T.~X., \& {Guseva}, N.~G. 2012, \aap, 546, A122, \dodoi{10.1051/0004-6361/201219733}

\bibitem[{{Izotov} {et~al.}(2021){Izotov}, {Thuan}, \& {Guseva}}]{Izotov2021}
---. 2021, \mnras, 508, 2556, \dodoi{10.1093/mnras/stab2798}

\bibitem[{{Jakobsen} {et~al.}(2022){Jakobsen}, {Ferruit}, {Alves de Oliveira}, {Arribas}, {Bagnasco}, {Barho}, {Beck}, {Birkmann}, {B{\"o}ker}, {Bunker}, {Charlot}, {de Jong}, {de Marchi}, {Ehrenwinkler}, {Falcolini}, {Fels}, {Franx}, {Franz}, {Funke}, {Giardino}, {Gnata}, {Holota}, {Honnen}, {Jensen}, {Jentsch}, {Johnson}, {Jollet}, {Karl}, {Kling}, {K{\"o}hler}, {Kolm}, {Kumari}, {Lander}, {Lemke}, {L{\'o}pez-Caniego}, {L{\"u}tzgendorf}, {Maiolino}, {Manjavacas}, {Marston}, {Maschmann}, {Maurer}, {Messerschmidt}, {Moseley}, {Mosner}, {Mott}, {Muzerolle}, {Pirzkal}, {Pittet}, {Plitzke}, {Posselt}, {Rapp}, {Rauscher}, {Rawle}, {Rix}, {R{\"o}del}, {Rumler}, {Sabbi}, {Salvignol}, {Schmid}, {Sirianni}, {Smith}, {Strada}, {te Plate}, {Valenti}, {Wettemann}, {Wiehe}, {Wiesmayer}, {Willott}, {Wright}, {Zeidler}, \& {Zincke}}]{Jakobsen2022}
{Jakobsen}, P., {Ferruit}, P., {Alves de Oliveira}, C., {et~al.} 2022, \aap, 661, A80, \dodoi{10.1051/0004-6361/202142663}

\bibitem[{{Kennicutt} \& {Evans}(2012)}]{Kennicutt2012}
{Kennicutt}, R.~C., \& {Evans}, N.~J. 2012, \araa, 50, 531, \dodoi{10.1146/annurev-astro-081811-125610}

\bibitem[{{Kewley} {et~al.}(2006){Kewley}, {Groves}, {Kauffmann}, \& {Heckman}}]{Kewley2006}
{Kewley}, L.~J., {Groves}, B., {Kauffmann}, G., \& {Heckman}, T. 2006, \mnras, 372, 961, \dodoi{10.1111/j.1365-2966.2006.10859.x}

\bibitem[{{Kewley} {et~al.}(2013){Kewley}, {Maier}, {Yabe}, {Ohta}, {Akiyama}, {Dopita}, \& {Yuan}}]{Kewley2013}
{Kewley}, L.~J., {Maier}, C., {Yabe}, K., {et~al.} 2013, \apjl, 774, L10, \dodoi{10.1088/2041-8205/774/1/L10}

\bibitem[{{Kewley} {et~al.}(2019{\natexlab{a}}){Kewley}, {Nicholls}, {Sutherland}, {Rigby}, {Acharya}, {Dopita}, \& {Bayliss}}]{Kewley2019a}
{Kewley}, L.~J., {Nicholls}, D.~C., {Sutherland}, R., {et~al.} 2019{\natexlab{a}}, \apj, 880, 16, \dodoi{10.3847/1538-4357/ab16ed}

\bibitem[{{Kewley} {et~al.}(2019{\natexlab{b}}){Kewley}, {Nicholls}, \& {Sutherland}}]{Kewley2019}
{Kewley}, L.~J., {Nicholls}, D.~C., \& {Sutherland}, R.~S. 2019{\natexlab{b}}, \araa, 57, 511, \dodoi{10.1146/annurev-astro-081817-051832}

\bibitem[{{Kocevski} {et~al.}(2023){Kocevski}, {Onoue}, {Inayoshi}, {Trump}, {Arrabal Haro}, {Grazian}, {Dickinson}, {Finkelstein}, {Kartaltepe}, {Hirschmann}, {Aird}, {Holwerda}, {Fujimoto}, {Juneau}, {Amor{\'\i}n}, {Backhaus}, {Bagley}, {Barro}, {Bell}, {Bisigello}, {Calabr{\`o}}, {Cleri}, {Cooper}, {Ding}, {Grogin}, {Ho}, {Hutchison}, {Inoue}, {Jiang}, {Jones}, {Koekemoer}, {Li}, {Li}, {McGrath}, {Molina}, {Papovich}, {P{\'e}rez-Gonz{\'a}lez}, {Pirzkal}, {Wilkins}, {Yang}, \& {Yung}}]{Kocevski2023}
{Kocevski}, D.~D., {Onoue}, M., {Inayoshi}, K., {et~al.} 2023, \apjl, 954, L4, \dodoi{10.3847/2041-8213/ace5a0}

\bibitem[{{Kocevski} {et~al.}(2025){Kocevski}, {Finkelstein}, {Barro}, {Taylor}, {Calabr{\`o}}, {Laloux}, {Buchner}, {Trump}, {Leung}, {Yang}, {Dickinson}, {P{\'e}rez-Gonz{\'a}lez}, {Pacucci}, {Inayoshi}, {Somerville}, {McGrath}, {Akins}, {Bagley}, {Bowler}, {Bisigello}, {Carnall}, {Casey}, {Cheng}, {Cleri}, {Costantin}, {Cullen}, {Davis}, {Donnan}, {Dunlop}, {Ellis}, {Ferguson}, {Fujimoto}, {Fontana}, {Giavalisco}, {Grazian}, {Grogin}, {Hathi}, {Hirschmann}, {Huertas-Company}, {Holwerda}, {Illingworth}, {Juneau}, {Kartaltepe}, {Koekemoer}, {Li}, {Lucas}, {Magee}, {Mason}, {McLeod}, {McLure}, {Napolitano}, {Papovich}, {Pirzkal}, {Rodighiero}, {Santini}, {Wilkins}, \& {Yung}}]{Kocevski2024}
{Kocevski}, D.~D., {Finkelstein}, S.~L., {Barro}, G., {et~al.} 2025, \apj, 986, 126, \dodoi{10.3847/1538-4357/adbc7d}

\bibitem[{{Kokorev} {et~al.}(2025){Kokorev}, {Ch{\'a}vez Ortiz}, {Taylor}, {Finkelstein}, {Arrabal Haro}, {Dickinson}, {Chisholm}, {Fujimoto}, {noz}, {Endsley}, {Hu}, {Napolitano}, {Wilkins}, {Akins}, {Amori{\'\i}n}, {Casey}, {Cheng}, {Cleri}, {Cole}, {Cullen}, {Daddi}, {Davis}, {Donnan}, {Dunlop}, {Fern{\'a}ndez}, {Giavalisco}, {Grogin}, {Hathi}, {Hirschmann}, {Kartaltepe}, {Koekemoer}, {Leung}, {Lucas}, {McLeod}, {Papovich}, {Pentericci}, {P{\'e}rez-Gonz{\'a}lez}, {Somerville}, {Wang}, {Yung}, \& {Zavala}}]{Kokorev2025}
{Kokorev}, V., {Ch{\'a}vez Ortiz}, {\'O}.~A., {Taylor}, A.~J., {et~al.} 2025, \apjl, 988, L10, \dodoi{10.3847/2041-8213/ade8f5}

\bibitem[{{Kroupa} \& {Weidner}(2003)}]{Kroupa}
{Kroupa}, P., \& {Weidner}, C. 2003, \apj, 598, 1076, \dodoi{10.1086/379105}

\bibitem[{{Larson} {et~al.}(2023){Larson}, {Finkelstein}, {Kocevski}, {Hutchison}, {Trump}, {Arrabal Haro}, {Bromm}, {Cleri}, {Dickinson}, {Fujimoto}, {Kartaltepe}, {Koekemoer}, {Papovich}, {Pirzkal}, {Tacchella}, {Zavala}, {Bagley}, {Behroozi}, {Champagne}, {Cole}, {Jung}, {Morales}, {Yang}, {Zhang}, {Zitrin}, {Amor{\'\i}n}, {Burgarella}, {Casey}, {Ch{\'a}vez Ortiz}, {Cox}, {Chworowsky}, {Fontana}, {Gawiser}, {Grazian}, {Grogin}, {Harish}, {Hathi}, {Hirschmann}, {Holwerda}, {Juneau}, {Leung}, {Lucas}, {McGrath}, {P{\'e}rez-Gonz{\'a}lez}, {Rigby}, {Seill{\'e}}, {Simons}, {de La Vega}, {Weiner}, {Wilkins}, {Yung}, \& {Ceers Team}}]{Larson2023}
{Larson}, R.~L., {Finkelstein}, S.~L., {Kocevski}, D.~D., {et~al.} 2023, \apjl, 953, L29, \dodoi{10.3847/2041-8213/ace619}

\bibitem[{{Llerena} {et~al.}(2024){Llerena}, {Amor{\'\i}n}, {Pentericci}, {Arrabal Haro}, {Backhaus}, {Bagley}, {Calabr{\`o}}, {Cleri}, {Davis}, {Dickinson}, {Finkelstein}, {Gawiser}, {Grogin}, {Hathi}, {Hirschmann}, {Kartaltepe}, {Koekemoer}, {McGrath}, {Mobasher}, {Napolitano}, {Papovich}, {Pirzkal}, {Trump}, {Wilkins}, \& {Yung}}]{Llerena2024}
{Llerena}, M., {Amor{\'\i}n}, R., {Pentericci}, L., {et~al.} 2024, \aap, 691, A59, \dodoi{10.1051/0004-6361/202449904}

\bibitem[{{Llerena} {et~al.}(2025){Llerena}, {Pentericci}, {Amor{\'\i}n}, {Ferrara}, {Dickinson}, {Arevalo}, {Calabr{\`o}}, {Napolitano}, {Mascia}, {Arrabal Haro}, {Begley}, {Cleri}, {Davis}, {Hu}, {Kartaltepe}, {Koekemoer}, {Lucas}, {McGrath}, {McLeod}, {Papovich}, {Stanton}, {Taylor}, {Tripodi}, {Wang}, \& {Yung}}]{Llerena2025}
{Llerena}, M., {Pentericci}, L., {Amor{\'\i}n}, R., {et~al.} 2025, arXiv e-prints, arXiv:2510.25647, \dodoi{10.48550/arXiv.2510.25647}

\bibitem[{{Madau} {et~al.}(2024){Madau}, {Giallongo}, {Grazian}, \& {Haardt}}]{Madau2024}
{Madau}, P., {Giallongo}, E., {Grazian}, A., \& {Haardt}, F. 2024, \apj, 971, 75, \dodoi{10.3847/1538-4357/ad5ce8}

\bibitem[{{Matharu} {et~al.}(2024){Matharu}, {Nelson}, {Brammer}, {Oesch}, {Allen}, {Shivaei}, {Naidu}, {Chisholm}, {Covelo-Paz}, {Fudamoto}, {Giovinazzo}, {Herard-Demanche}, {Kerutt}, {Kramarenko}, {Marchesini}, {Meyer}, {Prieto-Lyon}, {Reddy}, {Shuntov}, {Weibel}, {Wuyts}, \& {Xiao}}]{Matharu2024}
{Matharu}, J., {Nelson}, E.~J., {Brammer}, G., {et~al.} 2024, \aap, 690, A64, \dodoi{10.1051/0004-6361/202450522}

\bibitem[{{Matthee} {et~al.}(2023){Matthee}, {Mackenzie}, {Simcoe}, {Kashino}, {Lilly}, {Bordoloi}, \& {Eilers}}]{Mathee2023}
{Matthee}, J., {Mackenzie}, R., {Simcoe}, R.~A., {et~al.} 2023, \apj, 950, 67, \dodoi{10.3847/1538-4357/acc846}

\bibitem[{{Matthee} {et~al.}(2024){Matthee}, {Naidu}, {Brammer}, {Chisholm}, {Eilers}, {Goulding}, {Greene}, {Kashino}, {Labbe}, {Lilly}, {Mackenzie}, {Oesch}, {Weibel}, {Wuyts}, {Xiao}, {Bordoloi}, {Bouwens}, {van Dokkum}, {Illingworth}, {Kramarenko}, {Maseda}, {Mason}, {Meyer}, {Nelson}, {Reddy}, {Shivaei}, {Simcoe}, \& {Yue}}]{Mathee2024}
{Matthee}, J., {Naidu}, R.~P., {Brammer}, G., {et~al.} 2024, \apj, 963, 129, \dodoi{10.3847/1538-4357/ad2345}

\bibitem[{{Mazzolari} {et~al.}(2024){Mazzolari}, {Scholtz}, {Maiolino}, {Gilli}, {Traina}, {L{\'o}pez}, {{\"U}bler}, {Trefoloni}, {D'Eugenio}, {Ji}, {Mignoli}, {Vito}, \& {Brusa}}]{Mazzolari2024}
{Mazzolari}, G., {Scholtz}, J., {Maiolino}, R., {et~al.} 2024, arXiv e-prints, arXiv:2408.15615, \dodoi{10.48550/arXiv.2408.15615}

\bibitem[{{Mazzolari} {et~al.}(2025){Mazzolari}, {Scholtz}, {Maiolino}, {Gilli}, {Traina}, {L{\'o}pez}, {{\"U}bler}, {Trefoloni}, {D'Eugenio}, {Ji}, {Mignoli}, {Vito}, {Vignali}, \& {Brusa}}]{Mazzolari2025}
---. 2025, \aap, 700, A12, \dodoi{10.1051/0004-6361/202451860}

\bibitem[{{McClymont} {et~al.}(2025){McClymont}, {Tacchella}, {Smith}, {Kannan}, {Puchwein}, {Borrow}, {Garaldi}, {Keating}, {Vogelsberger}, {Zier}, {Shen}, \& {Popovic}}]{McClymont2025}
{McClymont}, W., {Tacchella}, S., {Smith}, A., {et~al.} 2025, arXiv e-prints, arXiv:2503.04894, \dodoi{10.48550/arXiv.2503.04894}

\bibitem[{McGrath {et~al.}(2026)McGrath, Finkelstein, Barro, Pandya, Ferguson, Kartaltepe, Kocevski, Amorín, Backhaus, Buitrago, Calabrò, Cheng, Costantin, Cox, Davis, Gandolfi, Guo, Hathi, Hirschmann, Holwerda, Huertas-Company, Koekemoer, Lucas, Mobasher, Pacucci, Papovich, Pérez-González, Trump, Yung, Arrabal~Haro, Bagley, Dickinson, Fontana, Grazian, Grogin, Kewley, Kirkpatrick, Lotz, Pentericci, Pirzkal, Ravindranath, Somerville, Wilkins, Yang, Seillé, \& Wang}]{McGrath2026}
McGrath, E.~J., Finkelstein, S.~L., Barro, G., {et~al.} 2026, The Astrophysical Journal Letters, 999, L6, \dodoi{10.3847/2041-8213/ae3da2}

\bibitem[{{McKaig} {et~al.}(2024){McKaig}, {Satyapal}, {Laor}, {Abel}, {Doan}, {Ricci}, \& {Cann}}]{McKaig2024}
{McKaig}, J.~D., {Satyapal}, S., {Laor}, A., {et~al.} 2024, \apj, 976, 130, \dodoi{10.3847/1538-4357/ad7a79}

\bibitem[{{Mingozzi} {et~al.}(2025){Mingozzi}, {Garcia del Valle-Espinosa}, {James}, {Rickards Vaught}, {Hayes}, {Amor{\'\i}n}, {Leitherer}, {Aloisi}, {Hunt}, {Law}, \& et~al.}]{Mingozzi2025}
{Mingozzi}, M., {Garcia del Valle-Espinosa}, M., {James}, B.~L., {et~al.} 2025, \apj, 985, 253, \dodoi{10.3847/1538-4357/adc996}

\bibitem[{{Molina} {et~al.}(2021){Molina}, {Reines}, {Greene}, {Darling}, \& {Condon}}]{Molina2021}
{Molina}, M., {Reines}, A.~E., {Greene}, J.~E., {Darling}, J., \& {Condon}, J.~J. 2021, \apj, 910, 5, \dodoi{10.3847/1538-4357/abe120}

\bibitem[{{Morishita} {et~al.}(2024){Morishita}, {Stiavelli}, {Chary}, {Trenti}, {Bergamini}, {Chiaberge}, {Leethochawalit}, {Roberts-Borsani}, {Shen}, \& {Treu}}]{Morishita2024}
{Morishita}, T., {Stiavelli}, M., {Chary}, R.-R., {et~al.} 2024, \apj, 963, 9, \dodoi{10.3847/1538-4357/ad1404}

\bibitem[{{Muzzin} {et~al.}(2025){Muzzin}, {Suess}, {Marchesini}, {Robbins}, {Willott}, {Alberts}, {Antwi-Danso}, {Asada}, {Brammer}, {Cutler}, {Iyer}, {Labbe}, {Martis}, {Miller}, {Mitsuhashi}, {Pope}, {Sajina}, {Sarrouh}, {Sharma}, {Stefanon}, {Whitaker}, {Abraham}, {Atek}, {Bradac}, {Berek}, {Bezanson}, {Brown}, {Burgasser}, {Chicoine}, {Cloonan}, {Cooper}, {Dayal}, {de Graaff}, {Desprez}, {Feldmann}, {Forrest}, {Franx}, {Fudamoto}, {Fujimoto}, {Furtak}, {Glazebrook}, {Goovaerts}, {Greene}, {Jagga}, {Jarvis}, {Kriek}, {Khullar}, {La Torre}, {Leja}, {Lin}, {Lorenz}, {Lyon}, {Markov}, {Maseda}, {McConachie}, {Merchant}, {Merida}, {Mowla}, {Myers}, {Naidu}, {Nanayakkara}, {Nelson}, {Noirot}, {Oesch}, {Omori}, {Pan}, {Porraz Barrera}, {Price}, {Ravindranath}, {Sawicki}, {Setton}, {Smit}, {Sok}, {Speagle}, {Taylor}, {Tan}, {Tripodi}, {van der Wel}, {Perez Vidal}, {Wang}, {Weaver}, {Williams}, {Withers}, \& {Zaidi}}]{Muzzin2025}
{Muzzin}, A., {Suess}, K.~A., {Marchesini}, D., {et~al.} 2025, arXiv e-prints, arXiv:2507.19706, \dodoi{10.48550/arXiv.2507.19706}

\bibitem[{{Navarro-Carrera} {et~al.}(2024){Navarro-Carrera}, {Rinaldi}, {Caputi}, {Iani}, {Kokorev}, {Kerutt}, \& {Cooper}}]{Navarro-Carrera2024}
{Navarro-Carrera}, R., {Rinaldi}, P., {Caputi}, K.~I., {et~al.} 2024, arXiv e-prints, arXiv:2410.23249, \dodoi{10.48550/arXiv.2410.23249}

\bibitem[{{Negus} {et~al.}(2023){Negus}, {Comerford}, {S{\'a}nchez}, {Revalski}, {Riffel}, {Bundy}, {Nevin}, \& {Rembold}}]{Negus2023}
{Negus}, J., {Comerford}, J.~M., {S{\'a}nchez}, F.~M., {et~al.} 2023, \apj, 945, 127, \dodoi{10.3847/1538-4357/acb772}

\bibitem[{{Nelson} {et~al.}(2012){Nelson}, {van Dokkum}, {Brammer}, {F{\"o}rster Schreiber}, {Franx}, {Fumagalli}, {Patel}, {Rix}, {Skelton}, {Bezanson}, {Da Cunha}, {Kriek}, {Labbe}, {Lundgren}, {Quadri}, \& {Schmidt}}]{Nelson2012}
{Nelson}, E.~J., {van Dokkum}, P.~G., {Brammer}, G., {et~al.} 2012, \apjl, 747, L28, \dodoi{10.1088/2041-8205/747/2/L28}

\bibitem[{{Nelson} {et~al.}(2016){Nelson}, {van Dokkum}, {F{\"o}rster Schreiber}, {Franx}, {Brammer}, {Momcheva}, {Wuyts}, {Whitaker}, {Skelton}, {Fumagalli}, {Hayward}, {Kriek}, {Labb{\'e}}, {Leja}, {Rix}, {Tacconi}, {van der Wel}, {van den Bosch}, {Oesch}, {Dickey}, \& {Ulf Lange}}]{Nelson2016}
{Nelson}, E.~J., {van Dokkum}, P.~G., {F{\"o}rster Schreiber}, N.~M., {et~al.} 2016, \apj, 828, 27, \dodoi{10.3847/0004-637X/828/1/27}

\bibitem[{{Oke} \& {Gunn}(1983)}]{Oke1983}
{Oke}, J.~B., \& {Gunn}, J.~E. 1983, \apj, 266, 713, \dodoi{10.1086/160817}

\bibitem[{{Olivier} {et~al.}(2022){Olivier}, {Berg}, {Chisholm}, {Erb}, {Pogge}, \& {Skillman}}]{Olivier2022}
{Olivier}, G.~M., {Berg}, D.~A., {Chisholm}, J., {et~al.} 2022, \apj, 938, 16, \dodoi{10.3847/1538-4357/ac8f2c}

\bibitem[{{Pacucci} {et~al.}(2026){Pacucci}, {Ferrara}, \& {Kocevski}}]{Pacucci_2026}
{Pacucci}, F., {Ferrara}, A., \& {Kocevski}, D.~D. 2026, arXiv e-prints, arXiv:2601.14368, \dodoi{10.48550/arXiv.2601.14368}

\bibitem[{pandas~development team(2020)}]{pandas1}
pandas~development team, T. 2020, pandas-dev/pandas: Pandas, 1.3.1,  Zenodo, \dodoi{10.5281/zenodo.3509134}

\bibitem[{{Pirie} {et~al.}(2024){Pirie}, {Best}, {Duncan}, {McLeod}, {Cochrane}, {Clausen}, {Dunlop}, {Flury}, {Geach}, {Hale}, {Ibar}, {Kondapally}, {Li}, {Matthee}, {McLure}, {Ossa-Fuentes}, {Patrick}, {Smail}, {Sobral}, {Stephenson}, {Stott}, \& {Swinbank}}]{Pirie2024}
{Pirie}, C.~A., {Best}, P.~N., {Duncan}, K.~J., {et~al.} 2024, arXiv e-prints, arXiv:2410.11808, \dodoi{10.48550/arXiv.2410.11808}

\bibitem[{{Planck Collaboration} {et~al.}(2020){Planck Collaboration}, {Aghanim}, {Akrami}, {Ashdown}, {Aumont}, {Baccigalupi}, {Ballardini}, {Banday}, {Barreiro}, {Bartolo}, {Basak}, {Battye}, {Benabed}, {Bernard}, {Bersanelli}, {Bielewicz}, {Bock}, {Bond}, {Borrill}, {Bouchet}, {Boulanger}, {Bucher}, {Burigana}, {Butler}, {Calabrese}, {Cardoso}, {Carron}, {Challinor}, {Chiang}, {Chluba}, {Colombo}, {Combet}, {Contreras}, {Crill}, {Cuttaia}, {de Bernardis}, {de Zotti}, {Delabrouille}, {Delouis}, {Di Valentino}, {Diego}, {Dor{\'e}}, {Douspis}, {Ducout}, {Dupac}, {Dusini}, {Efstathiou}, {Elsner}, {En{\ss}lin}, {Eriksen}, {Fantaye}, {Farhang}, {Fergusson}, {Fernandez-Cobos}, {Finelli}, {Forastieri}, {Frailis}, {Fraisse}, {Franceschi}, {Frolov}, {Galeotta}, {Galli}, {Ganga}, {G{\'e}nova-Santos}, {Gerbino}, {Ghosh}, {Gonz{\'a}lez-Nuevo}, {G{\'o}rski}, {Gratton}, {Gruppuso}, {Gudmundsson}, {Hamann}, {Handley}, {Hansen}, {Herranz}, {Hildebrandt}, {Hivon}, {Huang}, {Jaffe}, {Jones}, {Karakci}, {Keih{\"a}nen},
  {Keskitalo}, {Kiiveri}, {Kim}, {Kisner}, {Knox}, {Krachmalnicoff}, {Kunz}, {Kurki-Suonio}, {Lagache}, {Lamarre}, {Lasenby}, {Lattanzi}, {Lawrence}, {Le Jeune}, {Lemos}, {Lesgourgues}, {Levrier}, {Lewis}, {Liguori}, {Lilje}, {Lilley}, {Lindholm}, {L{\'o}pez-Caniego}, {Lubin}, {Ma}, {Mac{\'\i}as-P{\'e}rez}, {Maggio}, {Maino}, {Mandolesi}, {Mangilli}, {Marcos-Caballero}, {Maris}, {Martin}, {Martinelli}, {Mart{\'\i}nez-Gonz{\'a}lez}, {Matarrese}, {Mauri}, {McEwen}, {Meinhold}, {Melchiorri}, {Mennella}, {Migliaccio}, {Millea}, {Mitra}, {Miville-Desch{\^e}nes}, {Molinari}, {Montier}, {Morgante}, {Moss}, {Natoli}, {N{\o}rgaard-Nielsen}, {Pagano}, {Paoletti}, {Partridge}, {Patanchon}, {Peiris}, {Perrotta}, {Pettorino}, {Piacentini}, {Polastri}, {Polenta}, {Puget}, {Rachen}, {Reinecke}, {Remazeilles}, {Renzi}, {Rocha}, {Rosset}, {Roudier}, {Rubi{\~n}o-Mart{\'\i}n}, {Ruiz-Granados}, {Salvati}, {Sandri}, {Savelainen}, {Scott}, {Shellard}, {Sirignano}, {Sirri}, {Spencer}, {Sunyaev}, {Suur-Uski}, {Tauber}, {Tavagnacco},
  {Tenti}, {Toffolatti}, {Tomasi}, {Trombetti}, {Valenziano}, {Valiviita}, {Van Tent}, {Vibert}, {Vielva}, {Villa}, {Vittorio}, {Wandelt}, {Wehus}, {White}, {White}, {Zacchei}, \& {Zonca}}]{Planck2020}
{Planck Collaboration}, {Aghanim}, N., {Akrami}, Y., {et~al.} 2020, \aap, 641, A6, \dodoi{10.1051/0004-6361/201833910}

\bibitem[{{Reines} \& {Volonteri}(2015)}]{ReinesVolonteri2015}
{Reines}, A.~E., \& {Volonteri}, M. 2015, \apj, 813, 82, \dodoi{10.1088/0004-637X/813/2/82}

\bibitem[{{Robertson} {et~al.}(2010){Robertson}, {Ellis}, {Dunlop}, {McLure}, \& {Stark}}]{Robertson2010}
{Robertson}, B.~E., {Ellis}, R.~S., {Dunlop}, J.~S., {McLure}, R.~J., \& {Stark}, D.~P. 2010, \nat, 468, 49, \dodoi{10.1038/nature09527}

\bibitem[{{Sanders} {et~al.}(2024){Sanders}, {Shapley}, {Topping}, {Reddy}, \& {Brammer}}]{Sanders2024}
{Sanders}, R.~L., {Shapley}, A.~E., {Topping}, M.~W., {Reddy}, N.~A., \& {Brammer}, G.~B. 2024, \apj, 962, 24, \dodoi{10.3847/1538-4357/ad15fc}

\bibitem[{{Sanders} {et~al.}(2016){Sanders}, {Shapley}, {Kriek}, {Reddy}, {Freeman}, {Coil}, {Siana}, {Mobasher}, {Shivaei}, {Price}, \& {de Groot}}]{Sanders2016}
{Sanders}, R.~L., {Shapley}, A.~E., {Kriek}, M., {et~al.} 2016, \apj, 816, 23, \dodoi{10.3847/0004-637X/816/1/23}

\bibitem[{{Scholtz} {et~al.}(2025){Scholtz}, {Curti}, {D'Eugenio}, {{\"U}bler}, {Maiolino}, {Marconcini}, {Smit}, {Perna}, {Witstok}, {Arribas}, {B{\"o}ker}, {Bunker}, {Carniani}, {Charlot}, {Cresci}, {Lamperti}, {Parlanti}, {P{\'e}rez-Gonz{\'a}lez}, {Rodr{\'\i}guez Del Pino}, \& {Venturi}}]{Scholtz2025}
{Scholtz}, J., {Curti}, M., {D'Eugenio}, F., {et~al.} 2025, \mnras, 539, 2463, \dodoi{10.1093/mnras/staf518}

\bibitem[{{Shen} \& {Kelly}(2010)}]{Shen2010}
{Shen}, Y., \& {Kelly}, B.~C. 2010, \apj, 713, 41, \dodoi{10.1088/0004-637X/713/1/41}

\bibitem[{{Smith} {et~al.}(2025){Smith}, {Fries}, {Trump}, {Grier}, {Shen}, {Anderson}, {Brandt}, {Davis}, {Dwelly}, {Hall}, {Horne}, {Homayouni}, {McKaig}, {Morrison}, {Sharp}, {Assef}, {Bauer}, {Koekemoer}, {Schneider}, {Trakhtenbrot}, {Ibarra-Medel}, \& {Pe{\~n}aloza}}]{Smith2025}
{Smith}, T.~B., {Fries}, L.~B., {Trump}, J.~R., {et~al.} 2025, \apj, 995, 185, \dodoi{10.3847/1538-4357/ae1f18}

\bibitem[{{Sutherland} {et~al.}(2018){Sutherland}, {Dopita}, {Binette}, \& {Groves}}]{Sutherland2018}
{Sutherland}, R., {Dopita}, M., {Binette}, L., \& {Groves}, B. 2018, {MAPPINGS V: Astrophysical plasma modeling code}, Astrophysics Source Code Library, record ascl:1807.005.
\newblock \doeprint{1807.005}

\bibitem[{{Taylor} {et~al.}(2024){Taylor}, {Finkelstein}, {Kocevski}, {Jeon}, {Bromm}, {Amorin}, {Arrabal Haro}, {Backhaus}, {Bagley}, {Ba{\~n}ados}, {Bhatawdekar}, {Brooks}, {Calabro}, {Chavez Ortiz}, {Cheng}, {Cleri}, {Cole}, {Davis}, {Dickinson}, {Donnan}, {Dunlop}, {Ellis}, {Fernandez}, {Fontana}, {Fujimoto}, {Giavalisco}, {Grazian}, {Guo}, {Hathi}, {Holwerda}, {Hirschmann}, {Inayoshi}, {Kartaltepe}, {Khusanova}, {Koekemoer}, {Kokorev}, {Larson}, {Leung}, {Lucas}, {McLeod}, {Napolitano}, {Onoue}, {Pacucci}, {Papovich}, {P{\'e}rez-Gonz{\'a}lez}, {Pirzkal}, {Somerville}, {Trump}, {Wilkins}, {Yung}, \& {Zhang}}]{Taylor2024}
{Taylor}, A.~J., {Finkelstein}, S.~L., {Kocevski}, D.~D., {et~al.} 2024, arXiv e-prints, arXiv:2409.06772, \dodoi{10.48550/arXiv.2409.06772}

\bibitem[{{Taylor} {et~al.}(2025){Taylor}, {Kokorev}, {Kocevski}, {Akins}, {Cullen}, {Dickinson}, {Finkelstein}, {Arrabal Haro}, {Bromm}, {Giavalisco}, {Inayoshi}, {Juneau}, {Leung}, {P{\'e}rez-Gonz{\'a}lez}, {Somerville}, {Trump}, {Amor{\'\i}n}, {Barro}, {Burgarella}, {Brooks}, {Carnall}, {Casey}, {Cheng}, {Chisholm}, {Chworowsky}, {Davis}, {Donnan}, {Dunlop}, {Ellis}, {Fern{\'a}ndez}, {Fujimoto}, {Grogin}, {Gupta}, {Hathi}, {Jung}, {Hirschmann}, {Kartaltepe}, {Koekemoer}, {Larson}, {Leung}, {Llerena}, {Lucas}, {McLeod}, {McLure}, {Napolitano}, {Papovich}, {Stanton}, {Tripodi}, {Wang}, {Wilkins}, {Yung}, \& {Zavala}}]{Taylor2025}
{Taylor}, A.~J., {Kokorev}, V., {Kocevski}, D.~D., {et~al.} 2025, \apjl, 989, L7, \dodoi{10.3847/2041-8213/ade789}

\bibitem[{{Topping} {et~al.}(2024){Topping}, {Stark}, {Senchyna}, {Plat}, {Zitrin}, {Endsley}, {Charlot}, {Furtak}, {Maseda}, {Smit}, {Mainali}, {Chevallard}, {Molyneux}, \& {Rigby}}]{Topping2024}
{Topping}, M.~W., {Stark}, D.~P., {Senchyna}, P., {et~al.} 2024, \mnras, 529, 3301, \dodoi{10.1093/mnras/stae682}

\bibitem[{{Trump} {et~al.}(2023){Trump}, {Arrabal Haro}, {Simons}, {Backhaus}, {Amor{\'\i}n}, {Dickinson}, {Fern{\'a}ndez}, {Papovich}, {Nicholls}, {Kewley}, {Brunker}, {Salzer}, {Wilkins}, {Almaini}, {Bagley}, {Berg}, {Bhatawdekar}, {Bisigello}, {Buat}, {Burgarella}, {Calabr{\`o}}, {Casey}, {Ciesla}, {Cleri}, {Cole}, {Cooper}, {Cooray}, {Costantin}, {Croton}, {Ferguson}, {Finkelstein}, {Fujimoto}, {Gardner}, {Gawiser}, {Giavalisco}, {Grazian}, {Grogin}, {Hathi}, {Hirschmann}, {Holwerda}, {Huertas-Company}, {Hutchison}, {Jogee}, {Juneau}, {Jung}, {Kartaltepe}, {Kirkpatrick}, {Kocevski}, {Koekemoer}, {Lotz}, {Lucas}, {Magnelli}, {Matharu}, {P{\'e}rez-Gonz{\'a}lez}, {Pirzkal}, {Rafelski}, {Rose}, {Seill{\'e}}, {Somerville}, {Straughn}, {Tacchella}, {Vanderhoof}, {Weiner}, {Wuyts}, {Yung}, \& {Zavala}}]{Trump2023}
{Trump}, J.~R., {Arrabal Haro}, P., {Simons}, R.~C., {et~al.} 2023, \apj, 945, 35, \dodoi{10.3847/1538-4357/acba8a}

\bibitem[{{{\"U}bler} {et~al.}(2024){{\"U}bler}, {Maiolino}, {P{\'e}rez-Gonz{\'a}lez}, {D'Eugenio}, {Perna}, {Curti}, {Arribas}, {Bunker}, {Carniani}, {Charlot}, {Rodr{\'\i}guez Del Pino}, {Baker}, {B{\"o}ker}, {Cresci}, {Dunlop}, {Grogin}, {Jones}, {Kumari}, {Lamperti}, {Laporte}, {Marshall}, {Mazzolari}, {Parlanti}, {Rawle}, {Scholtz}, {Venturi}, \& {Witstok}}]{Ubler2024}
{{\"U}bler}, H., {Maiolino}, R., {P{\'e}rez-Gonz{\'a}lez}, P.~G., {et~al.} 2024, \mnras, 531, 355, \dodoi{10.1093/mnras/stae943}

\bibitem[{{van der Wel} {et~al.}(2011){van der Wel}, {Straughn}, {Rix}, {Finkelstein}, {Koekemoer}, {Weiner}, {Wuyts}, {Bell}, {Faber}, {Trump}, {Koo}, {Ferguson}, {Scarlata}, {Hathi}, {Dunlop}, {Newman}, {Dickinson}, {Jahnke}, {Salmon}, {de Mello}, {Kocevski}, {Lai}, {Grogin}, {Rodney}, {Guo}, {McGrath}, {Lee}, {Barro}, {Huang}, {Riess}, {Ashby}, \& {Willner}}]{vanderWel2011}
{van der Wel}, A., {Straughn}, A.~N., {Rix}, H.~W., {et~al.} 2011, \apj, 742, 111, \dodoi{10.1088/0004-637X/742/2/111}

\bibitem[{{Veilleux} \& {Osterbrock}(1987)}]{Veilleux1987}
{Veilleux}, S., \& {Osterbrock}, D.~E. 1987, \apjs, 63, 295, \dodoi{10.1086/191166}

\bibitem[{Virtanen {et~al.}(2020)Virtanen, Gommers, Oliphant, Haberland, Reddy, Cournapeau, Burovski, Peterson, Weckesser, Bright, {van der Walt}, Brett, Wilson, Millman, Mayorov, Nelson, Jones, Kern, Larson, Carey, Polat, Feng, Moore, {VanderPlas}, Laxalde, Perktold, Cimrman, Henriksen, Quintero, Harris, Archibald, Ribeiro, Pedregosa, {van Mulbregt}, \& {SciPy 1.0 Contributors}}]{scipy}
Virtanen, P., Gommers, R., Oliphant, T.~E., {et~al.} 2020, Nature Methods, 17, 261, \dodoi{10.1038/s41592-019-0686-2}

\bibitem[{{W}es {M}c{K}inney(2010)}]{pandas2}
{W}es {M}c{K}inney. 2010, in {P}roceedings of the 9th {P}ython in {S}cience {C}onference, ed. {S}t\'efan van~der {W}alt \& {J}arrod {M}illman, 56 -- 61, \dodoi{10.25080/Majora-92bf1922-00a}

\bibitem[{{Withers} {et~al.}(2023){Withers}, {Muzzin}, {Ravindranath}, {Sarrouh}, {Abraham}, {Asada}, {Brada{\v{c}}}, {Brammer}, {Desprez}, {Iyer}, {Martis}, {Mowla}, {Noirot}, {Sawicki}, {Strait}, \& {Willott}}]{Withers2023}
{Withers}, S., {Muzzin}, A., {Ravindranath}, S., {et~al.} 2023, \apjl, 958, L14, \dodoi{10.3847/2041-8213/ad01c0}

\bibitem[{{Yang} {et~al.}(2017){Yang}, {Malhotra}, {Rhoads}, \& {Wang}}]{Yang2017}
{Yang}, H., {Malhotra}, S., {Rhoads}, J.~E., \& {Wang}, J. 2017, \apj, 847, 38, \dodoi{10.3847/1538-4357/aa8809}

\end{thebibliography}

\section{Appendix}
\label{sec:app}

\begin{figure*}[h!]
    \centering
    \includegraphics[width=.4\linewidth]{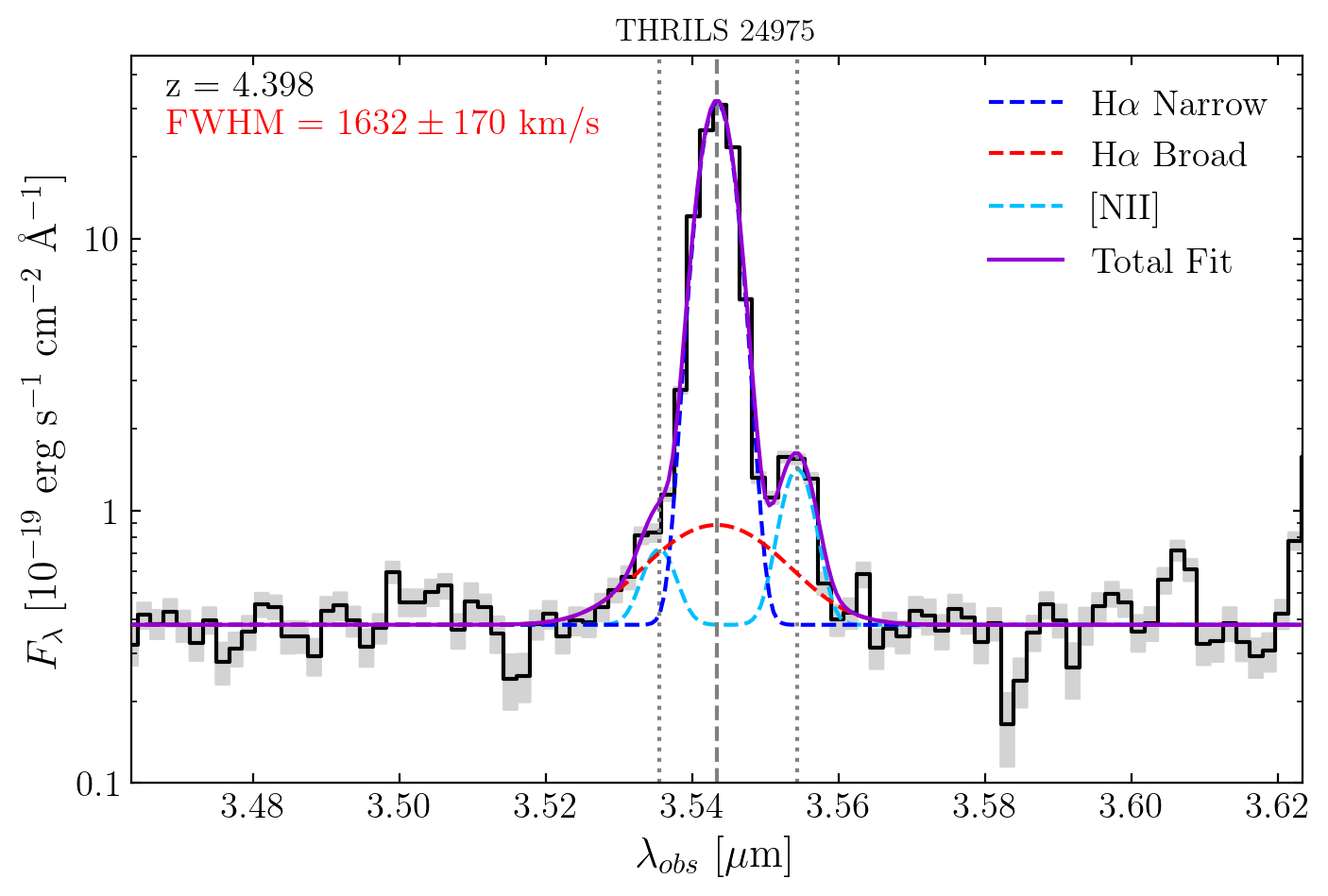} 
    \includegraphics[width=.4\linewidth]{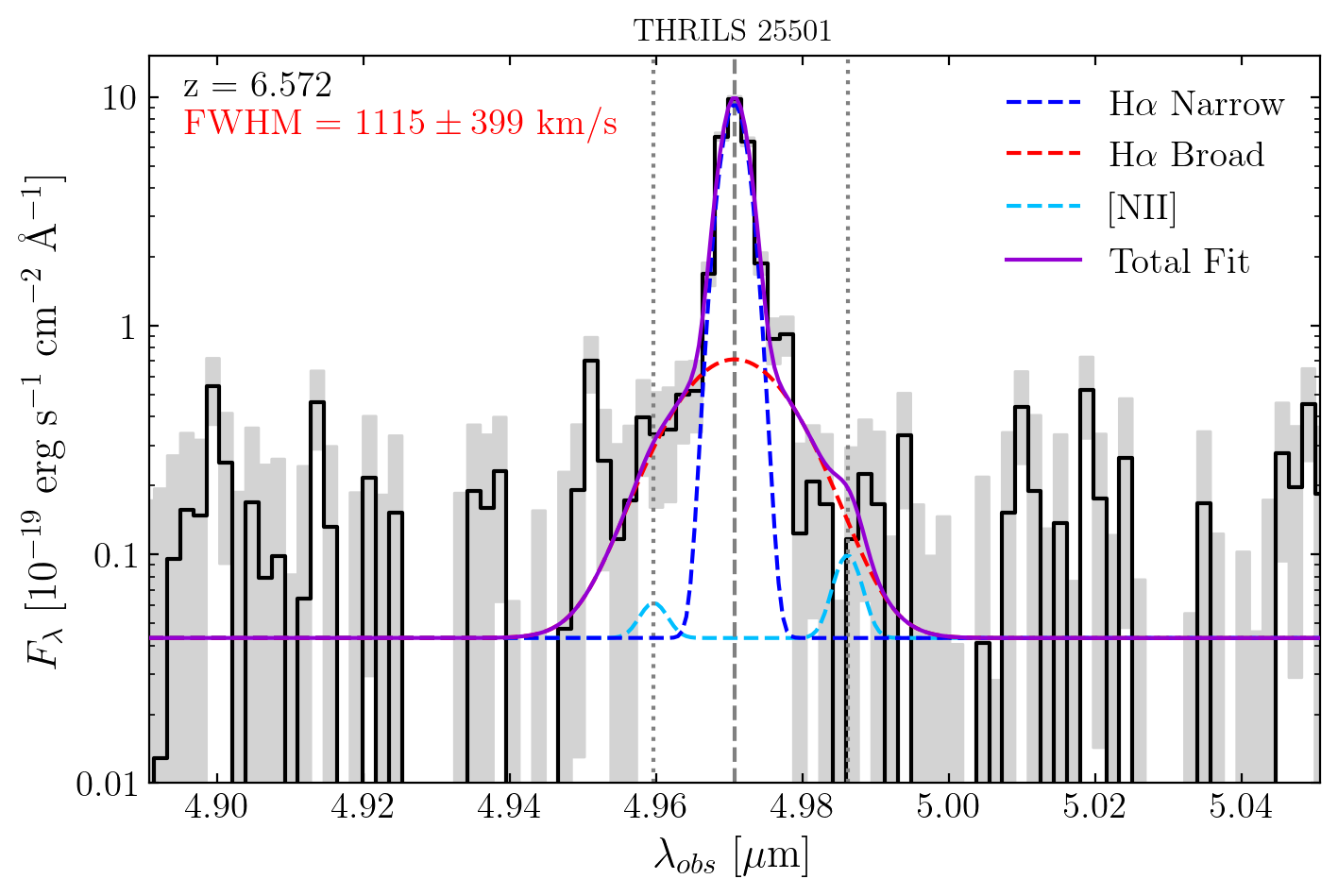}
    \includegraphics[width=.4\linewidth]{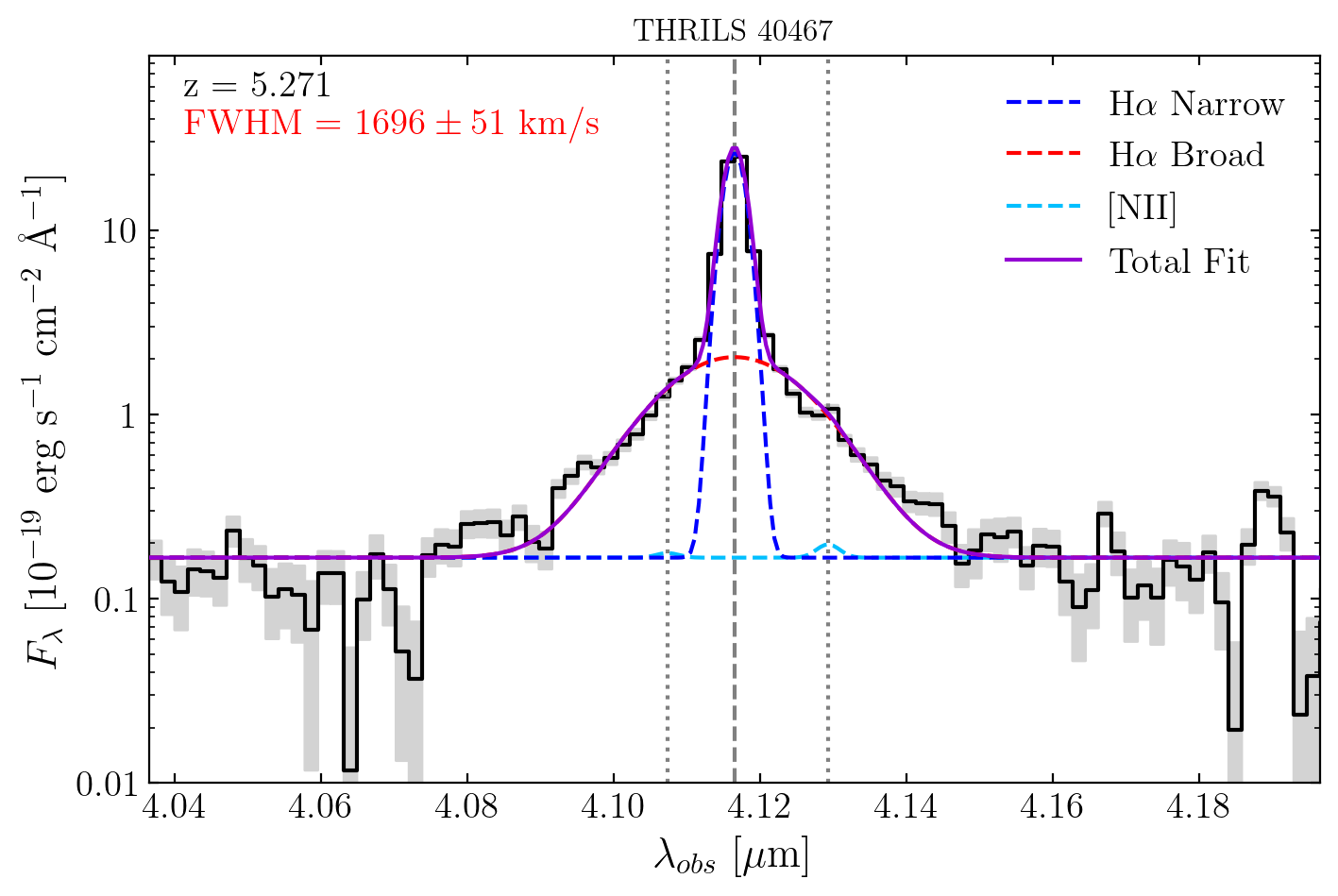}
    \includegraphics[width=.4\linewidth]{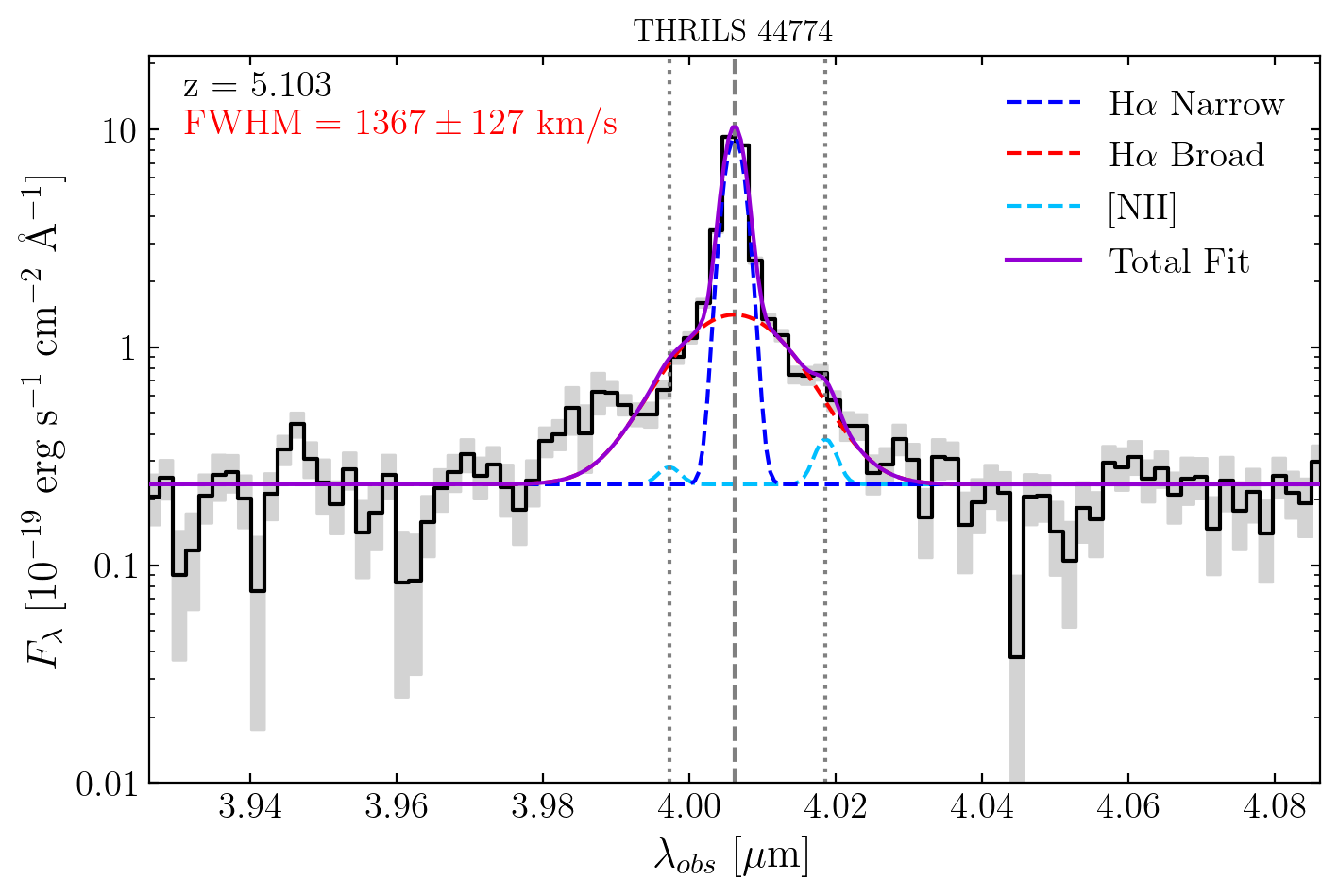}
    \includegraphics[width=.4\linewidth]{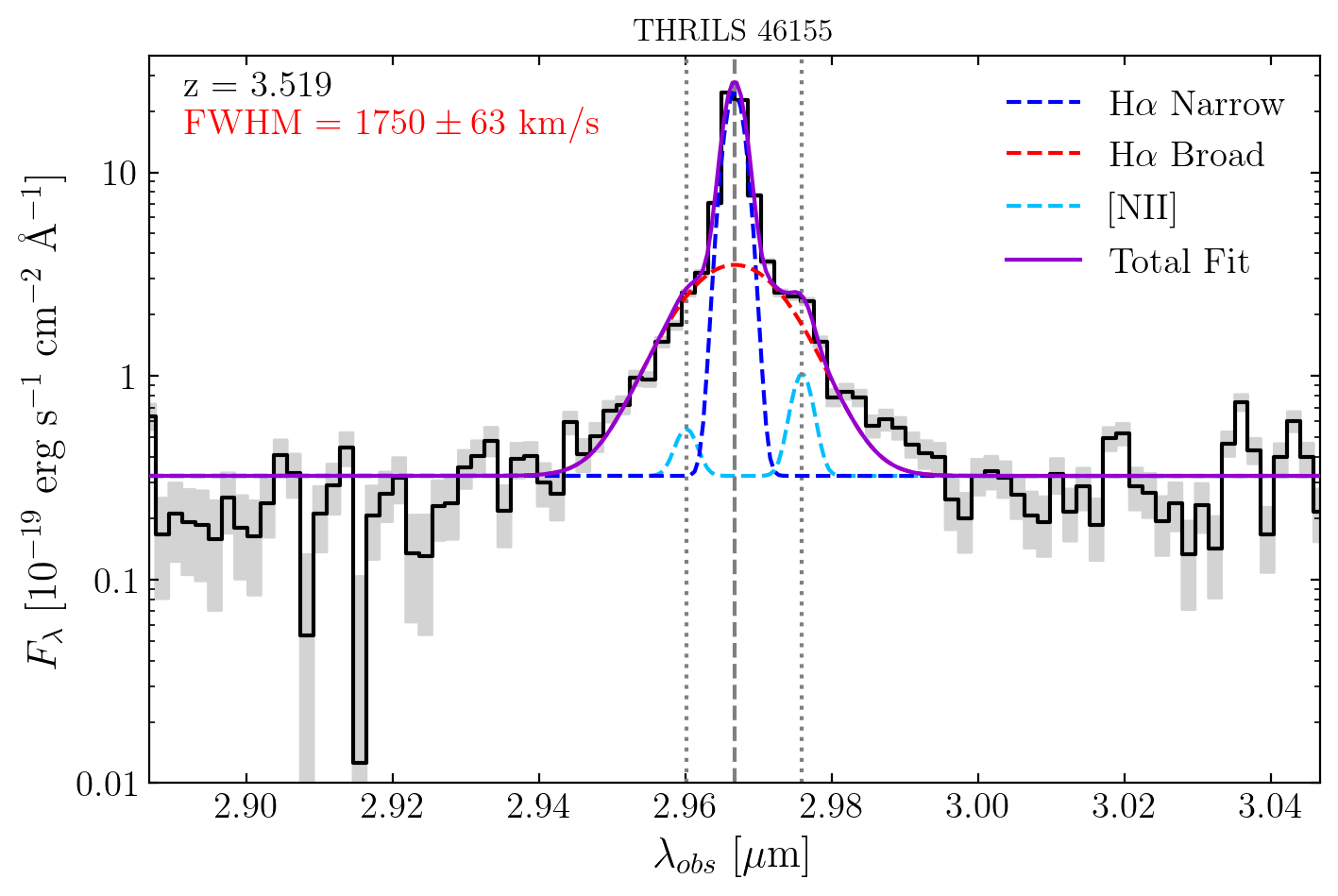}
    \includegraphics[width=.4\linewidth]{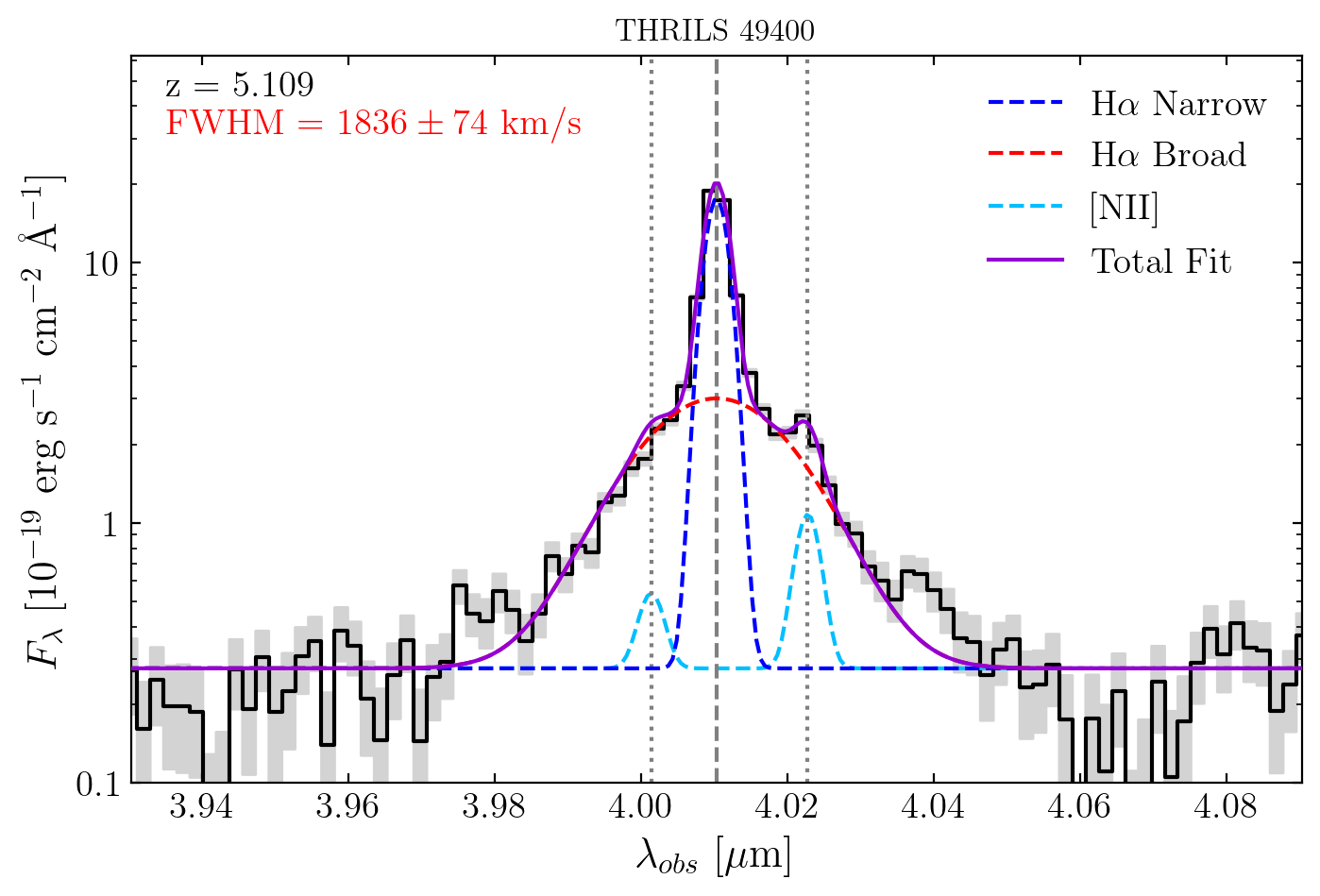}
    \includegraphics[width=.4\linewidth]{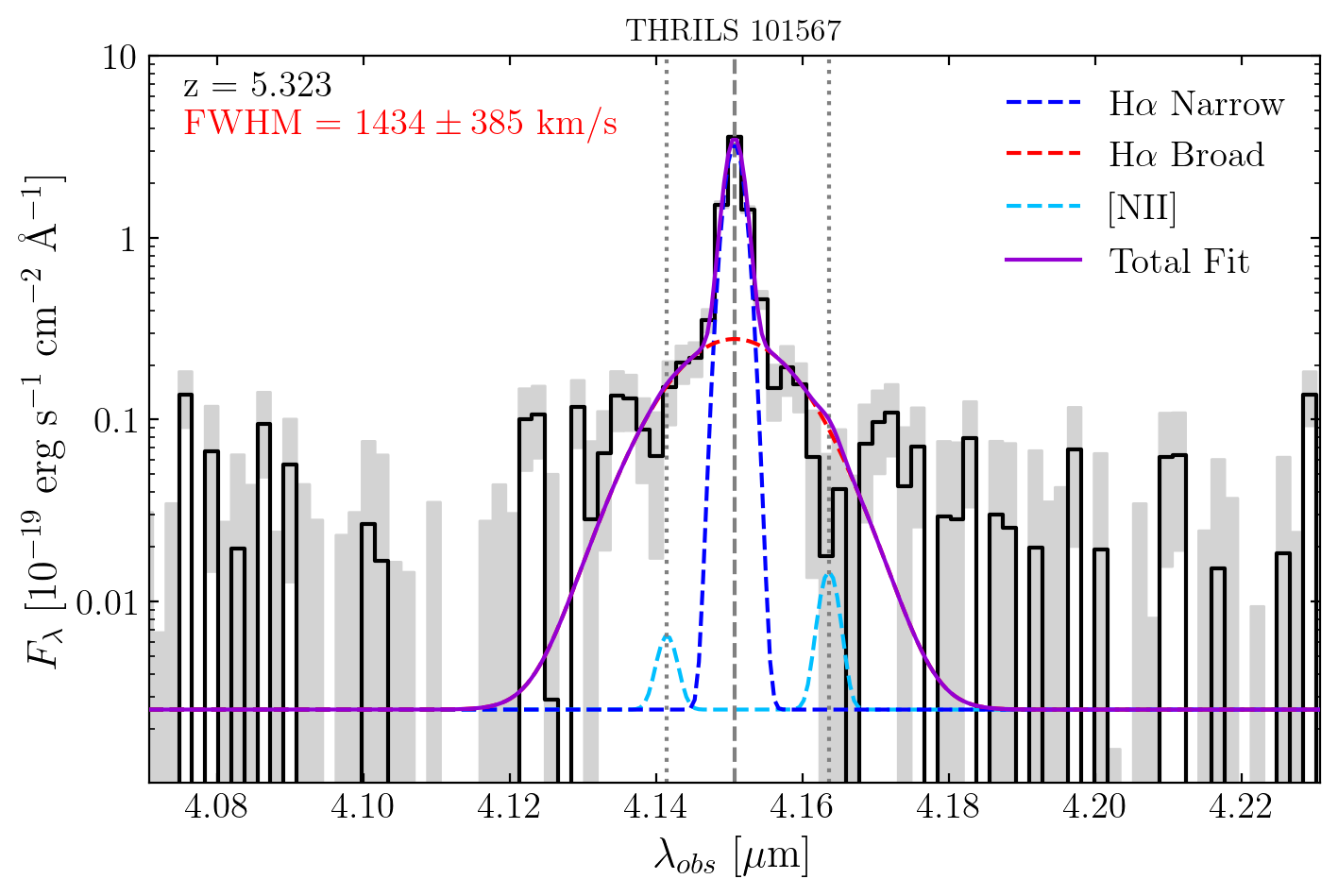}

    \caption{Broad line \Ha\ fits on log flux scales for the THRILS sample where we identify broad \Ha\ lines. We find 7/19, or 37\%, of THRILS EELG targets have broad \Ha\ components. They have a mean velocity width of $1,500~ \rm{km/s}$. 
    }
    \label{fig:thrils}
\end{figure*}

We identify 6 new BL AGN and confirm 1 (THRILS 46155) additional BL AGN in the deep NIRSpec G395M grating spectroscopy of EELGs from the THRILS program, presented in Figure \ref{fig:thrils}. Line fits for all AGN are presented in Table \ref{table:thrils}. We note only sources 40467, 25501, 101567, 49400, and 25501 pass our ``extreme'' line criteria as discussed in Section 5, but we present all BL AGN in THRILS data here for completeness. Masses are in agreement with those presented in Ganapathy et al (in prep).

\begin{table*}[h!]
\centering
\caption{Measurements for BL AGN in THRILS data}
\begin{adjustbox}{width=\textwidth}
\begin{tabular}{@{}ccccccc@{}}
\toprule
 $z_{\text{phot}}$ & $z_{\text{spec}}$ & THRILS ID & Broad \Ha\ Flux & Narrow \Ha\ Flux & CEERS Photometry ID & $\log(M_{\text{BH}})$  \\
 & & & [$\mathrm{erg } \, \mathrm{s}^{-1} \mathrm{cm}^{-2} $] & [$\mathrm{erg} \,\mathrm{s}^{-1} \mathrm{cm}^{-2} $] & & [$M_{\odot}$]  \\
 & & & ($\times 10^{-20}$) & ($\times 10^{-19}$) & &   \\
\midrule
 5.17 & 5.11 & 49400 & $2.73 \pm 0.16$ & $1.74 \pm 0.04$ & 51622 & $7.26 \pm 0.03$  \\
 7.48 & 5.33 & 101567 & $0.28 \pm 0.10$ & $0.33 \pm 0.02$ & 27384 & $6.55 \pm 0.17$ \\
 6.55 & 6.57 & 25501 & $0.69 \pm 0.40$ & $0.93 \pm 0.06$ & 16205 & $6.59 \pm 0.21$  \\
 5.29 & 5.27 & 40467 & $1.87 \pm 0.08$ & $2.62 \pm 0.03$ & 13539 & $7.12 \pm 0.02$  \\
 5.17 & 5.11 & 44774 & $1.18 \pm 0.15$ & $0.90 \pm 0.03$ & 29781 & $6.76 \pm 0.06$  \\
 3.58 & 3.52 & 46155 & $3.19 \pm 0.18$ & $2.45 \pm 0.04$ & 52773 & $7.00 \pm 0.02$  \\
     0.70 & 4.40 & 24975 & $0.51 \pm 0.12$ & $3.13 \pm 0.03$ & 16788 & $6.69 \pm 0.06$  \\
\bottomrule
\label{table:thrils}
\end{tabular}
\end{adjustbox}
\end{table*}

\begin{figure*}[h!]
    \includegraphics[width=1\linewidth]{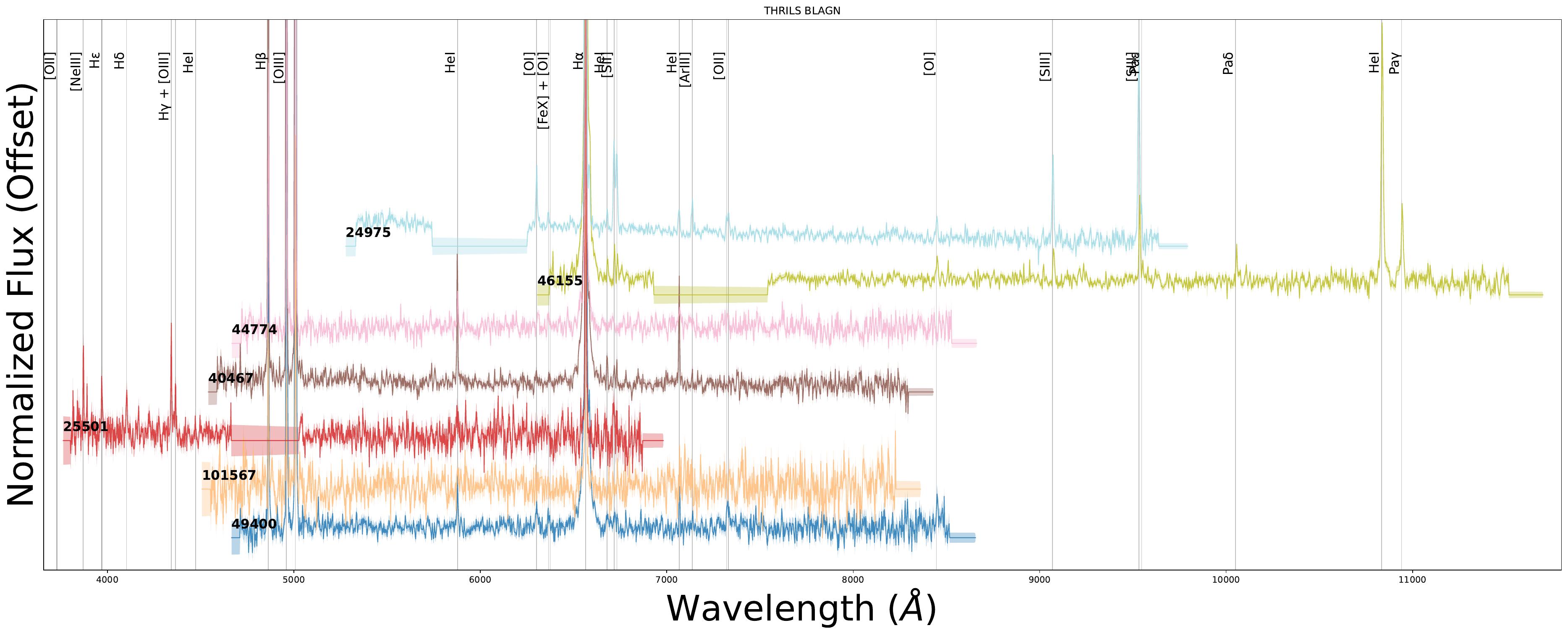}
    \caption{Full G395M spectra of the 7 THRILS BL AGN.}
\end{figure*}


\end{document}